\def\k{\kappa}

\def\m{\mu}
\def\n{\nu}

\newcommand{\kl}[3]{\mbox{$\rm #1$}^{\mu\nu , \alpha\beta}_{#2}(#3)}

\def\be{\begin{equation}}
\def\ee{\end{equation}}
\def\te{\end{equation}}
\def\bea{\begin{eqnarray}}
\def\nn{\nonumber}
\def\eea{\end{eqnarray}}
\def\tea{\end{eqnarray}}

\documentstyle[11pt]{article}

\makeatletter                    
\@addtoreset{equation}{section}  
\makeatother                     

\begin{document}

\title{Stochastic Gravity: Theory and Applications}

\author{B. L. Hu \\
         Maryland Center for Fundamental Physics, \\ Department of Physics,
         University of Maryland, \\ College Park, Maryland 20742-4111, U.S.A. \\
        e-mail:blhu@umd.edu\\
\\
E. Verdaguer \\
        Departament de F\'{\i}sica Fonamental \\
        and Institut de Ci\`encies del Cosmos,\\
Universitat de Barcelona,\\
             Av.~Diagonal 647, 08028 Barcelona, Spain\\
e-mail:enric.verdague@ub.edu \\
\\
}

\date{({\scriptsize February 5, 2008 update of the 2003 review in {\it Living Reviews in Relativity}})}
\maketitle

\begin{abstract}
Stochastic semiclassical gravity of the 90's is a theory naturally
evolved  from semiclassical gravity of the 80's and quantum field
theory in curved spacetimes of the 70's.  Whereas semiclassical
gravity is based on the semiclassical Einstein equation with sources
given by the expectation value of the stress-energy tensor of quantum
fields, stochastic semiclassical gravity is based on the
Einstein-Langevin equation, which has in addition sources due to the
noise kernel. The noise kernel is the vacuum expectation value of the
(operator-valued) stress-energy bi-tensor which describes the
fluctuations of quantum matter fields in curved spacetimes. A new
criterion for the validity of semiclassical gravity may also be
formulated from the viewpoint of this theory. In the first part, we
describe the fundamentals of this new theory via two approaches: the
axiomatic and the functional. The axiomatic approach is useful to see
the structure of the theory from the framework of semiclassical
gravity, showing the link from the mean value of the stress-energy
tensor to their correlation functions. The functional approach uses
the Feynman-Vernon influence functional and the Schwinger-Keldysh
closed-time-path effective action methods which are convenient for
computations. It also brings out the open systems concepts and the
statistical and stochastic contents of the theory such as
dissipation, fluctuations, noise and decoherence. We then focus on
the properties of the stress energy bi-tensor. We obtain a general
expression for the noise kernel of a quantum field defined at two
distinct points in an arbitrary curved spacetime as products of
covariant derivatives of the quantum field's Green function, and show
from this that the trace anomaly of the noise kernel is zero for
massless conformal scalar fields. In the second part, we describe
three applications of stochastic gravity theory. First, we consider
metric perturbations in a Minkowski spacetime. We offer an analytical
solution of the Einstein-Langevin equation and compute the two-point
correlation functions for the linearized Einstein tensor and for the
metric perturbations.
We also prove that Minkowski spacetime is a stable solution of
semiclassical gravity.
Second, we discuss structure formation from the stochastic gravity
viewpoint, which can go beyond the standard treatment by
incorporating the full quantum effect of the inflaton fluctuations.
Third, we discuss the backreaction of Hawking radiation in the
gravitational background of a quasi-static black hole (enclosed in a
box). We derive a fluctuation-dissipation relation between the
fluctuations in the radiation and the dissipative dynamics of metric
fluctuations. Finally we describe the behavior of metric fluctuations
near the event horizon of an evaporating black hole using the
Einstein-Langevin equation and point out directions for further
development.
\end{abstract}



\section{Overview}
\label{introduction}
Stochastic semiclassical gravity
is a theory developed in the Nineties using
semiclassical gravity (quantum fields in classical spacetimes, the
dynamics of both matter and spacetime are solved self-consistently)
as the starting point and aiming at a theory of quantum gravity as
the goal. While semiclassical gravity is based on the semiclassical
Einstein equation with the source given by the expectation value of
the stress-energy tensor of quantum fields, stochastic semiclassical
gravity, or stochastic gravity for short,
includes also its fluctuations in a new stochastic semiclassical
Einstein-Langevin equation (we
will often use the shortened term {\it stochastic gravity} as there
is no confusion as to the nature and source of stochasticity in
gravity being induced from the quantum fields and not a priori from
the classical spacetime). If the centerpiece in semiclassical
gravity theory is the vacuum expectation value of the stress-energy
tensor of a quantum field, and the central issues being how well the
vacuum is defined and how the divergences can be controlled by
regularization and renormalization, the centerpiece in stochastic
semiclassical gravity theory  is the stress-energy bi-tensor and its
expectation value known as the noise kernel. The mathematical
properties of this quantity and its physical content in relation to
the behavior of fluctuations of quantum fields in curved spacetimes
are the central issues of this new theory. How they induce metric
fluctuations and seed the structures of the universe, how they affect
the black hole horizons and the backreaction of Hawking radiance in
black hole dynamics, including implications on trans-Planckian
physics, are new horizons to explore. On the theoretical issues,
stochastic gravity is the necessary foundation to investigate the
validity of semiclassical gravity and the viability of inflationary
cosmology based on the appearance and sustenance of a vacuum
energy-dominated phase. It is also a useful beachhead supported by
well-established low energy (sub-Planckian) physics to explore the
connection with high energy (Planckian) physics in the realm of
quantum gravity.

In this review we summarize major work and results on this theory
since 1998. It is in the nature of a progress report rather than a
review. In fact we will have room only to discuss a handful of topics
of basic importance. A review of ideas leading to stochastic gravity
and further developments originating from it can be found in
Refs.~\cite{Physica,stogra}; a set of lectures which include a
discussion of the issue of the validity of semiclassical gravity in
Ref. ~\cite{HVErice}; a pedagogical introduction of stochastic
gravity theory with a more detailed treatment of backreaction
problems in cosmology and black holes in quasi-equilibrium in Ref.
~\cite{HuVer03a}. A comprehensive formal description of the
fundamentals is given in Refs.~\cite{MarVer99a,MarVer99} while that
of  the noise kernel in arbitrary spacetimes in
Refs.~\cite{MarVer99,PhiHu01,PhiHu03}. Here we will try to mention
related work so the reader can at least trace out the parallel and
sequential developments. The references at the end of each topic
below are representative work where one can seek out further
treatments.

Stochastic gravity theory is built on three pillars:  general
relativity, quantum fields and nonequilibrium statistical
mechanics. The first two uphold semiclassical gravity, the last
two span statistical field theory.  Strictly speaking one can
understand a great deal without appealing to statistical
mechanics, and we will try to do so here. But concepts such as
quantum open systems \cite{Dav76,LinWes90,Wei93} and techniques
such as the influence functional \cite{FeyVer63,FeyHib65} (which
is related to the
closed-time-path effective action \cite{Sch61,BakMah63,Kel64,ChoEtal85,SuEtal88,CalHu89,CooEtal94,%
DeW86,Jor86,CalHu87,Jor87,Paz90}) were a great help in our
understanding of the physical meaning of issues involved toward the
construction of this new theory, foremost because quantum
fluctuations and correlation have ascended the stage and in some new
paradigm become the focus. Quantum statistical field theory and the
statistical mechanics of quantum field theory
\cite{CalHu88,dch,cddn,CalHu00} also aided us in searching for the
connection with quantum gravity through the retrieval of correlations
and coherence. We show the scope of stochastic gravity as follows:

\begin{enumerate}
  \renewcommand{\labelenumi}{\bf \theenumi}
\item {\bf Ingredients:}
  \begin{enumerate}
  \item From General Relativity
\cite{MisThoWhe73,Wal84} to Semiclassical Gravity.
  \item Quantum Field Theory in Curved Spacetimes
\cite{BirDav82,Ful89,Wal94,GriMamMos94}:
    \begin{enumerate}
    \item Stress-energy tensor: Regularization and renormalization.
    \item Self-consistent solution: Backreaction problems in early universe and black holes \cite{LukSta74,Gri76,Har77,HuPar77,HuPar78,FisHarHu79,HarHu79,%
    HarHu80,Har81,And83,And84,HisLarAnd97,AndEtal06},
and analog gravity
\cite{BarLibVis05,BalEtal05,Sch07,SchEtal07,MaiSch07}.
    \item Effective action: Closed time path, initial value
          formulation \cite{Sch61,BakMah63,Kel64,ChoEtal85,SuEtal88,%
DeW86,CalHu87,Jor87,CalHu89,Paz90,CooEtal94}.
    \item  Equation of motion: Real and causal \cite{Jor86}.
    \end{enumerate}
  \item Nonequilibrium Statistical Mechanics (see \cite{CalHu08} and references therein) :
    \begin{enumerate}
    \item Open quantum systems \cite{Dav76,LinWes90,Wei93}.
    \item Influence Functional: Stochastic equations
\cite{FeyVer63,FeyHib65}.
    \item Noise and Decoherence: Quantum to
classical transition \cite{Zur81,Zur82,Zur86,Zur91,JooZeh85,CalLeg85,UnrZur89,%
Zur93,GiuEtal96,Gri84,Omn88a,Omn88b,Omn88c,Omn90,Omn92,Omn94,%
GelHar90,Har93,DowHal92,Hal93,Hal98,Bru93,PazZur93,Twa93,Ish94,%
IshLin94a,IshLin94b,Hal95,DowKen95,DowKen96,Ken96,Ken97,Ken98,IshLin98}.
    \end{enumerate}
  \item  Decoherence in Quantum Cosmology and
Emergence of Classical Spacetimes
\cite{Kie87,Hal89,Pad89,Hu90,Cal89,Cal91,HuPazSin93}.
  \end{enumerate}
\item {\bf Theory:}
  \begin{enumerate}
  \item Dissipation from  Particle Creation \cite{DeW86,Jor86,CalHu87,Jor87,Paz90,CamVer94};\\
     Backreaction as Fluctuation-Dissipation Relation (FDR) \cite{CanSci77,Mottola,HuSin95,CamHu98}.
  \item Noise from Fluctuations of Quantum Fields
\cite{Physica,Banff,CalHu94}.
  \item Einstein-Langevin
Equations
\cite{CalHu94,HuMat95,HuSin95,CamVer96,CamVer97,CalCamVer97,LomMaz97,%
MarVer99a,MarVer99,MarVer99b}.
  \item  Metric Fluctuations in Minkowski spacetime
\cite{MarVer00}.
  \end{enumerate}
\item {\bf Issues:}
  \begin{enumerate}
  \item Validity of Semiclassical Gravity
\cite{Hor80,Hor81,Jor87,Sim90,AndMolMot03,For82,KuoFor93,HuPhi00,%
PhiHu00,HuRouVer04a,HuRouVer04b}.
  \item Viability of Vacuum Dominance and Inflationary Cosmology.
  \item Stress-Energy Bitensor and Noise Kernel: Regularization Reassessed
\cite{PhiHu01,PhiHu03}.
  \end{enumerate}
\item {\bf Applications: Early Universe and Black Holes:}
  \begin{enumerate}
  \item Wave Propagation in Stochastic Geometry~\cite{HuShi98}.
  \item Black Hole Horizon Fluctuations: Spontaneous/Active
versus Induced/Passive
\cite{ForSva97,WuFor99,Sor96,SorSud99,BarFroPar99,BarFroPar00,%
MasPar00,Par01,PhiHu03}.
  \item Noise induced inflation~\cite{CalVer99}.
  \item Structure Formation \cite{CalHu95,Mat97a,Mat97b,CalGon97,RouVer07};\\
trace anomaly-driven inflation \cite{Sta80,Vil85,HawHerRea01}.
  \item Black Hole Backreaction and Fluctuations
\cite{CanSci77,Sci79,SciCanDeu81,Mottola,Vishu,CamHu98,CamHu99,SinRavHu03,%
HuRou06a,HuRou06b}.
  \end{enumerate}
\item {\bf Related Topics:}
  \begin{enumerate}
  \item  Metric Fluctuations and Trans-Planckian Problem
  \cite{BarFroPar99,BarFroPar00,MasPar00,Par01,NiePar01}.
  \item Spacetime Foam, Loop and Spin Foam \cite{Car97,Car98,Gar98a,Gar98b,Gar99,Ng03,ChrNgDam06,%
    BerKli07,NicPee06,FreKra07}.
  \item  Universal `Metric Conductance' Fluctuations  \cite{Shio}.
  \end{enumerate}
\item {\bf Ideas:}
  \begin{enumerate}
  \item General Relativity as Geometro-Hydrodynamics
\cite{grhydro,Hu05,Vol03,Jac95,EliJac06,Vol07,Her02};\\
Emergent Gravity \cite{Sei05,HorPol06,Wen05a,Wen06}.
  \item Semiclassical Gravity as Mesoscopic Physics \cite{meso,Hu07}.
  \item From Stochastic to Quantum Gravity:
    \begin{enumerate}
    \item Via Correlation hierarchy of interacting quantum fields
  \cite{stogra,dch,CalHu00,kinQG}.
    \item Possible relation to string theory and matrix theory.
    \item Other major approaches to quantum gravity \cite{Ori08}.
    \end{enumerate}
  \end{enumerate}
\end{enumerate}

{}For lack of space we list only the latest work in the respective
topics above describing ongoing research. The reader should consult
the references therein for earlier work and the background material.
We do not seek a complete coverage here, but will discuss only those
selected topics in theory, issues and applications. We use the
$(+,+,+)$ sign conventions of Refs.~\cite{MisThoWhe73,Wal84}, and
units in which $c=\hbar=1$.

\section{From Semiclassical to Stochastic Gravity }
\label{sec1}

There are three main steps that lead to the recent
development of stochastic gravity. The first step begins with
{\it{quantum field theory in curved spacetime}}
~\cite{DeW75,BirDav82,Ful89,Wal94,GriMamMos94}, which describes
the behavior of quantum matter fields propagating in a specified
(not dynamically determined by the quantum matter field as source)
background gravitational field. In this theory the gravitational
field is given by the classical spacetime metric determined from
classical sources by the classical Einstein equations, and the
quantum fields propagate as test fields in such a spacetime. An
important process described by quantum field theory in curved
spacetime is indeed particle creation from the vacuum, and effects
of vacuum fluctuations and polarizations, in the early universe
\cite{Par69,SexUrb69,Zel70,ZelSta71,Hu74,Ber74,Ber75a,Ber75b,%
DeW75,Fri89,CesVer90},
and Hawking radiation in black holes
\cite{Haw74,Haw75,Isr75,Par75,Wal75}.

The second step in the description of the interaction of gravity
with quantum fields is back-reaction, {\it i.e.}, the effect of
the quantum fields on the spacetime geometry. The source here is
the expectation value of the stress-energy operator for the matter
fields in some quantum state in the spacetime, a classical
observable. However, since this object is quadratic in the field
operators, which are only well defined as distributions on the
spacetime, it involves ill defined quantities. It contains
ultraviolet divergences the removal of which requires a
renormalization procedure \cite{DeW75,Chr76,Chr78}. The final
expectation value of the stress-energy operator using a reasonable
regularization technique is essentially unique, modulo some terms
which depend on the spacetime curvature and which are independent
of the quantum state. This uniqueness was proved by Wald
\cite{Wal77,Wal78} who investigated the criteria that a physically
meaningful expectation value of the  stress-energy tensor ought to
satisfy.

The theory obtained from a self-consistent solution of the geometry
of the spacetime and the quantum field is known as {\it semiclassical
gravity}. Incorporating the backreaction of the quantum matter field
on the spacetime is thus the central task in semiclassical gravity.
One assumes a general class of spacetime where the quantum fields
live in and act on, and seek a solution which satisfies
simultaneously the Einstein equation for the spacetime and the field
equations for the quantum fields. The Einstein equation which has the
expectation value of the stress-energy operator of the quantum matter
field as the source is known as the {\it semiclassical Einstein
equation}. Semiclassical gravity was first investigated in
cosmological backreaction problems
\cite{LukSta74,Gri76,Har77,HuPar77,HuPar78,FisHarHu79,HarHu79,HarHu80,%
Har81,And83,And84},
an example is the damping of anisotropy in Bianchi universes by the
backreaction of vacuum particle creation. Using the effect of quantum
field processes such as particle creation to explain why the universe
is so isotropic at the present was investigated in the context of
chaotic cosmology \cite{Mis69,BelKhaLif70,BelKhaLif82} in the late
seventies prior to the inflationary cosmology proposal of the
eighties \cite{Gut81,AlbSte82,Lin82,Lin85}, which assumes the vacuum
expectation value of an inflaton field as the source, another,
perhaps more well-known, example of semiclassical gravity.

\subsection{The importance of quantum fluctuations}

For a free quantum field semiclassical gravity is fairly well
understood. The theory is in some sense unique, since the only
reasonable c-number stress-energy tensor that one may construct
\cite{Wal77,Wal78} with the stress-energy operator is a
renormalized expectation value. However the scope and limitations
of the theory are not so well understood. It is expected that the
semiclassical theory would  break down at the Planck scale. One
can conceivably assume that it would also break down when the
fluctuations of the stress-energy operator are large
\cite{For82,KuoFor93}. Calculations of the fluctuations of the
energy density for Minkowski, Casimir and hot flat spaces as well
as Einstein and de Sitter universes are available
\cite{KuoFor93,PhiHu97,HuPhi00,PhiHu00,PhiHu01,PhiHu03,%
MarVer99,MarVer00,RouVer99a,RouVer07,OsbSho,CogGuiEli02}. It is less clear, however, how to
quantify what a large fluctuation is, and different criteria have
been proposed
\cite{KuoFor93,ForSCG,ForWu,HuPhi00,PhiHu00,AndMolMot02,AndMolMot03,YuFor99,YuFor00}.
The issue of the validity of the semiclassical gravity viewed in
the light of quantum fluctuations was discussed in our Erice
lectures \cite{HVErice}. More recently in Refs.~\cite{HuRouVer04a,HuRouVer04a} a new criterion has been proposed for the validity of semiclassical gravity.
It is based based on the quantum fluctuations of the semiclassical metric
and incorporates in a unified and self-consistent way previous criteria
that have been used \cite{Hor80,AndMolMot03,For82,KuoFor93}.
One can see the essence of the validity problem by
the following example inspired by Ford \cite{For82}.

Let us assume a quantum state formed by an isolated system which
consists of a superposition with equal amplitude of  one
configuration of mass $M$ with the center of mass at $X_1$ and
another configuration of the same mass with the center of mass at
$X_2$. The semiclassical theory as described by the semiclassical
Einstein equation predicts that the center of mass of the
gravitational field of the system is centered at $(X_1+X_2)/2$.
However, one would expect that if we send a succession of test
particles to probe the gravitational field of the above system
half of the time they would react to a gravitational field of
mass $M$ centered at $X_1$ and half of the time to the field
centered at $X_2$. The two predictions are clearly different,
note that the fluctuation in the position of the center of masses
is of the order of $(X_1-X_2)^2$. Although this example raises
the issue of how to place the importance of fluctuations to the
mean, a word of caution should be added to the effect that it
should not be taken too literally. In fact, if the previous
masses are macroscopic the quantum system decoheres very quickly
\cite{Zur91,Zur93} and instead of being described by a pure quantum
state it is described by a density matrix which diagonalizes in a
certain pointer basis. For observables associated to such a
pointer basis the density matrix description is equivalent to
that provided by a statistical ensemble. The results will differ,
in any case, from the semiclassical prediction.

In other words, one would expect that a stochastic source that
describes the quantum fluctuations should enter into the
semiclassical equations. A significant step in this direction was
made in Ref.~\cite{Physica} where it was proposed to view the
back-reaction problem in the framework of an open quantum system:
the quantum fields seen as the ``environment" and the
gravitational field as the ``system". Following this proposal a
systematic study of the connection between semiclassical gravity
and open quantum systems resulted in the development of a new
conceptual and technical framework where (semiclassical)
Einstein-Langevin equations were derived
\cite{CalHu94,HuMat95,HuSin95,CamVer96,CamVer97,CalCamVer97,LomMaz97}.
The key technical factor to most of these results was the use of
the influence functional method of Feynman and Vernon
\cite{FeyVer63} when only the coarse-grained effect of the
environment on the system is of interest. Note that the word
semiclassical put in parentheses refers to the fact that the noise
source in the Einstein-Langevin equation arises from the quantum
field, while the background spacetime is classical; generally we
will not carry this word since there is no confusion that the
source which contributes to the stochastic features of this theory
comes from quantum fields.

In the language of the consistent histories formulation of quantum
mechanics \cite{Gri84,Omn88a,Omn88b,Omn88c,Omn90,Omn92,Omn94,GelHar90,%
Har93,DowHal92,Hal93,Hal98,Bru93,PazZur93,Twa93,Ish94,IshLin94a,%
IshLin94b,Hal95,DowKen95,DowKen96,Ken96,Ken97,Ken98,IshLin98}
for the existence of a semiclassical
regime for the dynamics of the system one needs two requirements:
The first is decoherence, which guarantees that probabilities can
be consistently assigned to histories describing the evolution of
the system, and the second is that these probabilities should peak
near histories which correspond to solutions of classical
equations of motion. The effect of the environment is crucial, on
the one hand, to provide decoherence and, on the other hand, to
produce both dissipation and noise to the system through
back-reaction, thus inducing a semiclassical stochastic dynamics
on the system. As shown by different authors
\cite{GelHar93,Zur81,Zur82,Zur86,Zur91,JooZeh85,%
CalLeg85,UnrZur89,Zur93,GiuEtal96}, indeed over a long history
predating the current revival of decoherence, stochastic
semiclassical equations are obtained in an open quantum system
after a coarse graining of the environmental degrees of freedom
and a further coarse graining in the system variables. It is
expected but has not yet been shown that this mechanism could also
work for decoherence and classicalization of the metric field.
Thus far, the analogy could be made formally \cite{MarVer99b} or
under certain assumptions, such as adopting the Born-Oppenheimer
approximation in quantum cosmology \cite{Paz91,PazSin92}.

An alternative axiomatic approach to the Einstein-Langevin
equation without invoking the open system paradigm was later
suggested based on the formulation of a self-consistent dynamical
equation for a perturbative extension of semiclassical gravity
able to account for the lowest order stress-energy fluctuations of
matter fields \cite{MarVer99a}. It was shown that the same
equation could be derived, in this general case, from the
influence functional of Feynman and Vernon \cite{MarVer99}. The
field equation is deduced via an effective action which is
computed assuming that the gravitational field is a c-number. The
important new element in the derivation of the Einstein-Langevin
equation, and of the stochastic gravity theory, is the physical
observable that measures the stress-energy fluctuations, namely,
the expectation value of the symmetrized bi-tensor constructed
with the stress-energy tensor operator: the {\it noise kernel}. It
is interesting to note that the Einstein-Langevin equation can
also be understood as a useful intermediary tool to compute
symmetrized two-point correlations of the quantum metric
perturbations on the semiclassical background, independent of a
suitable classicalization mechanism \cite{RouVer03b}.

\section{The Einstein-Langevin equation: Axiomatic approach}
\label{sec2}

In this section we introduce {\it stochastic semiclassical
gravity}, or {\it stochastic gravity} for short,
in an axiomatic way. It is introduced as an extension of
semiclassical gravity motivated by the search of self-consistent
equations which describe the back-reaction of the quantum
stress-energy fluctuations on the gravitational field
\cite{MarVer99a}.

\subsection{Semiclassical gravity}

Semiclassical gravity describes the interaction of a classical
gravitational field with quantum matter fields. This theory can
be formally derived as the leading $1/N$ approximation of quantum
gravity interacting with $N$ independent and identical free
quantum fields \cite{HorWal80,HorWal82,HarHor81,Tom77}
which interact with gravity only.  By keeping the value of $NG$
finite, where $G$ is Newton's gravitational constant, one arrives
at a theory in which formally the gravitational field can be
treated as a c-number field (i.e. quantized at tree level) while
matter fields are fully quantized.
The semiclassical theory may be summarized as follows.

Let $({\cal M},g_{ab})$ be a globally hyperbolic four-dimensional
spacetime manifold ${\cal M}$ with metric $g_{ab}$ and consider a real
scalar quantum field $\phi$ of mass $m$ propagating on that manifold;
we just assume a scalar field for
simplicity.
The classical action $S_m$ for this matter field is given by
the functional
\begin{equation}
S_m[g,\phi]=-{1\over2}\int d^4x\sqrt{-g}\left[g^{ab}
\nabla_a\phi\nabla_b\phi+\left(m^2+\xi R\right)\phi^2\right],
\label{2.1}
\end{equation}
where $\nabla_a$ is the covariant derivative associated to the
metric $g_{ab}$, $\xi$ is a  coupling parameter between the field
and the scalar curvature of the underlying spacetime $R$, and
$g={\rm det} g_{ab}$.

The field may be quantized in the manifold using the standard
canonical quantization formalism \cite{BirDav82,Ful89,Wal94}. The
field operator in the Heisenberg representation $\hat\phi$ is an
operator-valued distribution solution of the Klein-Gordon
equation, the field equation derived from  Eq. (\ref{2.1}),
\begin{equation}
(\Box-m^2 -\xi R)\hat\phi=0.
\label{2.2}
\end{equation}
We may write the field operator as $\hat\phi[g;x)$
to indicate that it is a functional of the metric $g_{ab}$ and a function
of the spacetime point $x$. This notation will be used also for other
operators and tensors.

The classical stress-energy tensor is obtained by functional derivation
of this action in the usual way
$T^{ab}(x)=(2/\sqrt{-g})\delta S_m/\delta g_{ab}$, leading to
\begin{eqnarray}
T^{ab}[g,\phi]&=&\nabla^a\phi\nabla^b\phi-{1\over2}g^{ab}
\left(\nabla^c\phi\nabla_c\phi+m^2\phi^2\right)
\nonumber \\
&&+\xi\left(g^{ab}\Box-\nabla^a\nabla^b+G^{ab}
\right)\phi^2,
\label{2.3}
\end{eqnarray}
where $\Box=\nabla_a\nabla^a$ and $G_{ab}$ is the Einstein tensor.
With the notation $T^{ab}[g,\phi]$ we explicitly
indicate that the stress-energy tensor is
a functional of the metric $g_{ab}$  and the field $\phi$.

The next step is to define a stress-energy tensor operator $\hat
T^{ab}[g;x)$. Naively one would replace the classical field
$\phi[g;x)$ in the above functional by the quantum operator
$\hat\phi[g;x)$, but this procedure involves taking the product of
two distributions at the same spacetime point. This is ill-defined
and we need a regularization procedure. There are several
regularization methods which one may use, one is the
point-splitting or point-separation regularization method
\cite{Chr76,Chr78} in which one introduces a point $y$ in a
neighborhood of the point $x$ and then uses as the regulator the
vector tangent at the point $x$ of the geodesic joining $x$ and
$y$; this method is discussed for instance in Refs.
\cite{PhiHu00,PhiHu01,PhiHu03} and in section \ref{sec4}. Another
well known method is dimensional regularization in which one works
in arbitrary $n$ dimensions, where $n$ is not necessarily an
integer, and then uses as the regulator the parameter
$\epsilon=n-4$; this method is implicitly used in this section.
The regularized stress-energy operator using the Weyl ordering
prescription, {\it i.e.} symmetrical ordering, can be written as
\be \hat{T}^{ab}[g] = {1\over 2} \{
     \nabla^{a}\hat{\phi}[g]\, , \,
     \nabla^{b}\hat{\phi}[g] \}
     + {\cal D}^{ab}[g]\, \hat{\phi}^2[g],
\label{regul s-t 2} \ee where ${\cal D}^{ab}[g]$ is the
differential operator:
\begin{equation}
{\cal D}^{ab} \equiv
\left(\xi-1/4\right) g^{ab} \Box+ \xi \left( R^{ab}- \nabla^{a}
\nabla^{b}\right).
\label{diff operator}
\end{equation}
Note that if dimensional regularization is
used, the field operator $\hat \phi[g;x)$ propagates in a
$n$-dimensional spacetime. Once the regularization prescription
has been introduced a regularized and renormalized stress-energy
operator $\hat T^R_{ab}[g;x)$ may be defined as
\begin{equation}
\hat T^R_{ab}[g;x)= \hat T_{ab}[g;x)+F^C_{ab}[g;x)\hat I,
\label{2.4}
\end{equation}
which differs from the regularized $\hat T_{ab}[g;x)$ by the
identity operator times some tensor counterterms $F^C_{ab}[g;x)$,
which depend on the regulator and are local functionals of the
metric, see Ref.~\cite{MarVer99} for details. The field states can
be chosen in such a way that for any pair of physically acceptable
states, {\it i.e.}, Hadamard states in the sense of
Ref.~\cite{Wal94}, $|\psi\rangle$, and $|\varphi\rangle$ the
matrix element $\langle\psi|T^R_{ab}|\varphi\rangle$, defined as
the limit when the regulator takes the physical value, is finite
and satisfies Wald's axioms \cite{Ful89,Wal77}. These counterterms
can be extracted from the singular part of a Schwinger-DeWitt
series \cite{Ful89,Chr76,Chr78,Bun79}. The choice of these
counterterms is not unique but this ambiguity can be absorbed into
the renormalized coupling constants which appear in the equations
of motion for the gravitational field.

The {\it semiclassical Einstein equation} for the metric $g_{ab}$
can then be written as
\begin{equation}
G_{ab}[g]+\Lambda g_{ab}
-2(\alpha A_{ab}+\beta B_{ab})[g]=
8\pi G \langle \hat T_{ab}^R[g]\rangle ,
\label{2.5}
\end{equation}
where $\langle \hat T_{ab}^R[g]\rangle $ is the expectation value
of the operator $\hat T_{ab}^R[g,x)$ after the regulator takes
the physical value in some physically acceptable state of the field on
$({\cal M},g_{ab})$. Note that both the stress tensor and the
quantum state are functionals of the metric, hence the notation.
The parameters $G$, $\Lambda$, $\alpha$ and $\beta$ are the
renormalized coupling constants, respectively, the gravitational
constant, the cosmological constant and two dimensionless  coupling
constants which are zero in the classical Einstein equation.
These constants must be understood as the result of  ``dressing''
the bare constants which appear in the classical action
before renormalization. The values of these constants must be
determined by experiment.
The left hand side of Eq. (\ref{2.5}) may be derived
from the gravitational action
\begin{equation}
S_g[g]= {1\over 8\pi G}\int d^4 x \sqrt{-g}\left[ {1\over 2}
R-\Lambda +\alpha C_{abcd}C^{abcd}
+\beta R^2\right],
\label{2.6}
\end{equation}
where $C_{abcd}$ is the Weyl tensor. The tensors
$A_{ab}$ and $B_{ab}$ come from the functional
derivatives with respect to the metric of the terms quadratic
in the curvature in Eq. (\ref{2.6}), they are explicitly given by
\begin{eqnarray}
A^{ab}&=&\frac{1}{\sqrt{-g}}\frac{\delta}{\delta g_{ab}}
\int d^4 \sqrt{-g} C_{cdef}C^{cdef}
\nonumber\\
&=&{1\over2}g^{ab}C_{cdef}
C^{cdef}-2R^{acde}
R^{b}_{\ cde}+4R^{ac}R_c^{\ b}
-{2\over3}RR^{ab}
\nonumber\\
&& -2\Box R^{ab}+{2\over3}\nabla^a\nabla^b R+
{1\over3}g^{ab}\Box R,
\label{2.7a}\\
B^{ab}&=&\frac{1}{\sqrt{-g}}\frac{\delta}{\delta g_{ab}}
\int d^4 \sqrt{-g} R^2
\nonumber\\
&=&{1\over2}g^{ab}R^2-2R R^{ab}
+2\nabla^a\nabla^b R-2g^{ab}\Box R,
\label{2.7b}
\end{eqnarray}
where $R_{abcd}$  and $R_{ab}$ are the Riemann and Ricci tensors,
respectively. These two tensors are, like the Einstein and metric
tensors, symmetric and divergenceless: $\nabla^a A_{ab}=0=\nabla^a
B_{ab}$.

A solution of semiclassical gravity consists of a spacetime
(${\cal M},g_{ab}$), a quantum field operator $\hat\phi[g]$
which satisfies the evolution equation (\ref{2.2}), and a physically
acceptable state  $|\psi[g]\rangle $ for this field, such that Eq.
(\ref{2.5}) is satisfied when the expectation value of the renormalized
stress-energy operator is evaluated in this state.

For a free quantum field  this theory is robust in the sense that
it is self-consistent and fairly well understood.
As long as the gravitational field is assumed to be described by a
classical metric, the above semiclassical Einstein
equations seems to be the only plausible dynamical equation
for this metric: the metric couples to matter fields via the
stress-energy tensor and for a given quantum state the only
physically observable c-number  stress-energy tensor that one
can construct is the above renormalized expectation value.
However, lacking a full quantum gravity theory the scope and
limits of the theory are not so well understood. It is assumed
that the semiclassical theory should break down at Planck scales,
which is when simple order of magnitude estimates suggest that
the quantum effects of gravity should not be ignored because the
energy of a quantum fluctuation in
a Planck size region, as determined by the Heisenberg uncertainty
principle, is comparable to the gravitational energy of
that fluctuation.

The theory is expected to break down when the fluctuations of the
stress-energy operator are large \cite{For82}. A criterion based
on the ratio of the fluctuations to the mean was proposed by Kuo
and Ford \cite{KuoFor93} (see also work via zeta-function methods
\cite{PhiHu97,CogGuiEli02}).  This proposal was questioned by Phillips and Hu
\cite{HuPhi00,PhiHu00,PhiHu01} because it does not contain a scale
at which the theory is probed or how accurately the theory can be
resolved. They suggested the use of a smearing scale or
point-separation distance, for integrating over the bi-tensor
quantities, equivalent to a stipulation of the resolution level of
measurements; see also the response by Ford \cite{ForSCG,ForWu}. A
different criterion is recently suggested by Anderson et al.
\cite{AndMolMot02,AndMolMot03} based on linear response theory. A
partial summary of this issue can be found in our Erice Lectures
\cite{HVErice}.

More recently, in collaboration with A. Roura
\cite{HuRouVer04a,HuRouVer04b},
we have proposed a criterion for the validity of semiclassical gravity
which is based on the stability of the solutions of the semiclassical
Einstein equations with respect to quantum metric fluctuations. The
two-point correlations for the metric perturbations can be described
in the framework of stochastic gravity, which is closely related to
the quantum theory of gravity interacting with $N$ matter fields,
to leading order in a $1/N$ expansion.
We will describe these developments in the following sections.

\subsection{Stochastic gravity}

The purpose of stochastic gravity is to extend the
semiclassical theory to account for these fluctuations in a
self-consistent way. A physical observable that describes these
fluctuations to lowest order is the {\it noise kernel} bi-tensor,
which is defined through the two-point correlation of the
stress-energy operator as
\begin{equation}
N_{abcd}[g;x,y)={1\over2}\langle\{\hat t_{ab}[g;x),
\hat t_{cd}[g;y)\}\rangle,
\label{2.8}
\end{equation}
where the curly brackets mean anticommutator, and where
\begin{equation}
\hat t_{ab}[g;x)
\equiv \hat T_{ab}[g;x)-\langle \hat T_{ab}[g;x)\rangle.
\label{2.9}
\end{equation}
This bi-tensor can also be written $N_{ab,c^\prime
d^\prime}[g;x,y)$, or $N_{ab,c^\prime d^\prime}(x,y)$ as we do in
section \ref{sec4}, to emphasize that it is a tensor with respect
to the first two indices at the point $x$ and a tensor with
respect to the last two indices at the point $y$, but we shall not
follow this notation here. The noise kernel is defined in terms of
the unrenormalized stress-tensor operator $\hat T_{ab}[g;x)$ on a
given background metric $g_{ab}$, thus a regulator is implicitly
assumed on the right-hand side of Eq. (\ref{2.8}).  However, for a
linear quantum field the above kernel -- the expectation function
of a bi-tensor -- is free of ultraviolet divergences because the
regularized $T_{ab}[g;x)$ differs from the renormalized
$T_{ab}^R[g;x)$ by the identity operator times some tensor
counterterms, see Eq.~(\ref{2.4}), so that in the subtraction
(\ref{2.9}) the counterterms cancel. Consequently the ultraviolet
behavior of $\langle\hat T_{ab}(x)\hat T_{cd}(y)\rangle$ is the
same as that of $\langle\hat T_{ab}(x)\rangle \langle\hat
T_{cd}(y)\rangle$, and $\hat T_{ab}$ can be replaced by the
renormalized operator $\hat T_{ab}^R$ in Eq. (\ref{2.8}); an
alternative proof of this result is given in Ref.
\cite{PhiHu01,PhiHu03}. The noise kernel should be thought of as a
distribution function, the limit of coincidence points has meaning
only in the sense of distributions. The bi-tensor
$N_{abcd}[g;x,y)$, or $N_{abcd}(x,y)$ for short, is real and
positive semi-definite, as a consequence of $\hat T_{ab}^R$ being
self-adjoint. A simple proof is given in Ref. \cite{HuVer03a}.

Once the fluctuations of the stress-energy operator have been
characterized we can  perturbatively extend the semiclassical
theory to account for such fluctuations. Thus we will assume that
the background spacetime metric $g_{ab}$ is a solution of the
semiclassical Einstein Eqs.~(\ref{2.5}) and we will write the new
metric for the extended theory as $g_{ab}+h_{ab}$, where we will
assume that $h_{ab}$ is a perturbation to the background solution.
The renormalized stress-energy operator and the state of the
quantum field may now be denoted by $\hat T_{ab}^R[g+h]$ and
$|\psi[g+h]\rangle$, respectively, and $\langle\hat
T_{ab}^R[g+h]\rangle$ will be the corresponding expectation value.

Let us now introduce a Gaussian stochastic tensor field
$\xi_{ab}[g;x)$ defined by the following correlators:
\begin{equation}
\langle\xi_{ab}[g;x)\rangle_s=0,\ \ \
\langle\xi_{ab}[g;x)\xi_{cd}[g;y)\rangle_s=
N_{abcd}[g;x,y),
\label{2.10}
\end{equation}
where $\langle\dots\rangle_s$ means statistical average. The
symmetry and positive semi-definite property of the noise kernel
guarantees that the stochastic field tensor $\xi_{ab}[g,x)$, or
$\xi_{ab}(x)$ for short, just introduced is well defined. Note
that this stochastic tensor captures only partially the quantum
nature of the fluctuations of the stress-energy operator since it
assumes that cumulants of higher order are zero.

An important property of this stochastic tensor is that it is
covariantly conserved in the background spacetime
$\nabla^a\xi_{ab}[g;x)=0$. In fact, as a consequence of the
conservation of $\hat T_{ab}^R[g]$ one can see that $\nabla_x^a
N_{abcd}(x,y)=0$. Taking the divergence in Eq.~(\ref{2.10}) one
can then show that $\langle\nabla^a\xi_{ab}\rangle_s=0$ and
$\langle\nabla_x^a\xi_{ab}(x) \xi_{cd}(y)\rangle_s=0$ so that
$\nabla^a\xi_{ab}$ is deterministic and represents with certainty
the zero vector field in $\cal{M}$.

For a conformal field, {\it i.e.}, a field whose classical action
is conformally invariant, $\xi_{ab}$ is traceless:
$g^{ab}\xi_{ab}[g;x)=0$; so that, for a conformal matter field the
stochastic source gives no correction to the trace anomaly. In
fact, from the trace anomaly result which states that $g^{ab}\hat
T^R_{ab}[g]$ is, in this case, a local c-number functional of
$g_{ab}$ times the identity operator, we have that
$g^{ab}(x)N_{abcd}[g;x,y)=0$. It then follows from Eq.
(\ref{2.10}) that $\langle g^{ab}\xi_{ab}\rangle_s=0$ and $\langle
g^{ab}(x)\xi_{ab}(x) \xi_{cd}(y)\rangle_s=0$; an alternative proof
based on the point-separation method is given in Ref.
\cite{PhiHu01,PhiHu03}, see also section \ref{sec4}.

All these properties make it quite natural to incorporate into the
Einstein equations the stress-energy fluctuations by using the
stochastic tensor $\xi_{ab}[g;x)$ as the source
of the metric perturbations.
Thus we will write the following equation.
\begin{equation}
G_{ab}[g\!+\!h]\!+\! \Lambda (g_{ab}\!+\!h_{ab}) -2(\alpha
A_{ab}+\beta B_{ab})[g\!+\!h]\!=\!8\pi G\!\left( \!\langle \hat
T_{ab}^R[g\!+\!h]\rangle \!+ \!\xi_{ab}[g]\!\right). \label{2.11}
\end{equation}
This equation is in the form of a {\it (semiclassical)
Einstein-Langevin equation}, it is a dynamical equation for the
metric perturbation $h_{ab}$ to linear order. It describes the
back-reaction of the metric to the quantum fluctuations of the
stress-energy tensor of matter fields, and gives a first order
extension to semiclassical gravity as described by the
semiclassical Einstein equation (\ref{2.5}).

Note that we refer to the  Einstein-Langevin equation as a first
order extension to semiclassical Einstein equation of
semiclassical gravity and the lowest level representation of
stochastic gravity. However, stochastic gravity has a much broader
meaning, it refers to the range of theories based on second and
higher order correlation functions. Noise can be defined in
effectively open systems (e.g. correlation noise \cite{CalHu00} in
the Schwinger-Dyson equation hierarchy) to some degree but one
should not expect the Langevin form to prevail. In this sense we
say stochastic gravity is the intermediate theory between
semiclassical gravity (a mean field theory based on the
expectation values of the energy momentum tensor of quantum
fields) and quantum gravity (the full hierarchy of correlation
functions retaining complete quantum coherence
\cite{stogra,kinQG}.

The renormalization of the operator $\hat T_{ab}[g+h]$ is carried
out exactly as in the previous case, now in the perturbed metric
$g_{ab}+h_{ab}$. Note that the stochastic source $\xi_{ab}[g;x)$
is not dynamical, it is independent of $h_{ab}$ since it describes
the fluctuations of the stress tensor on the semiclassical
background $g_{ab}$.

An important property of the Einstein-Langevin equation is that it is
gauge invariant under the change of $h_{ab}$ by
$h_{ab}^\prime =h_{ab} +\nabla_a\zeta_b+\nabla_b\zeta_a$, where
$\zeta^a$ is a stochastic vector field on the background manifold ${\cal
M}$. Note that a tensor such as
$R_{ab}[g+h]$, transforms as
$R_{ab}[g+h^\prime]=R_{ab}[g+h]+{\cal L}_\zeta R_{ab}[g]$ to linear order
in the perturbations, where ${\cal L}_\zeta $ is the Lie derivative with
respect to $\zeta^a$. Now, let us write the source tensors in
Eqs.~(\ref{2.11}) and (\ref{2.5}) to the left-hand sides of these
equations. If we substitute  $h$ by $h^\prime$ in this new version of Eq.
(\ref{2.11}), we get the same expression, with $h$ instead of $h^\prime$,
plus the Lie derivative of the combination of tensors which appear on
the left-hand side of the new Eq. (\ref{2.5}). This last combination
vanishes when Eq. (\ref{2.5}) is satisfied, {\it i.e.}, when the
background metric $g_{ab}$ is a solution of semiclassical gravity.

{}From the statistical average
of equation (\ref{2.11}) we have that $g_{ab}+\langle h_{ab}\rangle_s$
must be a solution of the semiclassical Einstein equation linearized
around the background  $g_{ab}$; this solution has been proposed as a test
for the validity of the semiclassical approximation
\cite{AndMolMot02,AndMolMot03} a point that will be
further discussed in section \ref{s3.3}.

The stochastic equation (\ref{2.11}) predicts that the
gravitational field has stochastic fluctuations over the
background $g_{ab}$. This equation is linear in $h_{ab}$, thus its solutions
can be written as follows,
\begin{equation}
h_{ab}(x)=h_{ab}^0(x)+8\pi G\int d^4x^\prime
\sqrt{-g(x^\prime)}G_{abcd}^{\mathrm{ret}}(x,x^\prime)\xi^{cd}(x^\prime),
\label{1.5b}
\end{equation}
where $h_{ab}^{0}(x)$ is the solution of the homogeneous equation
containing information on the initial conditions and
$G_{abcd}^{\mathrm{ret}}(x,x^\prime)$ is the retarded propagator of
equation (\ref{2.11}) with vanishing initial conditions. Form this
we obtain the two-point correlation functions for the metric perturbations:
\begin{eqnarray}
\langle h_{ab}(x)h_{cd}(y)\rangle_s &=& \langle
h_{ab}^0(x)h_{cd}^0(y)\rangle_s +\nonumber\\
&&(8\pi G)^2\int d^4x^\prime
d^4y^\prime \sqrt{g(x^\prime)g(y^\prime)}
G_{abef}^{\mathrm{ret}}(x,x^\prime) N^{efgh}(x^\prime,y^\prime)
G_{cdgh}^{\mathrm{ret}}(y,y^\prime)\nonumber\\
&\equiv& \langle h_{ab}(x)h_{cd}(y)\rangle_{\mathrm{int}}+
\langle h_{ab}(x)h_{cd}(y)\rangle_{\mathrm{ind}}.
\label{1.5c}
\end{eqnarray}
There are two different contributions to the two-point
correlations, which we have distinguished in the second equality.
The first one is connected to the fluctuations of
the initial state of the metric perturbations, and we will refer to them as
\textit{intrinsic fluctuations}. The second contribution is
proportional to the noise kernel and is thus connected with the
fluctuations of the quantum fields, we will refer to them as
\textit{induced fluctuations}. To find these two-point stochastic
correlation functions one needs to know
the noise kernel $N_{abcd}(x,y)$.
Explicit expressions of this kernel in terms of the
two-point Wightman functions is given in \cite{MarVer99},
expressions based on point-splitting methods have also been given
in \cite{RouVer00,PhiHu01}.
 Note that
the noise kernel should be thought of as a distribution function,
the limit of coincidence points has meaning only in the sense of
distributions.

The two-point stochastic correlation functions for the metric perturbations
of Eq.~(\ref{1.5c}) satisfy
a very important property. In fact, it can be shown that they
correspond exactly to the symmetrized two-point
correlation functions for the quantum metric
perturbations in the large $N$ expansion, \emph{i.e.}
the quantum theory describing the interaction of the gravitational field with $N$ arbitrary free fields and expanded in powers of $1/N$. To leading order for the
graviton propagator one finds that
\begin{equation}
\langle\{\hat h_{ab}(x),\hat h_{cd}(y)\}\rangle =2\,
\langle h_{ab}(x)h_{cd}(y)\rangle_s, \label{1.5d}
\end{equation}
where $\hat h_{ab}(x)$ mean the quantum operator corresponding to
the metric perturbations
and the statistical
average in Eq.~(\ref{1.5c}) for the homogeneous solutions
is now taken with respect to the Wigner distribution that
describes the initial quantum state of
the metric perturbations.
The Lorentz gauge condition
$\nabla^a(h_{ab}-(1/2)\eta_{ab}h_c^c)=0$ as well as some initial
condition to completely fix the gauge of the initial state should be
implicitly understood, moreover since there are now $N$ scalar fields
the stochastic
source has been rescaled so that the two-point correlation defined by
Eq. (\ref{2.10}) should be $1/N$ times the noise kernel of a single field.
This result was implicitly obtained in
the Minkowski background in Ref. \cite{MarVer00} where the
two-point correlation in the stochastic context was computed for
the linearized metric perturbations. This stochastic correlation
exactly agrees with the symmetrized part of the graviton
propagator computed by Tomboulis \cite{Tom77} in the quantum
context of gravity interacting with $N$ Fermion fields,
where the graviton propagator is of order $1/N$. This result can
be extended to an arbitrary background in the context of the large
$N$ expansion, a sketch of the proof with explicit
details in the Minkowski background can be found in Ref.
\cite{HuRouVer04a}. This connection between the stochastic
correlations and the quantum correlations was noted and studied in
detail in the context of simpler open quantum systems
\cite{CalRouVer03}. Stochastic gravity goes
beyond semiclassical gravity in the following sense. The
semiclassical theory, which is based on the expectation value of
the stress energy tensor, carries information on the field
two-point correlations only, since $\langle \hat T_{ab}\rangle$ is
quadratic in the field operator $\hat\phi$. The stochastic theory
on the other hand, is based on the noise kernel (\ref{2.8})  which
is quartic in the field operator. However, it does not carry
information on the graviton-graviton interaction, which in the
context of the large $N$ expansion gives diagrams of order
$1/N^2$. This will be illustrated in section \ref{largeN}.
Furthermore the retarded propagator gives also information
on the commutator
\begin{equation}
\langle [\hat h_{ab}(x),\hat h_{cd}(y)]\rangle =
16\pi iG\left(G_{abcd}^{\mathrm{ret}}(y,x)-
G_{abcd}^{\mathrm{ret}}(x,y) \right),
\label{1.5e}
\end{equation}
so that combining the commutator with the anticommutator the quantum two-point correlation functions are determined. Moreover, assuming a Gaussian initial state with vanishing expectation value for the metric perturbations any $n$-point quantum correlation function is determined
by the two-point quantum correlations and thus
by the stochastic approach. Consequently, one may
regard the Einstein-Langevin equation as a useful intermediary tool to
compute the correlation functions for the quantum metric perturbations.

We should, however, emphasize also that Langevin like equations are obtained
to describe the quantum to classical
transition in open quantum systems,
when quantum  decoherence takes place by
coarse graining of the environment as well as by suitable coarse graining
of the system variables \cite{GelHar93,hartle,DowHal92,Hal93,Hal98,Whe98}.
In those cases the stochastic correlation functions describe
actual classical correlations of the system variables.
Examples can be found in the case of a moving charged particle in an
electromagnetic field in quantum electrodynamics \cite{JohHu02}
and in several quantum Brownian models
\cite{CalRouVer03,CalRouVer01,CalRouVer02}.

\subsection{Validity of semiclassical gravity}
\label{s3.3}

As we have emphasized earlier the scope and limits of
semiclassical gravity are not well
understood because we still lack a fully well understood quantum
theory of gravity.
{}From the semiclassical Einstein equations it seems also clear that
the semiclassical theory should break down when the quantum
fluctuations of the stress tensor are large. Ford \cite{For82}
was among the first to have emphasized the importance of these
quantum fluctuations. It is less clear, however, how to quantify
the size of these fluctuations. Kuo and Ford \cite{KuoFor93}
used the variance of the fluctuations of the stress tensor
operator compared to the mean value as a measure of the validity
of semiclassical gravity. Hu and Phillips pointed out
\cite{HuPhi00,PhiHu00} that such a criterion should be refined by
considering the back-reaction of those fluctuations on the metric.
Ford and collaborators also noticed that the metric fluctuations
associated to the matter fluctuations can be meaningfully
classified as \emph{active}
\cite{ForSva97,YuFor99,YuFor00} and \emph{passive}
\cite{For82,KuoFor93,For99,ForWu},
which correspond to our \emph{intrinsic} and \emph{induced} fluctuations, respectively, and have studied their properties in different contexts
\cite{BorFor04a,BorFor04b,BorFor05}.
However, these fluctuations are not treated in a unified way and
their precise relation to the quantum correlation
function for the metric perturbations is not discussed. Furthermore,
the full averaged back-reaction of the matter fields is not included
self-consistently, and the contribution from the vacuum
fluctuations in Minkowski space is discarded.

A different approach to the validity of semiclassical gravity was
pioneered by Horowitz \cite{Hor80,Hor81} who studied the
stability of a semiclassical solution with respect to linear
metric perturbations. In the case of a free quantum matter field
in its Minkowski vacuum state, flat spacetime is a solution of
semiclassical gravity. The equations describing those metric
perturbations involve higher order derivatives, and Horowitz found
unstable \emph{runaway} solutions that grow exponentially with
characteristic timescales comparable to the Planck time; see also
the analysis by Jordan \cite{Jor87}. Later, Simon
\cite{Sim90,Sim91}, argued that those unstable solutions lie
beyond the expected domain of validity of the theory and
emphasized that only those solutions which resulted from
truncating perturbative expansions in terms of the square of the
Planck length are physically acceptable \cite{Sim90,Sim91}.
Further discussion was provided by Flanagan and Wald
\cite{FlaWal96}, who advocated the use of an \emph{order reduction}
prescription first introduced by Parker and Simon \cite{ParSim93}.
More recently Anderson, Molina-Par\'\i s and
Mottola have taken up the issue of the validity of semiclassical
gravity \cite{AndMolMot02,AndMolMot03} again.
Their starting point is the fact
that the semiclassical Einstein equation will fail to provide a
valid description of the dynamics of the mean spacetime geometry
whenever the higher order radiative corrections to the effective
action, involving loops of gravitons or internal graviton
propagators, become important. Next, they argue qualitatively that such
higher order radiative corrections cannot be neglected if the
metric fluctuations grow without bound. Finally, they propose a
criterion to characterize the growth of the metric fluctuations,
and hence the validity of semiclassical gravity, based on the
stability of the solutions of the linearized semiclassical
equation. Following these approaches the Minkowski
metric is shown to be a
stable solution of semiclassical gravity with respect to small
metric perturbations.

As emphasized in Refs.
\cite{AndMolMot02,AndMolMot03} the above criteria may be
understood as criteria based on semiclassical gravity itself. It is
certainly true that stability is a necessary condition for the
validity of a semiclassical solution, but one may also look for
criteria within extensions of semiclassical gravity. In the
absence of a quantum theory of gravity such criteria may be found
in some more modest extensions. Thus, Ford \cite{For82} considered
graviton production in linearized quantum gravity and compared the
results with the production of gravitational waves in
semiclassical gravity. Ashtekar \cite{Ash96} and Beetle
\cite{Bee98} found large quantum gravity effects in
three-dimensional quantum gravity models. In a more recent paper
\cite{HuRouVer04a} (see also Ref. \cite{HuRouVer04b}), we advocate
for a criteria within the stochastic gravity approach, and since
stochastic gravity extends semiclassical gravity by
incorporating the quantum stress tensor fluctuations of the matter
fields, this criteria is structurally the most complete to date.

It turns out that this validity criteria is equivalent to the
validity criteria that one might advocate within the large $N$
expansion, that is the quantum theory describing the interaction of the
gravitational field with $N$ identical free matter fields. In the
leading order, namely the limit in which $N$ goes to infinity and
the gravitational constant is appropriately rescaled, the theory
reproduces semiclassical gravity. Thus, a natural extension of
semiclassical gravity is provided by the next to leading order. It
turns out that the symmetrized two-point quantum correlation functions of
the metric perturbations in the large $N$ expansion are equivalent
to the two-point stochastic metric correlation functions predicted by
stochastic gravity. Our validity criterion can then be
formulated as follows: a solution of semiclassical gravity is
valid when it is stable with respect to quantum metric
perturbations.  This criterion involves the consideration of quantum
correlation functions of the metric perturbations,
since the quantum field describing the metric perturbations
$\hat h_{ab}(x)$ is characterized not only by its expectation value but also
by its $n$-point correlation functions.

It is important to emphasize that the above validity criterion
incorporates in a unified and self-consistent way the two main
ingredients of the criteria exposed above. Namely, the criteria
based on the quantum stress tensor fluctuations of
the matter fields, and the criteria based on the stability of
semiclassical solutions against classical metric perturbations.
The former is incorporated through the induced metric fluctuations,
and the later through the intrinsic
fluctuations introduced in Eq. (\ref{1.5c}).
Whereas information on the stability of the intrisic metric
fluctuations can be obtained from an analysis of the solutions
of the perturbed semiclassical Einstein equation,
the homogeneous part of Eq.~(\ref{2.11}), the effect of the induced
metric fluctuations is accounted only in stochastic gravity, the full
inhomogeneous Eq.~(\ref{2.11}).
We will illustrate this criteria in section \ref{s6.5} by
studying the stability
of Minkowski spacetime as a solution of semiclassical gravity.

\subsubsection{The large $N$ expansion}
\label{largeN}

To illustrate
the relation between the semiclassical, stochastic and quantum
theories, a simplified model of scalar gravity interacting with
$N$ scalar fields is considered here.

The large $N$ expansion has been successfully used in quantum
chromodynamics to compute some non-perturbative results. This
expansion re-sums and rearranges Feynman perturbative series
including self-energies. For gravity interacting with $N$ matter
fields it shows that graviton loops are of higher order than matter
loops. To illustrate the large $N$ expansion let us, first, consider the
following toy model of gravity, which we will simplify as a scalar
field $h$, interacting with a single scalar field $\phi$ described by the
action
\begin{eqnarray}
S&=&\frac{1}{\kappa}\int d^4 x\left(\partial_a h\partial^a
h+h\,\partial_a h\partial^a h+\dots\right)\nonumber\\
&&-\int d^4 x\left(\partial_a \phi\partial^a
\phi+m^2\phi^2\right)+\int d^4 x\left( h\,\partial_a \phi\partial^a \phi+\dots\right),
 \label{N1}
\end{eqnarray}
where $\kappa=8\pi G$, and we have assumed that the
interaction is linear in the (dimensionless) scalar gravitational
field $h$ and quadratic in the matter field $\phi$ to simulate in
a simplified way the coupling of the metric with the stress tensor
of the matter fields.
We have also included a self coupling graviton term of $O(h^3)$
which also appears in perturbative gravity beyond the linear
approximation.

We may now compute the dressed graviton propagator perturbatively
as the following series of Feynman diagrams. The first diagram is
just the free graviton propagator which is of $O(\kappa)$, as one
can see from the kinetic term for the graviton in equation
(\ref{N1}). The next diagram is one loop of matter with two
external legs which are the graviton propagators. This diagram has
two vertices with one graviton propagator and two matter field
propagators. Since the vertices and the matter propagators
contribute with 1 and each graviton propagator contributes with a
$\kappa$ this diagram is of order $O(\kappa^2)$. The next diagram
contains two loops of matter and three gravitons, and consequently
it is of order $O(\kappa^3)$. There will also be terms with one
graviton loop and two graviton propagators as external legs, with
three graviton propagators at the two vertices due to the $O(h^3)$
term in the action (\ref{N1}). Since there are four graviton
propagators which carry a $\kappa^4$ but two vertices which have
$\kappa^{-2}$ this diagram is of order $O(\kappa^2)$, like the
term with one matter loop. Thus,
in this perturbative expansion a graviton loop and a matter loop
both contribute at the same order to the dressed graviton propagator.

Let us now consider the large $N$ expansion. We assume that the
gravitational field is coupled with a large number of identical
fields $\phi_j$, $j=1,\dots,N$ which couple only with $h$. Next we
rescale the gravitational coupling in such a way that
$\bar\kappa=\kappa N$ is finite even when $N$ goes to infinity.
The action of this system is:
\begin{eqnarray}
S&=&\frac{N}{\bar\kappa}\int d^4 x\left(\partial_a h\partial^a
h+h\,\partial_a h \partial^a h+\dots\right)\nonumber\\
&&-\sum_j^N\int d^4 x\left(\partial_a \phi_j\partial^a
\phi_j+m^2\phi^2\right)+\sum_j^N
\int d^4 x\left( h\,\partial_a \phi_j \partial^a \phi_j+\dots\right).
 \label{N2}
\end{eqnarray}

Now an expansion in powers of $1/N$ of the dressed graviton
propagator is given by the following series of Feynman diagrams.
The first diagram is the free graviton propagator which is now of
order $O(\bar\kappa/N)$ the following diagrams are $N$ identical
Feynman diagrams with one loop of matter and two graviton
propagators as external legs, each diagram due to the two graviton
propagators is of order $O(\bar\kappa^2/N^2)$ but since there are
$N$ of them the sum can be represented by a single diagram with a
loop of matter of weight $N$, and therefore this diagram is of
order $O(\bar\kappa^2/N)$. This means that it is of the same order
as the first diagram in an expansion in $1/N$. Then there are the
diagrams with two loops of matter and three graviton propagators,
as before we can assign a weight of $N$ to each loop and taking
into account the three graviton propagators this diagram is of
order $O(\bar\kappa^3/N)$, and so on. This means that to order
$1/N$ the dressed graviton propagator contains all the
perturbative series in powers of $\bar\kappa$ of the matter loops.

Next, there is a diagram with one graviton loop and two graviton
legs. Let us count the order of this diagram: it contains four
graviton propagators and two vertices, the propagators contribute
as $(\bar\kappa/N)^4$ and the vertices as $(N/\bar\kappa)^2$, thus
this diagram is of $O(\bar\kappa^2/N^2)$. Therefore graviton loop
contributes to higher order in the $1/N$ expansion than matter
loops. Similarly there are $N$ diagrams with one loop of matter
with an internal graviton propagator and two external graviton
legs. Thus we have three graviton propagators and since there are
$N$ of them, their sum is of order $O(\bar\kappa^3/N^2)$. To
summarize, we have that when $N\to\infty$ there are no graviton
propagators and gravity is classical yet the matter fields are
quantized, this is semiclassical
gravity as was first described in Ref. \cite{HarHor81}.
Then we go to the next to leading order, $1/N$, now the graviton propagator
includes all matter loop contributions, but no contributions from
graviton loops or internal graviton propagators in matter loops.
This is what stochastic gravity reproduces.

That stochastic gravity is connected to the large $N$ expansion
can be seen from the stochastic correlations of linear metric
perturbations on the Minkowski background computed in Ref.
\cite{MarVer00}. These correlations are in exact agreement with
the imaginary part of the graviton propagator found by Tomboulis
in the large $N$ expansion for the quantum theory of gravity
interacting with $N$ Fermion fields \cite{Tom77}. This has been
proved in detail in Ref. \cite{HuRouVer04a}, see also
\cite{HuRouVer04b}, where the case of a general background is
also briefly discussed.

\section{The Einstein-Langevin equation: Functional approach}
\label{sec3}

The Einstein-Langevin equation (\ref{2.11}) may also be derived by a
method based on functional techniques  \cite{MarVer99}. Here we will
summarize these techniques starting with semiclassical gravity.

In semiclassical gravity functional methods were used to study the
back-reaction of quantum fields in cosmological models
\cite{Har77,FisHarHu79,HarHu79}. The primary advantage of the
effective action approach is, in addition to the well-known fact that
it is easy to introduce perturbation schemes like loop expansion and
nPI formalisms, that it yields a {\it fully} self-consistent
solution. (For a general discussion in the semiclassical context of
these two approaches, equation of motion versus effective action,
contrast e.g., \cite{LukSta74,Gri76,HuPar77,HuPar78} with the above
references and \cite{HarHu80,Har81,And83,And84}. See also comments in
Sec. 5.6 on the black hole backreaction problem comparing the
approach by York et al. \cite{Yor83,Yor85,Yor86} versus that of
Sinha, Raval and Hu \cite{SinRavHu03}.

The well known in-out effective action method treated in
textbooks, however, led to equations of motion which were not real
because they were tailored to compute transition elements of
quantum operators rather than expectation values. The correct
technique to use for the backreaction problem is the
Schwinger-Keldysh
\cite{Sch61,BakMah63,Kel64,ChoEtal85,SuEtal88,CalHu89,CooEtal94}
closed-time-path (CTP) or `in-in'
effective action. These techniques were adapted to the
gravitational context
\cite{DeW86,Jor86,CalHu87,Jor87,Paz90,CamVer94} and applied to
different problems in cosmology. One could deduce the
semiclassical Einstein equation from the CTP effective action for
the gravitational field (at tree level) with quantum matter
fields.

Furthermore, in this case the CTP functional formalism turns out
to be related
\cite{SuEtal88,CalHu94,CamVer96,LomMaz96,GreMul97,CamHu98,CamHu99,%
Mor86,LeeBoy93,MarVer99,MarVer99b} to the influence functional
formalism of Feynman and Vernon \cite{FeyVer63} since the full
quantum system may be understood as consisting of a distinguished
subsystem (the ``system'' of interest) interacting with the
remaining degrees of freedom (the environment). Integrating out
the environment variables in a CTP path integral yields the
influence functional, from which one can define an effective
action for the dynamics of the system
\cite{CalHu94,HuSin95,HuMat94,GreMul97}. This approach to
semiclassical gravity is motivated by the observation
\cite{Physica} that in some open quantum systems classicalization
and decoherence
\cite{Zur81,Zur82,Zur86,Zur91,JooZeh85,CalLeg85,UnrZur89,Zur93,GiuEtal96}
on the system may be brought about by interaction with an
environment, the environment being in this case the matter fields
and some ``high-momentum'' gravitational modes
\cite{Kie87,Hal89,Pad89,Hu90,Cal89,Cal91,HuPazSin93,Whe98}.
Unfortunately, since the form of a complete quantum theory of
gravity interacting with matter is unknown, we do not know what
these ``high-momentum'' gravitational modes are. Such a
fundamental quantum theory might not even be a field theory, in
which case the metric and scalar fields would not be fundamental
objects \cite{stogra}. Thus, in this case, we cannot attempt to
evaluate the influence action of Feynman and Vernon starting from
the fundamental quantum theory and performing the path
integrations in the environment variables. Instead, we introduce
the influence action for an effective quantum field theory of
gravity and matter
\cite{Don94a,Don94b,Don96a,Don96b,SinHu91,Paz91,PazSin92}, in
which such ``high-momentum'' gravitational modes are assumed to
have already been ``integrated out.''


\subsection{Influence action for semiclassical gravity}


Let us formulate semiclassical gravity in this functional
framework. Adopting the usual procedure of effective field
theories \cite{Wei95,Wei96,Don94a,Don94b,Don96a,Don96b,CalKan97}, one
has to take the effective action for the metric and the scalar
field of the most general local form compatible with general
covariance: $S[g,\phi] \equiv S_g[g]+S_m[g,\phi]+ \cdots ,$ where
$S_g[g]$ and $S_m[g,\phi]$ are given by Eqs. (\ref{2.6}) and
(\ref{2.1}), respectively, and the dots stand for terms of order
higher than two in the curvature and in the number of derivatives
of the scalar field. Here, we shall neglect the higher order terms
as well as self-interaction terms for the scalar field. The second
order terms are necessary to renormalize one-loop ultraviolet
divergences of the scalar field stress-energy tensor, as we have
already seen. Since ${\cal M}$ is a globally hyperbolic manifold,
we can foliate it by a family of $t\!=\! {\rm constant}$ Cauchy
hypersurfaces $\Sigma_{t}$, and we will indicate the initial and
final times by $t_i$ and $t_f$, respectively.

The {\it influence functional} corresponding to the action
(\ref{2.1}) describing a scalar field in a spacetime (coupled to
a metric field) may be introduced as a functional of two copies
of the metric, denoted by $g_{ab}^+$ and $g_{ab}^-$, which
coincide at some final time $t=t_f$. Let us assume that, in the
quantum effective theory, the state of the full system (the
scalar and the metric fields) in the Schr\"{o}dinger picture at
the initial time $t\! =\! t_{i}$ can be described by a density
operator which can be written as the tensor product of two
operators on the Hilbert spaces of the metric and of the scalar
field. Let $\rho_i(t_i)\equiv
\rho_i \left[\phi_+(t_i),\phi_-(t_i) \right] $ be the
matrix element of the density operator $\hat{\rho}^{\rm
\scriptscriptstyle S}(t_{i})$ describing the initial state of the
scalar field. The Feynman-Vernon influence functional is defined
as the following path integral over the two copies of the scalar
field:
\begin{equation}
{\cal F}_{\rm IF}[g^\pm] \equiv
\int\! {\cal D}\phi_+\;
{\cal D}\phi_- \;
\rho_i (t_i)
\delta\!\left[\phi_+(t_f)\!-\!\phi_-(t_f)  \right]\:
e^{i\left(S_m[g^+,\phi_+]-S_m[g^-,\phi_-]\right) }.
\label{path integral}
\end{equation}
Alternatively, the  above double path integral can be rewritten
as a closed time path (CTP) integral, namely, as a single path
integral in a complex time contour with two different time
branches, one going forward in time from $t_i$ to $t_f$, and the
other going backward in time from $t_f$ to $t_i$ (in practice one
usually takes $t_i\to -\infty$). {}From this influence functional,
the {\it influence action} $S_{\rm IF}[g^+,g^-]$, or $S_{\rm
IF}[g^\pm]$ for short,  defined  by \be {\cal F}_{\rm IF}[g^\pm]
\equiv e^{i S_{\rm IF}[g^\pm]}, \label{influence functional} \ee
carries all the information about the environment (the matter
fields) relevant to the system (the gravitational field). Then we
can define the CTP {\it effective action} for the gravitational
field, $S_{\rm eff}[g^\pm]$, as
\begin{equation}
S_{\rm eff}[g^\pm]\equiv S_{g}[g^+]-S_{g}[g^-] +S_{\rm
IF}[g^\pm]. \label{ctpif}
\end{equation}
This is the effective action for the classical gravitational
field in the CTP formalism. However, since the gravitational
field is treated only at the tree level, this is also the
effective classical action from which the classical equations of
motion can be derived.

Expression (\ref{path integral}) contains ultraviolet divergences
and must be regularized. We shall assume that dimensional
regularization can be applied, that is, it makes sense to
dimensionally continue all the quantities that appear in Eq.
(\ref{path integral}).  For this we need to work with the
$n$-dimensional actions corresponding to $S_m$ in (\ref{path
integral}) and $S_g$ in (\ref{2.6}). For example,  the parameters
$G$, $\Lambda$ $\alpha$ and $\beta$ of Eq. (\ref{2.6}) are the
bare parameters $G_B$, $\Lambda_B$, $\alpha_B$ and $\beta_B$, and
in $S_g[g]$, instead of the square of the Weyl tensor in Eq.
(\ref{2.6}),  one must use $(2/3)(R_{abcd}R^{abcd}- R_{ab}R^{ab})$
which by the Gauss-Bonnet theorem leads to the same equations of
motion as the action (\ref{2.6}) when $n \!=\! 4$. The form of
$S_g$ in $n$ dimensions is suggested by the Schwinger-DeWitt
analysis of the ultraviolet divergences in the matter
stress-energy tensor using dimensional regularization. One can
then write the Feynman-Vernon effective action $S_{\rm
eff}[g^\pm]$ in Eq. (\ref{ctpif}) in a form suitable for
dimensional regularization. Since both $S_m$ and $S_g$ contain
second order derivatives of the metric, one should also add some
boundary terms \cite{Wal84,HuSin95}. The effect of these terms is
to cancel out the boundary terms which appear when taking
variations of $S_{\rm eff}[g^\pm]$ keeping the value of $g^+_{ab}$
and $g^-_{ab}$ fixed at $\Sigma_{t_i}$ and $\Sigma_{t_f}$.
Alternatively, in order to obtain the equations of motion for the
metric in the semiclassical regime, we can work with the action
terms  without boundary terms and neglect all boundary terms when
taking variations with respect to $g^{\pm}_{ab}$. From now on, all
the functional derivatives with respect to the metric will be
understood in this sense.

The semiclassical Einstein equation (\ref{2.5}) can now be derived.
Using the definition of the stress-energy tensor
$T^{ab}(x)=(2/\sqrt{-g})\delta S_m/\delta g_{ab}$
and the definition
of the influence functional, Eqs.
(\ref{path integral}) and (\ref{influence functional}), we see that
\begin{equation}
\langle \hat{T}^{ab}[g;x) \rangle =
\left. {2\over\sqrt{- g(x)}} \,
 \frac{\delta S_{\rm IF}[g^\pm]}
{\delta g^+_{ab}(x)} \right|_{g^\pm=g},
\label{s-t expect value}
\end{equation}
where the expectation value is taken in the $n$-dimensional
spacetime generalization of the state described by
$\hat{\rho}^{\rm \scriptscriptstyle S}(t_i)$. Therefore,
differentiating $S_{\rm eff}[g^\pm]$ in Eq. (\ref{ctpif}) with
respect to $g^+_{ab}$, and then setting
$g^+_{ab}=g^-_{ab}=g_{ab}$, we get the semiclassical Einstein
equation in $n$ dimensions. This equation is then renormalized by
absorbing the divergences in the regularized $\langle\hat
T^{ab}[g]\rangle$ into the bare parameters. Taking the limit
$n\to 4$ we obtain the physical semiclassical Einstein equation
(\ref{2.5}).


\subsection{Influence action for stochastic gravity}


In the spirit of the previous derivation of the Einstein-Langevin
equation, we now seek a dynamical equation for a linear
perturbation $h_{ab}$ to the semiclassical metric $g_{ab}$,
solution of Eq. (\ref{2.5}). Strictly speaking if we use
dimensional regularization we must consider the $n$-dimensional
version of that equation. {}From
the results just described, if such an equation were simply a
linearized semiclassical Einstein equation, it could be obtained
from an expansion of the effective action $S_{\rm eff}[g+h^\pm]$.
In particular, since, from Eq. (\ref{s-t expect value}), we have
that
\begin{equation}
\langle \hat{T}^{ab}[g+h;x) \rangle =
\left. {2\over\sqrt{-\det (g\!+\!h)(x)}} \,
 \frac{\delta S_{\rm IF}
   [g\!+\!h^\pm]}{\delta h^+_{ab}(x)}
 \right|_{h^\pm=h},
\label{perturb s-t expect value}
\end{equation}
the expansion of $\langle \hat{T}^{ab}[g\!+\!h]\rangle $
to linear order in $h_{ab}$ can be obtained from an expansion of the
influence action $S_{\rm IF}[g+h^\pm]$ up to second order
in $h^{\pm}_{ab}$.

To perform the expansion of the influence action,
we have to compute the first and second order
functional derivatives of $S_{\rm IF}[g+h^\pm]$
and then set $h^+_{ab}\!=\!h^-_{ab}\!=\!h_{ab}$.
If we do so using the path integral representation
(\ref{path integral}), we can interpret these derivatives as
expectation values of operators.
The relevant second order derivatives are
\begin{eqnarray}
\left. {4\over\sqrt{\!- g(x)}\sqrt{\!- g(y)} }
 \frac{\delta^2 S_{\rm IF}[g+h^\pm]}
{\delta h^+_{ab}\!(x)\delta h^+_{cd}\!(y)}
 \right|_{h^\pm=h} \!\!\!\!\!\!
&=& \!\!\!\!\!-H_{\scriptscriptstyle \! {\rm S}}^{abcd}[g;x,y)
\!-\!K^{abcd}[g;x,y)
\nonumber\\
&&+i N^{abcd}[g;x,y),      \nonumber \\
\left. {4\over\sqrt{\!- g(x)}\sqrt{\!- g(y)} }
 \frac{\delta^2 S_{\rm IF}[g^\pm]}
{\delta h^+_{ab}\!(x)\delta h^-_{cd}\!(y)}
 \right|_{h^\pm=h} \!\!\!\!\!\!
&=& \!\!\!\!\!-H_{\scriptscriptstyle \! {\rm A}}^{abcd} [g;x,y)
\!-\! i N^{abcd}[g;x,y)\!, \label{derivatives}
\end{eqnarray}
where
$$
N^{abcd}[g;x,y) \equiv
{1\over 2} \left\langle  \bigl\{
 \hat{t}^{ab}[g;x) , \,
 \hat{t}^{cd}[g;y)
 \bigr\} \right\rangle ,
$$
$$
H_{\scriptscriptstyle \!
{\rm S}}^{abcd}
[g;x,y) \equiv
{\rm Im} \left\langle {\rm T}^*\!\!
\left( \hat{T}^{ab}[g;x) \hat{T}^{cd}[g;y)
\right) \right\rangle \!,
$$
$$
H_{\scriptscriptstyle \!
{\rm A}}^{abcd}
[g;x,y) \equiv
-{i\over 2} \left\langle
\bigl[ \hat{T}^{ab}[g;x), \, \hat{T}^{cd}[g;y)
\bigr] \right\rangle \!,\,
$$
$$
K^{abcd}[g;x,y) \equiv
\left. {-4\over\sqrt{- g(x)}\sqrt{- g(y)} } \, \left\langle
\frac{\delta^2 S_m[g+h,\phi]}
{\delta h_{ab}(x)\delta h_{cd}(y)}
\right|_{\phi=\hat{\phi}}\right\rangle \!,
$$
with $\hat{t}^{ab}$ defined in Eq. (\ref{2.9}), $[ \; , \: ]$
denotes the commutator and $\{ \; , \: \}$ the anti-commutator.
Here we use a Weyl ordering prescription for the operators.
The symbol ${\rm T}^*$ denotes the
following ordered operations: First, time order the field
operators $\hat{\phi}$ and then apply the derivative operators
which appear in each term of the product $T^{ab}(x) T^{cd}(y)$,
where $T^{ab}$ is the functional (\ref{2.3}). This ${\rm T}^{*}$
``time ordering'' arises because we have path integrals
containing products of derivatives of the field, which can be
expressed as derivatives of the path integrals which do not
contain such derivatives. Notice, from their definitions, that
all the kernels which appear in expressions (\ref{derivatives})
are real and also $H_{\scriptscriptstyle \!{\rm A}}^{abcd}$ is
free of ultraviolet divergences in the limit $n \to 4$.

{}From (\ref{s-t expect value}) and
(\ref{derivatives}), since
$S_{\rm IF}[g,g]=0$ and
$S_{\rm IF}[g^-,g^+]=
-S^{ {\displaystyle \ast}}_{\rm IF}[g^+,g^-]$, we can write the
expansion for the influence action
$S_{\rm IF}[g\!+\!h^\pm]$ around a background
metric $g_{ab}$ in terms of the previous kernels.
Taking into account that
these kernels satisfy the symmetry relations
\begin{equation}
H_{\scriptscriptstyle \!{\rm S}}^{abcd}(x,y)\!=\!
H_{\scriptscriptstyle \!{\rm S}}^{cdab}(y,x),
H_{\scriptscriptstyle \!{\rm A}}^{abcd}(x,y)\!=\!
-H_{\scriptscriptstyle \!{\rm A}}^{cdab}(y,x),
K^{abcd}(x,y) \!= \! K^{cdab}(y,x),
\label{symmetries}
\end{equation}
and introducing the new kernel
\begin{equation}
H^{abcd}(x,y)\equiv
H_{\scriptscriptstyle \!{\rm S}}^{abcd}(x,y)
+H_{\scriptscriptstyle \!{\rm A}}^{abcd}(x,y),
\label{H}
\end{equation}
the expansion of $S_{\rm IF}$ can be finally written as
\begin{eqnarray}
S_{\rm IF}[g\!+\!h^\pm]
&=& {1\over 2} \int\! d^4x\, \sqrt{- g(x)}\:
\langle \hat{T}^{ab}[g;x) \rangle  \,
\left[h_{ab}(x) \right] \nonumber\\
&&-{1\over 8} \int\! d^4x\, d^4y\, \sqrt{- g(x)}\sqrt{- g(y)}\,
\nonumber  \\
&& \ \ \ \ \times\left[h_{ab}(x)\right]
\left(H^{abcd}[g;x,y)\!
+\!K^{abcd}[g;x,y) \right)
\left\{ h_{cd}(y) \right\}  \nonumber  \\
&&
+{i\over 8} \int\! d^4x\, d^4y\, \sqrt{- g(x)}\sqrt{- g(y)}\,
\nonumber  \\
&& \ \ \ \ \times\left[h_{ab}(x) \right]
N^{abcd}[g;x,y)
\left[h_{cd}(y) \right]+0(h^3),
\label{expansion 2}
\end{eqnarray}
where we have used the notation
\begin{equation}
\left[h_{ab}\right] \equiv h^+_{ab}\!-\!h^-_{ab},
\hspace{5 ex}
\left\{ h_{ab}\right\} \equiv h^+_{ab}\!+\!h^-_{ab}.
\label{notation}
\end{equation}
{}From Eqs.~(\ref{expansion 2}) and
(\ref{perturb s-t expect value})
it is clear that the imaginary part of the
influence action does not contribute to the perturbed
semiclassical Einstein equation (the expectation value of the
stress-energy tensor is real), however, as it depends on the noise kernel,
it contains information on the fluctuations of the operator
$\hat{T}^{ab}[g]$.

We are now in a position to carry out the derivation of the
semiclassical Einstein-Langevin equation. The procedure is well
known
\cite{CalHu94,HuSin95,CamVer96,GleRam94,BoyEtal95,YamYok97,RamHuSty98}:
it consists of deriving a new ``stochastic'' effective action from
the observation that the effect of the imaginary part of the
influence action (\ref{expansion 2}) on the corresponding
influence functional is equivalent to the averaged effect of the
stochastic source $\xi^{ab}$ coupled linearly to the perturbations
$h_{ab}^{\pm}$. This observation follows from the identity first
invoked by Feynman and Vernon for such purpose:
\begin{eqnarray}
&&\exp\left(-{1\over 8} \!\int\! d^4x\, d^4y \, \sqrt{- g(x)}\sqrt{- g(y)}\,
\left[h_{ab}(x) \right]\,
N^{abcd}(x,y)\, \left[h_{cd}(y)\right] \right)
\nonumber  \\
&&\quad\quad =
\int\! {\cal D}\xi \: {\cal P}[\xi]\, \exp\left({i\over 2} \!\int\! d^4x \,
\sqrt{- g(x)}\,\xi^{ab}(x)\,\left[h_{ab}(x) \right] \right),
\label{Gaussian path integral}
\end{eqnarray}
where ${\cal P}[\xi]$ is the probability distribution
functional of a Gaussian stochastic tensor $\xi^{ab}$
characterized by the correlators (\ref{2.10})
with $N^{abcd}$ given by Eq.~(\ref{2.8}),
and where
the path integration measure is assumed to be a scalar under
diffeomorphisms of $({\cal M},g_{ab})$. The above identity follows
from the identification of the right-hand side of
(\ref{Gaussian path integral}) with the characteristic functional for
the stochastic field $\xi^{ab}$. The
probability distribution functional for $\xi^{ab}$ is explicitly
given by
\begin{equation}
{\cal P}[\xi]\!=\! {\rm det}(2\pi N)^{-1/2}\!\!
 \exp\!\left[\!-{1\over2}\!\int\!\! d^4x d^4y \!
\sqrt{\!-g(x)}\sqrt{\!-g(y)}\!
 \xi^{ab}\!(x) \! N^{-1}_{abcd}(x,y)\! \xi^{cd}\!(y)\!\right]\!.
\end{equation}

We may now introduce the {\it stochastic effective action} as
\begin{equation}
S^s_{\rm eff}[g+h^\pm,\xi] \equiv S_{g}[g+h^+]-S_{g}[g+h^-]+
S^s_{\rm IF}[g+h^\pm,\xi],
\label{stochastic eff action}
\end{equation}
where the ``stochastic'' influence action is defined as
\begin{equation}
S^s_{\rm IF}[g+h^\pm,\xi] \equiv {\rm Re}\, S_{\rm
IF}[g\!+\!h^\pm]+\! {1\over 2} \int\! d^4x \, \sqrt{-
g(x)}\,\xi^{ab}(x)\left[h_{ab}(x) \right]+ O(h^3). \label{eff
influence action}
\end{equation}
Note that, in fact, the influence functional can now be written as a
statistical average over $\xi^{ab}$:
$
{\cal F}_{\rm IF}[g+h^\pm]= \left\langle
\exp\left(i S^s_{\rm IF}[g+h^\pm,\xi]\right)
\right\rangle_{\! s}.
$
The stochastic equation of motion for $h_{ab}$ reads
\begin{equation}
\left.
\frac{\delta S^s_{\rm eff}[g+h^\pm,\xi]}{\delta h^+_{ab}(x)}
\right|_{h^\pm=h}=0,
\label{eq of motion}
\end{equation}
which is the Einstein-Langevin equation (\ref{2.11}); notice that only the
real part of $S_{IF}$ contributes to the expectation value
(\ref{perturb s-t expect value}).
To be precise we get
first the regularized
$n$-dimensional equations with the bare parameters,
and where instead of the tensor $A^{ab}$ we get
$(2/3)D^{ab}$ where
\begin{eqnarray}
D^{ab} &\equiv & {1\over\sqrt{- g}}   \frac{\delta}{\delta g_{ab}}
          \int \! d^n x \,\sqrt{- g}
\left(R_{cdef}R^{cdef}-
                         R_{cd}R^{cd}  \right)
\nonumber\\
   &=& {1\over2}\, g^{ab} \!
\left(  R_{cdef} R^{cdef}-
         R_{cd}R^{cd}+\Box  R \right)
      -2R^{acde}{R^b}_{cde}
\nonumber \\
&&
      -2 R^{acbd}R_{cd}
      +4R^{ac}{R_c}^b
      -3 \Box  R^{ab}
  +\nabla^{a}\nabla^{b} R.
\label{D}
\end{eqnarray}
Of course, when $n=4$ these tensors are related,
$A^{ab}=(2/3) D^{ab}$. After that
we renormalize and
take the limit $n\to 4$ to obtain the Einstein-Langevin
equations in the physical spacetime.


\subsection{Explicit form of the Einstein-Langevin equation}


We can write the Einstein-Langevin equation in a more explicit
form by working out the expansion of $\langle
\hat{T}^{ab}[g\!+\!h]\rangle $ up to linear order in the
perturbation $h_{ab}$. {}From Eq. (\ref{perturb s-t expect
value}), we see that this expansion can be easily obtained from
(\ref{expansion 2}). The result is
\begin{eqnarray}
\langle \hat{T}_n^{ab}[g\!+\!h;x) \rangle
\! &=&\!
\langle \hat{T}_n^{ab}[g,x) \rangle
 + \langle
\hat{T}_n^{{\scriptscriptstyle (1)}\hspace{0.1ex} ab} [g,h;x)
\hspace{-0.1ex} \rangle
\nonumber  \\
&&- \frac{1}{2} \!\int\! \hspace{-0.2ex}
d^ny \, \sqrt{- g(y)} \hspace{0.2ex}  H_n^{abcd}[g;x,y) h_{cd}(y)
+ 0(h^2). \label{s-t expect value expansion}
\end{eqnarray}
Here we use a subscript $n$ on a given tensor to indicate that we
are explicitly working in $n$-dimensions, as we use dimensional
regularization, and we also use the superindex ${\scriptstyle
(1)}$ to generally indicate that the tensor is the first order
correction, linear in $h_{ab}$, in a perturbative expansion around
the background $g_{ab}$.

Using the Klein-Gordon equation (\ref{2.2}), and expressions
(\ref{2.3})  for the stress-energy tensor and the corresponding
operator, we can write \be \hat{T}_n^{{\scriptscriptstyle
(1)}\hspace{0.1ex} ab} [g,h]=\left({1\over 2}\, g^{ab}h_{cd}-
\delta^a_c h^b_d- \delta^b_c h^a_d  \right) \hat{T}_{n}^{cd}[g]
+{\cal F}^{ab}[g,h]\, \hat{\phi}_{n}^2[g], \label{T(1) operator}
\ee where ${\cal F}^{ab}[g;h]$ is the differential operator \bea
{\cal F}^{ab}\!\!\!\!\!&\equiv&\!\!\!\! \!\left(\xi\!-\!{1\over
4}\right)\!\! \left(h^{ab}\!-\!{1\over 2}\, g^{ab} h^c_c \right)\!
\Box
\nonumber  \\
&&\!\!\!\!\!+\!{\xi \over 2} \left[ \nabla^{c} \nabla^{a} h^b_c+
\nabla^{c} \nabla^{b} h^a_c\!- \!\Box h^{ab}\!-\! \nabla^{a}
\nabla^{b}  h^c_c\!-\! g^{ab} \nabla^{c} \nabla^{d} h_{cd}+g^{ab}
\Box h^c_c
\right.   \nonumber \\
&&\left. \!\!\!\!\!+\!\left(\! \nabla^{a} h^b_c\!+\! \nabla^{b}
h^a_c\!-\!\nabla_{c} h^{ab}\!\!-\! 2 g^{ab} \nabla^{d} h_{cd} \!+
\!g^{ab} \nabla_{c}  h^d_d \right)\! \nabla^{c} \!\!-\!g^{ab}
h_{cd} \nabla^{c} \nabla^{d} \right]\!\!. \label{diff operator F}
\eea It is understood that indices are raised with the background
inverse metric $g^{ab}$ and that all the covariant derivatives are
associated to the metric $g_{ab}$.

Substituting (\ref{s-t expect value expansion}) into
the $n$-dimensional version of the Einstein-Langevin
Eq. (\ref{2.11}),
taking into account that
$g_{ab}$ satisfies the semiclassical Einstein equation
(\ref{2.5}), and substituting expression (\ref{T(1) operator})
we can write the Einstein-Langevin
equation in dimensional regularization as
\bea
&&{1\over 8 \pi G_{B}}\Biggl[
G^{{\scriptscriptstyle (1)}\hspace{0.1ex} ab}\!-\!
{1\over 2}\, g^{ab} G^{cd}
h_{cd}+ G^{ac} h^b_c+G^{bc} h^a_c+
\Lambda_{B} \left( h^{ab}\!-\!{1\over 2}\,
g^{ab} h^c_c \right)
\Biggr]
   \nn \\
&&
- \,
{4\alpha_{B}\over 3}  \left( D^{{\scriptscriptstyle
(1)}ab}
-{1\over 2} g^{ab} D^{cd} h_{cd}+
D^{ac} h^b_c+D^{bc} h^a_c
\right)\!
\nonumber  \\
&&-2\beta_{B}\!\left( B^{{\scriptscriptstyle (1)}ab}\!-\!
{1\over 2} g^{ab} B^{cd}
h_{cd}+ B^{ac} h^b_c+B^{bc} h^a_c
\right)   \nn \\
&&- \, \mu^{-(n-4)}\, {\cal F}^{ab}_x \langle
\hat{\phi}_{n}^2[g;x) \rangle +{1\over 2} \!\int\! d^ny  \sqrt{-
g(y)}\, \mu^{-(n-4)} H_n^{abcd}[g;x,y) h_{cd}(y)
\nonumber  \\
 &&= \mu^{-(n-4)}
\xi^{ab}_n, \label{Einstein-Langevin eq 3} \eea where the tensors
$G^{ab}$, $D^{ab}$ and $B^{ab}$ are computed from the
semiclassical metric $g_{ab}$, and where we have omitted the
functional dependence on $g_{ab}$ and $h_{ab}$ in
$G^{{\scriptscriptstyle (1)}ab}$, $D^{{\scriptscriptstyle
(1)}ab}$, $B^{{\scriptscriptstyle (1)}ab}$ and ${\cal F}^{ab}$ to
simplify the notation. The parameter $\mu$ is a mass scale which
relates the dimensions of the physical field $\phi$ with the
dimensions of the corresponding field in $n$-dimensions,
$\phi_n=\mu^{(n-4)/2}\phi$. Notice that, in Eq.
(\ref{Einstein-Langevin eq 3}), all the ultraviolet divergences
in the limit $n\to 4$, which must be removed by renormalization of
the coupling constants, are in $\langle \hat{\phi}_{n}^2(x)
\rangle$ and the symmetric part $H_{\scriptscriptstyle \! {\rm
S}_{\scriptstyle n}}^{abcd}(x,y)$ of the kernel $H_n^{abcd}(x,y)$,
whereas the kernels $N_n^{abcd}(x,y)$ and $H_{\scriptscriptstyle
\! {\rm A}_{\scriptstyle n}}^{abcd}(x,y)$ are free of ultraviolet
divergences. If we introduce the bi-tensor $F_{n}^{abcd}[g;x,y)$
defined by
\begin{equation}
F_{n}^{abcd}[g;x,y) \equiv
\left\langle \hat{t}_n^{ab}[g;x)\,
\hat{t}_n^{\rho\sigma}[g;y)
  \right\rangle
\label{bitensor F}
\end{equation}
where $\hat t^{ab}$ is defined by Eq. (\ref{2.9}), then the
kernels $N$ and $H_A$ can be written as \be N_n^{abcd}[g;x,y)=
{\rm Re} \, F_{n}^{abcd}[g;x,y), \hspace{7ex}
H_{\scriptscriptstyle \! {\rm A}_{\scriptstyle n}}^{abcd}[g;x,y)=
{\rm Im} \, F_{n}^{abcd}[g;x,y), \label{finite kernels} \ee where
we have used that $2 \langle \hat{t}^{ab}(x)\, \hat{t}^{cd}(y)
\rangle= \langle \{ \hat{t}^{ab}(x), \, \hat{t}^{cd}(y) \}\rangle
+ \langle [ \hat{t}^{ab}(x), \, \hat{t}^{cd}(y)]\rangle$, and the
fact that the first term on the right hand side of this identity
is real, whereas the second one is pure imaginary. Once we perform
the renormalization procedure in Eq.~(\ref{Einstein-Langevin eq
3}), setting $n = 4$ will yield the physical Einstein-Langevin
equation. Due to the presence of the kernel $H_n^{abcd}(x,y)$,
this equation will be usually non-local in the metric
perturbation. In section \ref{sec:flucminspa} we will carry out an
explicit evaluation of the physical Einstein-Langevin equation
which will illustrate the procedure.


\subsubsection{The kernels for the vacuum state}


When the expectation values in the Einstein-Langevin equation are
taken in a vacuum state $|0 \rangle$, such as, for instance, an
``in'' vacuum, we can be more explicit, since we can write the
expectation values in terms of the Wightman and Feynman functions,
defined as \be G_n^+[g;x,y)\! \equiv\! \langle 0| \!
   \hat{\phi}_{n}[g;x)  \hat{\phi}_{n}[g;y) \,
   \!|0 \rangle ,
i G\!_{\scriptscriptstyle F_{\scriptstyle \hspace{0.1ex}  n}}
 \hspace{-0.2ex}[g;x,y)
  \!\equiv\! \langle 0| \!
  {\rm T}\! \left(\!\hat{\phi}_{n}[g;x)  \hat{\phi}_{n}[g;y)\! \right)
  \!
  |0 \rangle.
\label{Wightman and Feynman functions}
\ee
These expressions for the kernels in the Einstein-Langevin
equation will be very useful for explicit
calculations.
To simplify the notation, we omit the functional
dependence on the semiclassical metric $g_{ab}$, which will be
understood in all the expressions below.

{}From Eqs. (\ref{finite kernels}), we see that the kernels
$N_n^{abcd}(x,y)$ and
$H_{\scriptscriptstyle \!
{\rm A}_{\scriptstyle n}}^{abcd}(x,y)$
are the real and imaginary parts,
respectively, of the bi-tensor
$F_{n}^{abcd}(x,y)$.
{}From the expression (\ref{regul s-t 2})
we see that
the stress-energy operator $\hat{T}_n^{ab}$
can be written as a
sum of terms of the form $\left\{ {\cal A}_x \hat{\phi}_{n}(x),
\,{\cal B}_x \hat{\phi}_{n}(x)\right\}$, where ${\cal A}_x$ and
${\cal B}_x$ are some differential operators. It  then follows
that we can express the bi-tensor
$F_{n}^{abcd}(x,y)$ in
terms of the Wightman function as
\bea
F_{n}^{abcd}(x,y)
\!\!\!\!&=&\!\!\!\!\nabla^{a}_x
 \nabla^{c}_y G_n^+(x,y)
 \nabla^{b}_x
 \nabla^{d}_y G_n^+(x,y)
+\nabla^{a}_x
 \nabla^{d}_y G_n^+(x,y)
 \nabla^{b}_x
 \nabla^{c}_y G_n^+(x,y)
   \nn \\
&&
+\, 2 {\cal D}^{ab}_{x}  \bigl(
  \nabla^{c}_y G_n^+(x,y)
  \nabla^{d}_y G_n^+(x,y) \bigr)
\nonumber  \\
&&
+2 {\cal D}^{cd}_{y} \bigl(
  \nabla^{a}_x G_n^+(x,y)
  \nabla^{b}_x G_n^+(x,y) \bigr)
+2 {\cal D}^{ab}_{x}
   {\cal D}^{cd}_{y}  \bigl(
 G_n^{+ 2}(x,y)  \bigr),
\label{Wightman expression 2}
\eea
where ${\cal D}^{ab}_{x}$ is the differential
operator (\ref{diff operator}).
{}From this expression and the relations
(\ref{finite kernels}), we get expressions for the kernels
$N_n$ and
$H_{\scriptscriptstyle \!{\rm A}_{\scriptstyle n}}$ in
terms of the Wightman function $G_n^+(x,y)$.

Similarly the kernel $H_{\scriptscriptstyle \! {\rm
S}_{\scriptstyle n}}^{abcd}(x,y)$, can be written in terms of the
Feynman function as \bea H_{\scriptscriptstyle \! {\rm
S}_{\scriptstyle n}}^{abcd}(x,y)&\!\!\!\!\!\!=\!\!\!\!\!\!& -
{\rm Im} \Bigl[
 \nabla^{a}_{{x}}
 \nabla^{c}_{{y}}
     G\!_{\scriptscriptstyle F_{\scriptstyle \hspace{0.1ex}  n}}
 \hspace{-0.2ex}(x,y)
 \nabla^{b}_{{x}}
 \nabla^{d}_{{y}}
     G\!_{\scriptscriptstyle F_{\scriptstyle \hspace{0.1ex}  n}}
 \hspace{-0.2ex}(x,y)
\nonumber  \\
&&
+\nabla^{a}_{{x}}
 \nabla^{d}_{{y}}
     G\!_{\scriptscriptstyle F_{\scriptstyle \hspace{0.1ex}  n}}
 \hspace{-0.2ex}(x,y)
 \nabla^{b}_{{x}}
 \nabla^{c}_{{y}}
     G\!_{\scriptscriptstyle F_{\scriptstyle \hspace{0.1ex}  n}}
 \hspace{-0.2ex}(x,y)   \nn \\
&&
-\,g^{ab}(x) \nabla^{e}_{{x}}
 \nabla^{c}_{{y}}
     G\!_{\scriptscriptstyle F_{\scriptstyle \hspace{0.1ex}  n}}
 \hspace{-0.2ex}(x,y)
 \nabla_{\!\!e}^{{x}}
 \nabla^{d}_{{y}}
     G\!_{\scriptscriptstyle F_{\scriptstyle \hspace{0.1ex}  n}}
 \hspace{-0.2ex}(x,y)
\nonumber  \\
&&
-g^{cd}(y) \nabla^{a}_{{x}}
 \nabla^{e}_{{y}}
     G\!_{\scriptscriptstyle F_{\scriptstyle \hspace{0.1ex}  n}}
 \hspace{-0.2ex}(x,y)
 \nabla^{b}_{{x}}
 \nabla_{\!\!e}^{{y}}
     G\!_{\scriptscriptstyle F_{\scriptstyle \hspace{0.1ex}  n}}
 \hspace{-0.2ex}(x,y)    \nn  \\
&&
+\,{1 \over 2}\, g^{ab}(x) g^{cd}(y)
 \nabla^{e}_{{x}}\!
 \nabla^{f}_{{y}}
     G\!_{\scriptscriptstyle F_{\scriptstyle \hspace{0.1ex}  n}}
 \hspace{-0.2ex}(x,y)
 \nabla_{\!\!e}^{{x} }
 \nabla_{\!\!f}^{{y} }
     G\!_{\scriptscriptstyle F_{\scriptstyle \hspace{0.1ex}  n}}
 \hspace{-0.2ex}(x,y)
\nonumber  \\
&&
+{\cal K}^{ab}_{ x}  \bigl(
 2 \hspace{-0.2ex} \nabla^{c}_{{y}}
   G\!_{\scriptscriptstyle F_{\scriptstyle \hspace{0.1ex}  n}}
   \hspace{-0.2ex}(x,y)
 \nabla^{d}_{{y}}
   G\!_{\scriptscriptstyle F_{\scriptstyle \hspace{0.1ex}  n}}
   \hspace{-0.2ex}(x,y)
     \nn   \\
&&
  -\, g^{cd}(y) \nabla^{e}_{{y}}
   G\!_{\scriptscriptstyle F_{\scriptstyle \hspace{0.1ex}  n}}
   \hspace{-0.2ex}(x,y)
\nabla_{\!\!e}^{{y} }
     G\!_{\scriptscriptstyle F_{\scriptstyle \hspace{0.1ex}  n}}
     \hspace{-0.2ex}(x,y) \bigr)
\nonumber  \\
&&
+{\cal K}^{cd}_{y}  \bigl(
 2 \hspace{-0.2ex} \nabla^{a}_{{x}}
   G\!_{\scriptscriptstyle F_{\scriptstyle \hspace{0.1ex}  n}}
   \hspace{-0.2ex}(x,y)
 \nabla^{b}_{{x}}
   G\!_{\scriptscriptstyle F_{\scriptstyle \hspace{0.1ex}  n}}
   \hspace{-0.2ex}(x,y)
     \nn   \\
&&
-\! g^{ab}(x) \nabla^{e}_{{x}}
   G\!_{\scriptscriptstyle F_{\scriptstyle \hspace{0.1ex}  n}}
   \hspace{-0.2ex}(x,y)
\nabla_{\!\!e}^{{x} }
     G\!_{\scriptscriptstyle F_{\scriptstyle \hspace{0.1ex}  n}}
     \hspace{-0.2ex}(x,y) \bigr)
\!+\!2 {\cal K}^{ab}_{x}
   {\cal K}^{cd}_{y} \! \bigl(
   G\!_{\scriptscriptstyle F_{\scriptstyle \hspace{0.1ex}  n}}^{\;\: 2}
   \hspace{-0.2ex}(x,y)  \bigr) \Bigr],
\label{Feynman expression 2}
\eea
where ${\cal K}^{ab}_{x}$ is the differential
operator
\be
{\cal K}^{ab}_{x} \equiv
\xi \left( g^{ab}(x) \Box_{x}
  -\nabla^{a}_{{x}}
   \nabla^{b}_{{x}}+
G^{ab}(x) \right)
-{1 \over 2}\, m^2 g^{ab}(x).
\label{diff operator K}
\ee
Note that, in the vacuum state
$|0 \rangle$, the term
$\langle \hat{\phi}_{n}^2 (x) \rangle$ in
equation (\ref{Einstein-Langevin eq 3}) can also be written as
$\langle \hat{\phi}_{n}^2(x) \rangle=
i G\!_{\scriptscriptstyle F_{\scriptstyle \hspace{0.1ex}  n}}
      \hspace{-0.2ex}(x,x)=G_n^+(x,x)$.

Finally, the causality of the Einstein-Langevin equation
(\ref{Einstein-Langevin eq 3})
can be explicitly seen as follows. The non-local terms in that
equation are due to the kernel $H(x,y)$ which is defined in Eq.
(\ref{H}) as the sum of $H_S(x,y)$ and $H_A(x,y)$. Now,
when the points $x$ and $y$ are spacelike
separated, $\hat{\phi}_{n}(x)$ and $\hat{\phi}_{n}(y)$ commute and,
thus, $G_n^+(x,y) \!=\!
i G\!_{\scriptscriptstyle F_{\scriptstyle \hspace{0.1ex}  n}}
 \hspace{-0.2ex}(x,y) \!=\!
(1/2) \langle 0| \, \{ \hat{\phi}_{n}(x) , \hat{\phi}_{n}(y) \} \,
|0 \rangle$, which is real. Hence, from the above expressions, we
have that $H_{\scriptscriptstyle \! {\rm A}_{\scriptstyle
n}}^{abcd}(x,y) \!=\! H_{\scriptscriptstyle \! {\rm
S}_{\scriptstyle n}}^{abcd}(x,y) \!=\! 0$, and thus
$H_n^{abcd}(x,y)=0$. This fact is expected since, from the
causality of the expectation value of the stress-energy operator
\cite{Wal77}, we know that the non-local dependence on the metric
perturbation in the Einstein-Langevin equation, see Eq.
(\ref{2.11}), must be causal. See Ref.~\cite{HuVer03a} for an
alternative proof of the causal nature of the Einstein-Langevin
equation.

\section{Noise Kernel and Point-Separation}
\label{sec4}

In this section we explore further the properties  of the noise
kernel and the stress energy bi-tensor. Similar to what was done
for the stress energy tensor it is desirable to relate the noise
kernel defined at separated points to the Green function of a
quantum field. We pointed out earlier \cite{stogra} that field
quantities defined at two separated points may possess important
information which could be the starting point for probes into
possible extended structures of spacetime. Of more practical
concern is how one can define a finite quantity at one point or in
some small region around it from the noise kernel defined at two
separated points. When we refer to, say, the fluctuations of
energy density in ordinary (point-wise) quantum field theory, we
are in actuality asking such a question. This is essential for
addressing fundamental issues like
\begin{itemize}
\item the validity of semiclassical gravity~\cite{KuoFor93} -- e.g.,
  whether the fluctuations to mean ratio is a correct
  criterion~\cite{HuPhi00,PhiHu00,ForSCG,ForWu,AndMolMot02,AndMolMot03}
\item whether the fluctuations in the vacuum energy density which
  drives some models of inflationary cosmology violates the positive
  energy condition;
\item physical effects of black hole horizon fluctuations and Hawking
  radiation backreaction -- to begin with, is the fluctuations finite
  or infinite?
\item general relativity as a low energy effective theory in the
  geometro-hydrodynamic limit towards a kinetic theory approach to
  quantum gravity~\cite{grhydro,stogra,kinQG}.
\end{itemize}

Thus, for comparison with ordinary phenomena at low energy we need
to find a reasonable prescription for obtaining a finite quantity
of the noise kernel in the limit of ordinary (point-defined)
quantum field theory. It is well-known that several
regularization methods can work equally well for the removal of
ultraviolet divergences in the stress energy tensor of quantum
fields  in curved spacetime. Their mutual relations are known, and
discrepancies explained. This formal structure of regularization
schemes for quantum fields in curved spacetime should remain
intact when applied to the regularization of the noise kernel in
general curved spacetimes; it is the meaning and relevance of
regularization of the noise kernel which is more of a concern (see
comments below).  Specific considerations will of course enter for
each method. But for the methods employed so far, such as
zeta-function, point separation, dimensional, smeared-field,
applied to simple cases (Casimir, Einstein, thermal fields) there
is no new inconsistency or discrepancy.

Regularization schemes used in obtaining a
finite expression for the stress energy tensor  have been applied
to the noise kernel.
This includes the simple
normal ordering \cite{KuoFor93,WuFor01} and smeared field operator
\cite{PhiHu00} methods applied to the Minkowski and Casimir
spaces, zeta-function \cite{EliEtal94,Kir00,Cam90} for spacetimes
with an Euclidean section, applied to the Casimir effect
\cite{CogGuiEli02} and the Einstein Universe \cite{PhiHu97}, or the covariant
point-separation methods applied to the Minkowski \cite{PhiHu00},
hot flat space and the Schwarzschild spacetime \cite{PhiHu03}.
There are differences and deliberations on
whether it is meaningful to seek a
point-wise expression for the noise kernel, and if so what is the
correct way to proceed -- e.g., regularization by a subtraction
scheme or by integrating over a test-field. Intuitively the smear
field method \cite{PhiHu00} may better preserve the
integrity of the noise kernel as it provides a sampling of the two
point function rather than using
a subtraction scheme which alters its innate
properties by forcing a nonlocal quantity into a local one. More
investigation is needed to clarify these points, which bear on
important issues like the validity of semiclassical gravity. We
shall set a more modest goal here, to derive a general expression
for the noise kernel for quantum fields in an arbitrary curved
spacetime in terms of Green functions and leave the discussion of
point-wise limit to a later date. For this purpose the covariant
point-separation method which highlights the bi-tensor features,
when used not as a regularization scheme, is perhaps closest to
the spirit of stochastic gravity.

The task of finding a general expression of the noise-kernel for
quantum fields in curved spacetimes was carried out by Phillips
and Hu in two papers using the ``modified'' point separation
scheme \cite{Wal75,AdlLieNg77,Wal78}. Their first paper
\cite{PhiHu01} begins with a discussion of the procedures for
dealing with the quantum stress tensor bi-operator at two
separated points, and ends with a general expression of the noise
kernel defined at separated points expressed as products of
covariant derivatives up to the fourth order of the quantum
field's Green function. (The stress tensor involves up to two
covariant derivatives.) This result holds for $x\ne y$ without
recourse to renormalization of the Green function, showing that
$N_{abc'd'}(x,y)$ is always finite for $x\ne y$ (and off the light
cone for massless theories). In particular, for a massless
conformally coupled free scalar field on a four dimensional
manifold they computed the trace of the noise kernel at both
points and found this double trace vanishes identically. This
implies that there is no stochastic correction to the trace
anomaly  for massless conformal fields, in agreement with results
arrived at in Refs. \cite{CalHu94,CamVer96,MarVer99} (see also
section \ref{sec2}). In their second paper \cite{PhiHu03} a
Gaussian approximation for the Green function (which is what
limits the accuracy of the results) is used to derive finite
expressions for two specific classes of spacetimes, ultrastatic
spacetimes, such as the hot flat space, and the conformally-
ultrastatic spacetimes, such as the Schwarzschild spacetime.
Again, the validity of these results may depend on how we view the
relevance and meaning of regularization. We will only report the
result of their first paper here.


\subsection{Point Separation}

The point separation scheme introduced in the 60's by DeWitt
\cite{DeW65}  was brought to more popular use in the 70's in the
context of quantum field theory in curved spacetimes
\cite{DeW75,Chr76,Chr78} as a means for obtaining a finite
quantum stress tensor.  Since the stress-energy tensor is built
from the product of a pair of field operators evaluated at a
single point, it is not well-defined. In this scheme, one
introduces an artificial separation of the single point $x$ to a
pair of closely separated points $x$ and $x'$. The problematic
terms involving field products such as $\hat\phi(x)^2$ becomes
$\hat\phi(x)\hat\phi(x')$, whose expectation value is well
defined. If one is interested in the low energy behavior captured
by the point-defined quantum field theory -- as the effort in the
70's was directed -- one takes the coincidence limit. Once the
divergences present are identified, they may be  removed
(regularization) or moved (by renormalizing the coupling
constants), to produce  a well-defined, finite stress tensor at a
single point.

Thus the first order of business is  the construction of the
stress tensor and then derive the symmetric stress-energy tensor
two point function, the noise kernel, in terms of the Wightman
Green function. In this section we will use the traditional
notation for index tensors in the point-separation context.

\subsubsection{n-tensors and end-point expansions}

An object like the Green function $G(x,y)$ is an example of a
{\em bi-scalar}: it transforms as scalar at both points $x$ and
$y$. We can also define a {\em bi-tensor}\, $T_{a_1\cdots
a_n\,b'_1\cdots b'_m}(x,y)$: upon a coordinate transformation,
this transforms as a rank $n$ tensor at $x$ and a rank $m$ tensor
at $y$. We will extend this up to a {\em quad-tensor}\,
$T_{a_1\cdots a_{n_1}\,b'_1\cdots b'_{n_2}\,
    c''_1\cdots c''_{n_3}\,d'''_1\cdots d'''_{n_4}}$
which has support at four points $x,y,x',y'$, transforming as
rank $n_1,n_2,n_3,n_4$ tensors at each of the four points. This
also sets the notation we will use: unprimed indices referring to
the tangent space constructed above $x$, single primed indices to
$y$, double primed to $x'$ and triple primed to $y'$. For each
point, there is the covariant derivative $\nabla_a$ at that point.
Covariant derivatives at different points commute and the
covariant derivative at, say, point $x'$, does not act on a
bi-tensor defined at, say,  $x$ and $y$:
\begin{equation}
T_{ab';c;d'} = T_{ab';d';c} \quad {\rm and } \quad T_{ab';c''} =
0.
\end{equation}
To simplify notation, henceforth we will eliminate the semicolons
after the first one for multiple covariant derivatives at
multiple points.

Having objects defined at different points, the {\rm coincident
limit} is defined as evaluation ``on the diagonal'', in the sense
of the spacetime support of the function or tensor, and the usual
shorthand $\left[ G(x,y) \right] \equiv G(x,x)$ is used. This
extends to $n$-tensors as
\begin{equation}
\left[ T_{a_1\cdots a_{n_1}\,b'_1\cdots b'_{n_2}\,
    c''_1\cdots c''_{n_3}\,d'''_1\cdots d'''_{n_4}} \right] =
T_{a_1\cdots a_{n_1}\,b_1\cdots b_{n_2}\,
    c_1\cdots c_{n_3}\,d_1\cdots d_{n_4}},
\end{equation}
{\it i.e.}, this becomes a rank $(n_1+n_2+n_3+n_4)$ tensor at $x$.
The multi-variable chain rule relates covariant derivatives
acting at different points, when we are interested in the
coincident limit:
\begin{equation}
\left[ T_{a_1\cdots a_m \,b'_1\cdots b'_n} \right]\!{}_{;c} =
\left[ T_{a_1\cdots a_m \,b'_1\cdots b'_n;c} \right] + \left[
T_{a_1\cdots a_m \,b'_1\cdots b'_n;c'} \right].
\label{ref-Synge's}
\end{equation}
This result is referred to as {\em Synge's theorem} in this
context; we  follow Fulling's \cite{Ful89} discussion.

The bi-tensor of {\em parallel transport}\, $g_a{}^{b'}$ is
defined such that when it acts on a vector $v_{b'}$ at $y$, it
parallel transports the vector along the geodesics connecting $x$
and $y$. This allows us to add vectors and tensors defined at
different points. We cannot directly add a vector $v_a$ at $x$
and vector $w_{a'}$ at $y$. But by using $g_a{}^{b'}$, we can
construct the sum $v^a + g_a{}^{b'} w_{b'}$. We will also need
the obvious property $\left[ g_a{}^{b'} \right] = g_a{}^b$.

The main bi-scalar we need  is the {\em world function}
$\sigma(x,y)$. This is defined as a half of the square of the
geodesic distance between the points $x$ and $y$. It satisfies
the equation
\begin{equation}
\sigma = \frac{1}{2} \sigma^{;p} \sigma_{;p} \label{define-sigma}
\end{equation}
Often in the literature, a covariant derivative is implied when
the world function appears with indices: $\sigma^a \equiv
\sigma^{;a}$, {\it i.e.}taking the covariant derivative at $x$,
while $\sigma^{a'}$ means the covariant derivative at $y$. This
is done since the vector $-\sigma^a$ is the tangent vector to the
geodesic with length equal the distance between $x$ and $y$. As
$\sigma^a$ records information about distance and direction for
the two points  this makes it ideal for constructing a series
expansion of a bi-scalar.  The {\em end point} expansion of a
bi-scalar $S(x,y)$ is of the form
\begin{equation}
S(x,y) = A^{(0)} + \sigma^p A^{(1)}_p + \sigma^p \sigma^q
A^{(2)}_{pq} + \sigma^p \sigma^q  \sigma^r A^{(3)}_{pqr} +
\sigma^p \sigma^q  \sigma^r \sigma^s A^{(4)}_{pqrs} + \cdots
\label{general-endpt-series}
\end{equation}
where, following our convention, the expansion tensors
$A^{(n)}_{a_1\cdots a_n}$ with  unprimed indices have support at
$x$ and hence the name end point expansion. Only the symmetric
part of these tensors  contribute to the expansion. For the
purposes of multiplying series expansions it is convenient to
separate the distance dependence from the direction dependence.
This is done by introducing the unit vector $p^a =
\sigma^a/\sqrt{2\sigma}$. Then the series expansion can be written
\begin{equation}
S(x,y) = A^{(0)} + \sigma^{\frac{1}{2}} A^{(1)} + \sigma  A^{(2)}
+ \sigma^{\frac{3}{2}} A^{(3)} + \sigma^2 A^{(4)} + \cdots
\end{equation}
The expansion scalars are related, via
$A^{(n)} = 2^{n/2} A^{(n)}_{p_1\cdots p_n} p^{p_1}\cdots p^{p_n}$,
to the expansion tensors.

The last object we need  is the {\em VanVleck-Morette}
determinant $D(x,y)$, defined as $D(x,y) \equiv -\det\left(
-\sigma_{;ab'} \right)$. The related bi-scalar
\begin{equation}
{\Delta\!^{1/2}} = \left( \frac{D(x,y)}{\sqrt{g(x)
g(y)}}\right)^\frac{1}{2}
\end{equation}
satisfies the equation
\begin{equation}
{\Delta\!^{1/2}}\left(4-\sigma_{;p}{}^p\right) -
2{\Delta\!^{1/2}}_{\,\,;p}\sigma^{;p} = 0 \label{define-VanD}
\end{equation}
with the boundary condition $\left[{\Delta\!^{1/2}}\right] = 1$.

Further details on these objects and discussions of the
definitions and properties are contained in \cite{Chr76,Chr78}
and \cite{NPsc}. There it is shown how the defining equations for
$\sigma$ and ${\Delta\!^{1/2}}$ are used to determine the
coincident limit expression for the various covariant derivatives
of the world function ($\left[ \sigma_{;a}\right]$, $\left[
\sigma_{;ab}\right]$, {\it etc.}) and how the defining
differential equation for ${\Delta\!^{1/2}}$ can be used to
determine the series expansion of ${\Delta\!^{1/2}}$. We show how
the expansion tensors $A^{(n)}_{a_1\cdots a_n}$ are determined in
terms of the coincident limits of covariant derivatives of the
bi-scalar $S(x,y)$.  (Ref.  \cite{NPsc} details how point
separation can be implemented on the computer to provide easy
access to a wider range of applications involving higher
derivatives of the curvature tensors.)

\subsection{Stress Energy Bi-Tensor Operator and Noise Kernel}

Even though we believe that the point-separated results are more
basic in the sense that it reflects a deeper structure of the
quantum theory of spacetime, we will nevertheless start with
quantities defined at one point because they are what enter in
conventional quantum field theory. We will use point separation
to introduce the bi-quantities. The key issue here is thus the
distinction between point-defined ({\it pt}) and point-separated
({\it bi}) quantities.

For  a free classical scalar field $\phi$ with the action
$S_m[g,\phi]$ defined in Eq. (\ref{2.1}), the classical
stress-energy tensor is
\begin{eqnarray}
T_{ab} &=& \left( 1 - 2\,\xi  \right) \,{\phi {}_;{}_{a}}\,{\phi
{}_;{}_{b}}
 + \left(2\,\xi -{1\over 2} \right) \,{\phi {}_;{}_{p}}
\,{\phi {}^;{}^{p}}\,{g{}_{a}{}_{b}} + 2\xi\,\phi \,
\,\left({\phi {}_;{}_{p}{}^{p} -{\phi {}_;{}_{a}{}_{b}}}
\,{g{}_{a}{}_{b}} \right)  \cr &&+ {{\phi }^2}\,\xi \,
\left({R{}_{a}{}_{b}} - {1\over 2}{ R\,{g{}_{a}{}_{b}}  }
 \right)
 - \frac{1}{2}{{m^2}\,{{\phi }^2}\,{g{}_{a}{}_{b}}},
\label{ref-define-classical-emt}
\end{eqnarray}
which is equivalent to the tensor of Eq. (\ref{2.3}) but written
in a slightly different form for convenience. When we make the
transition to quantum field theory, we promote the field
$\phi(x)$ to a field operator $\hat\phi(x)$. The fundamental
problem of defining a quantum operator for the stress tensor is
immediately visible: the field operator appears quadratically.
Since $\hat\phi(x)$ is an operator-valued distribution, products
at a single point are not well-defined. But if the product is
point separated ($\hat\phi^2(x) \rightarrow
\hat\phi(x)\hat\phi(x')$), they are finite and well-defined.

Let us first seek a point-separated extension of these classical
quantities and then consider the quantum field operators. Point
separation is symmetrically extended to products of covariant
derivatives of the field according to
\begin{eqnarray}
\left({\phi {}_;{}_{a}}\right)\left({\phi {}_;{}_{b}}\right)
&\rightarrow &\frac{1}{2}\left(
g_a{}^{p'}\nabla_{p'}\nabla_{b}+g_b{}^{p'}\nabla_a\nabla_{p'}
\right)\phi(x)\phi(x'),
\nonumber\\
\phi \,\left({\phi {}_;{}_{a}{}_{b}}\right) &\rightarrow&
\frac{1}{2}\left(
\nabla_a\nabla_b+g_a{}^{p'}g_b{}^{q'}\nabla_{p'}\nabla_{q'}
\right)\phi(x)\phi(x').
\nonumber
\end{eqnarray}
The bi-vector of parallel displacement $g_a{}^{a'}(x,x')$ is
included so that we may have objects that are rank 2 tensors at
$x$ and scalars at $x'$.

To carry out  point separation on
(\ref{ref-define-classical-emt}), we first define the
differential operator
\begin{eqnarray}
{\cal T}_{ab} &=&
  \frac{1}{2}\left(1-2\xi\right)
   \left(g_a{}^{a'}\nabla_{a'}\nabla_{b}+g_b{}^{b'}\nabla_a\nabla_{b'}\right)
+ \left(2\xi-\frac{1}{2}\right)
     g_{ab}g^{cd'}\nabla_c\nabla_{d'} \cr
&& - \xi
     \left(\nabla_a\nabla_b+g_a{}^{a'}g_b{}^{b'}\nabla_{a'}\nabla_{b'}\right)
+ \xi g_{ab}
     \left(\nabla_c\nabla^c+\nabla_{c'}\nabla^{c'}\right) \cr
&& +\xi\left(R_{ab} - \frac{1}{2}g_{ab}R\right)
    -\frac{1}{2}m^2 g_{ab}
\label{PSNoise-emt-diffop}
\end{eqnarray}
from which we obtain the classical stress tensor as
\begin{equation}
T_{ab}(x) = \lim_{x' \rightarrow x} {\cal T}_{ab}\phi(x)\phi(x').
\end{equation}
That the classical tensor field no longer appears as a product of
scalar fields at a single point allows a smooth transition to the
quantum tensor field. From the viewpoint of the stress tensor,
the separation of points is an artificial construct so when
promoting the classical field to a quantum one, neither point
should be favored. The product of field configurations is taken
to be the symmetrized operator product, denoted by curly brackets:
\begin{equation}
\phi(x)\phi(y) \rightarrow \frac{1}{2}
 \left\{{\hat\phi(x)},{\hat\phi(y)}\right\}
= \frac{1}{2}\left( {\hat\phi(x)} {\hat\phi(y)} +
                    {\hat\phi(y)} {\hat\phi(x)}
\right)
\end{equation}
With this, the point separated stress energy tensor operator is
defined as
\begin{equation}
\hat T_{ab}(x,x') \equiv \frac{1}{2} {\cal
T}_{ab}\left\{\hat\phi(x),\hat\phi(x')\right\}.
\label{PSNoise-emt-define}
\end{equation}
While the classical stress tensor was defined at the coincidence
limit $x'\rightarrow x$, we cannot attach any physical meaning to
the quantum stress tensor at one point until the issue of
regularization is dealt with, which will happen in the next
section. For now,  we will maintain point separation so as to
have a mathematically meaningful operator.

The expectation value of the point-separated stress tensor can
now be taken. This amounts to replacing the field operators by
their expectation value, which is given by the Hadamard (or
Schwinger) function
\begin{equation}
{G^{(1)}}(x,x') =
     \langle\left\{{\hat\phi(x)},{\hat\phi(x')}\right\}\rangle.
\end{equation}
and the point-separated stress tensor is defined as
\begin{equation}
\langle \hat T_{ab}(x,x') \rangle = \frac{1}{2} {\cal
T}_{ab}{G^{(1)}}(x,x') \label{ref-emt-PSdefine}
\end{equation}
where, since ${\cal T}_{ab}$ is a differential operator, it can
be taken ``outside'' the expectation value. The expectation value
of the point-separated quantum stress tensor for a free, massless
($m=0$)  conformally coupled ($\xi=1/6$) scalar field on a four
dimension spacetime with scalar curvature $R$ is
\begin{eqnarray}
\langle \hat T_{ab}(x,x') \rangle &=&
  \frac{1}{6}\left( {g{}^{p'}{}_{b}}\,{{G^{(1)}}{}_;{}_{p'}{}_{a}}
 + {g{}^{p'}{}_{a}}\,{{G^{(1)}}{}_;{}_{p'}{}_{b}} \right)
 -\frac{1}{12} {g{}^{p'}{}_{q}}\,{{G^{(1)}}{}_;{}_{p'}{}^{q}}
\,{g{}_{a}{}_{b}} \cr && -\frac{1}{12}\left(
{g{}^{p'}{}_{a}}\,{g{}^{q'}{}_{b}}
\,{{G^{(1)}}{}_;{}_{p'}{}_{q'}} + {{G^{(1)}}{}_;{}_{a}{}_{b}}
\right)
\nonumber  \\
&&
 +\frac{1}{12}\left( \left( {{G^{(1)}}{}_;{}_{p'}{}^{p'}}
 + {{G^{(1)}}{}_;{}_{p}{}^{p}} \right) \,{g{}_{a}{}_{b}} \right) \cr
&& +\frac{1}{12} {G^{(1)}}\, \left({R{}_{a}{}_{b}} -{1\over 2}
R\,{g{}_{a}{}_{b}} \right)
\end{eqnarray}

\subsubsection{Finiteness of Noise Kernel}

We now turn our attention to the noise kernel introduced in Eq.
(\ref{2.8}), which  is the symmetrized product of the (mean
subtracted) stress tensor operator:
\begin{eqnarray}
8 N_{ab,c'd'}(x,y) &=& \langle \left\{
        \hat T_{ab}(x)-\langle \hat T_{ab}(x)\rangle,
    \hat T_{c'd'}(y)-\langle \hat T_{c'd'}(y) \rangle
\right\} \rangle \cr &=& \langle \left\{ \hat T_{ab}(x),\hat
T_{c'd'}(y) \right\} \rangle -2 \langle \hat
T_{ab}(x)\rangle\langle \hat T_{c'd'}(y) \rangle
\end{eqnarray}
Since $\hat T_{ab}(x)$ defined at one point can be ill-behaved as
it is generally divergent, one can question the soundness of
these quantities. But as will be shown later, the noise kernel is
finite for $y\neq x$. All field operator products present in the
first expectation value that could be divergent are canceled by
similar products in the second term. We will replace each of the
stress tensor operators in the above expression for the noise
kernel by their point separated versions, effectively separating
the two points $(x,y)$ into the four points $(x,x',y,y')$. This
will allow us to express the noise kernel in terms of a pair of
differential operators acting on a combination of four and two
point functions. Wick's theorem will allow the four point
functions to be re-expressed in terms of two point functions.
{}From this we see that all possible divergences for $y\neq x$ will
cancel. When the coincidence limit is taken divergences do occur.
The above procedure will allow us to isolate the divergences and
obtain a finite result.

Taking the point-separated quantities as more basic, one should
replace each of the stress tensor operators in the above with the
corresponding point separated version (\ref{PSNoise-emt-define}),
with ${\cal T}_{ab}$ acting at $x$ and $x'$ and ${\cal T}_{c'd'}$
acting at $y$ and $y'$. In this framework the noise kernel is
defined as
\begin{equation}
8 N_{ab,c'd'}(x,y) =
   \lim_{x'\rightarrow x}\lim_{y'\rightarrow y}
   {\cal T}_{ab} {\cal T}_{c'd'}\, G(x,x',y,y')
\end{equation}
where the four point function is
\begin{eqnarray}
G(x,x',y,y') &=& \frac{1}{4}\left[
\langle\left\{\left\{{\hat\phi(x)},{\hat\phi(x')}\right\},\left\{{\hat\phi(y)}
,{\hat\phi(y')}\right\}\right\}\rangle \right.
\cr\cr&&\hspace{1cm}\left.
  -2\,\langle\left\{{\hat\phi(x)},{\hat\phi(x')}\right\}\rangle
  \langle\left\{{\hat\phi(y)},{\hat\phi(y')}\right\}\rangle \right].
\label{PSNoise-G4a}
\end{eqnarray}
We assume the pairs $(x,x')$ and $(y,y')$ are each within their
respective Riemann normal coordinate neighborhoods so as to avoid
problems that possible geodesic caustics might be present. When
we later turn our attention to computing the limit $y\rightarrow
x$, after issues of regularization are addressed, we will want to
assume all four points are within the same Riemann normal
coordinate neighborhood.

Wick's theorem, for the case of free fields which we are
considering, gives the simple product four point function in terms
of a sum of products of Wightman functions (we use the shorthand
notation $G_{xy}\equiv G_{+}(x,y) =
\langle{\hat\phi(x)}\,{\hat\phi(y)}\rangle$):
\begin{equation}
\langle{\hat\phi(x)}\,{\hat\phi(y)}\,{\hat\phi(x')}\,{\hat\phi(y')}\rangle
= {G_{xy'}}\,{G_{yx'}} + {G_{xx'}}\,{G_{yy'}} +
{G_{xy}}\,{G_{x'y'}}
\end{equation}
Expanding out the anti-commutators in (\ref{PSNoise-G4a}) and
applying Wick's theorem, the four point function becomes
\begin{equation}
G(x,x',y,y')  = {G_{xy'}}\,{G_{x'y}} + {G_{xy}}\,{G_{x'y'}} +
{G_{yx'}}\,{G_{y'x}} + {G_{yx}} \,{G_{y'x'}}.
\end{equation}
We can now easily see that the noise kernel defined via this
function is indeed well defined for the limit $(x',y')\rightarrow
(x,y)$:
\begin{equation}
G(x,x,y,y) = 2\,\left( {{{G^2_{xy}}}} + {{{G^2_{yx}}}} \right) .
\end{equation}
{}From this we can see that the noise kernel is also well defined
for $y \neq x$; any divergence present in the first expectation
value of (\ref{PSNoise-G4a}) have been cancelled by those present
in the pair of Green functions in the second term, in agreement
with the results of section \ref{sec2}.

\subsubsection{Explicit Form of the Noise Kernel}

We will let the points  separated for a while so we can keep
track of which covariant derivative acts on which arguments of
which Wightman function. As an example (the complete calculation
is quite long), consider the result of the first set of covariant
derivative operators in the differential operator
(\ref{PSNoise-emt-diffop}), from both ${\cal T}_{ab}$ and ${\cal
T}_{c'd'}$, acting on $G(x,x',y,y')$:
\begin{eqnarray}
&&\frac{1}{4}\left(1-2\xi\right)^2
   \left(g_a{}^{p''}\nabla_{p''}\nabla_{b}+
         g_b{}^{p''}\nabla_{p''}\nabla_{a}\right)\cr
&&\hspace{17mm}\times
   \left(g_{c'}{}^{q'''}\nabla_{q'''}\nabla_{d'}
        +g_{d'}{}^{q'''}\nabla_{q'''}\nabla_{c'}\right)
    G(x,x',y,y').
\end{eqnarray}
(Our notation is that $\nabla_a$ acts at $x$, $\nabla_{c'}$ at
$y$, $\nabla_{b''}$ at $x'$ and $\nabla_{d'''}$ at $y'$).
Expanding out the differential operator above, we can determine
which derivatives act on which Wightman function:
\begin{eqnarray}
{{{{\left( 1 - 2\,\xi  \right) }^2}}\over 4} &\times & \left[
    {g{}_{c'}{}^{p'''}}\,{g{}^{q''}{}_{a}}
 \left( {{G_{xy'}}{}_;{}_{b}{}_{p'''}}\,{{G_{x'y}}{}_;{}_{q''}{}_{d'}}
 + {{G_{xy}}{}_;{}_{b}{}_{d'}}\,{{G_{x'y'}}{}_;{}_{q''}{}_{p'''}}
 \right.\right. \cr
&&  \hspace{18mm} + \left. {{G_{yx'}}{}_;{}_{q''}{}_{d'}}
\,{{G_{y'x}}{}_;{}_{b}{}_{p'''}} + {{G_{yx}}{}_;{}_{b}{}_{d'}}
\,{{G_{y'x'}}{}_;{}_{q''}{}_{p'''}} \right) \cr &&+
{g{}_{d'}{}^{p'''}}\,{g{}^{q''}{}_{a}}
 \left( {{G_{xy'}}{}_;{}_{b}{}_{p'''}}\,{{G_{x'y}}{}_;{}_{q''}{}_{c'}}
 + {{G_{xy}}{}_;{}_{b}{}_{c'}}\,{{G_{x'y'}}{}_;{}_{q''}{}_{p'''}} \right. \cr
&&  \hspace{18mm} + \left. {{G_{yx'}}{}_;{}_{q''}{}_{c'}}
\,{{G_{y'x}}{}_;{}_{b}{}_{p'''}} + {{G_{yx}}{}_;{}_{b}{}_{c'}}
\,{{G_{y'x'}}{}_;{}_{q''}{}_{p'''}} \right) \cr &&+
{g{}_{c'}{}^{p'''}}\,{g{}^{q''}{}_{b}}
 \left( {{G_{xy'}}{}_;{}_{a}{}_{p'''}}\,{{G_{x'y}}{}_;{}_{q''}{}_{d'}}
 + {{G_{xy}}{}_;{}_{a}{}_{d'}}\,{{G_{x'y'}}{}_;{}_{q''}{}_{p'''}} \right. \cr
&&  \hspace{18mm} + \left. {{G_{yx'}}{}_;{}_{q''}{}_{d'}}
\,{{G_{y'x}}{}_;{}_{a}{}_{p'''}} + {{G_{yx}}{}_;{}_{a}{}_{d'}}
\,{{G_{y'x'}}{}_;{}_{q''}{}_{p'''}} \right) \cr &&+
{g{}_{d'}{}^{p'''}}\,{g{}^{q''}{}_{b}}
 \left( {{G_{xy'}}{}_;{}_{a}{}_{p'''}}\,{{G_{x'y}}{}_;{}_{q''}{}_{c'}}
 + {{G_{xy}}{}_;{}_{a}{}_{c'}}\,{{G_{x'y'}}{}_;{}_{q''}{}_{p'''}} \right. \cr
&&  \hspace{18mm} +\! \left.\left. {{G_{yx'}}{}_;{}_{q''}{}_{c'}}
{{G_{y'x}}{}_;{}_{a}{}_{p'''}}\! +\! {{G_{yx}}{}_;{}_{a}{}_{c'}}
{{G_{y'x'}}{}_;{}_{q''}{}_{p'''}} \right) \right]\! .
\end{eqnarray}
If we now  let $x'\rightarrow x$ and $y' \rightarrow y$ the
contribution to the noise kernel is (including the factor of
$\frac{1}{8}$ present in the definition of the noise kernel):
\begin{eqnarray}
&&\frac{1}{8}\left\{ {{\left( 1 - 2\,\xi  \right) }^2} \,\left(
{{G_{xy}}{}_;{}_{a}{}_{d'}}\,{{G_{xy}}{}_;{}_{b}{}_{c'}}
 + {{G_{xy}}{}_;{}_{a}{}_{c'}}\,{{G_{xy}}{}_;{}_{b}{}_{d'}}
 \right)  \right. \cr
&&\hspace{20mm} \left. + {{\left( 1 - 2\,\xi  \right) }^2}
\,\left( {{G_{yx}}{}_;{}_{a}{}_{d'}}\,{{G_{yx}}{}_;{}_{b}{}_{c'}}
 + {{G_{yx}}{}_;{}_{a}{}_{c'}}\,{{G_{yx}}{}_;{}_{b}{}_{d'}} \right)
 \right\}.
\end{eqnarray}
That this term can be written as the sum of a part involving
$G_{xy}$ and one involving $G_{yx}$ is a general property of the
entire noise kernel. It thus takes the form
\begin{equation}
N_{abc'd'}(x,y) = N_{abc'd'}\left[ G_{+}(x,y)\right]
                + N_{abc'd'}\left[ G_{+}(y,x)\right].
\end{equation}
We will present the form of the functional $N_{abc'd'}\left[ G
\right]$ shortly. First we note, for $x$ and $y$ time-like
separated, the above split of the noise kernel allows us to
express it in terms of the Feynman (time ordered) Green function
$G_F(x,y)$ and the Dyson (anti-time ordered) Green function
$G_D(x,y)$:
\begin{equation}
N_{abc'd'}(x,y) = N_{abc'd'}\left[ G_F(x,y)\right]
                + N_{abc'd'}\left[ G_D(x,y)\right].
\label{noiker}
\end{equation}
This can be connected  with the zeta function approach
to this problem \cite{PhiHu97} as follows: Recall when the quantum
stress tensor fluctuations determined in the Euclidean section is
analytically continued back to Lorentzian signature ($\tau
\rightarrow i t$), the time ordered product results. On the other
hand, if the continuation is $\tau \rightarrow -i t$, the
anti-time ordered product results. With this in mind, the noise
kernel is seen to be related to the quantum stress tensor
fluctuations derived via the effective action as
\begin{equation}
16 N_{abc'd'} =
   \left.\Delta T^2_{abc'd'}\right|_{t=-i\tau,t'=-i\tau'}
 + \left.\Delta T^2_{abc'd'}\right|_{t= i\tau,t'= i\tau'}.
\end{equation}
The complete form of the functional $N_{abc'd'}\left[ G \right]$
is
\begin{equation}
 N_{abc'd'}\left[ G \right]  =
    \tilde N_{abc'd'}\left[ G \right]
  + g_{ab}   \tilde N_{c'd'}\left[ G \right]
 + g_{c'd'} \tilde N'_{ab}\left[ G \right]
 + g_{ab}g_{c'd'} \tilde N\left[ G \right].
\label{general-noise-kernel}
\end{equation}
with
\begin{eqnarray}
8 \tilde N_{abc'd'} \left[ G \right]\!\!\!\!&=&\!\!\!\!
{{\left(1\!-\!2\xi\right)}^2}\!\left( G{}\!\,_{;}{}_{c'}{}_{b}
     G{}\!\,_{;}{}_{d'}{}_{a}
+
    G{}\!\,_{;}{}_{c'}{}_{a}\,G{}\!\,_{;}{}_{d'}{}_{b} \right)
\nonumber  \\
&&\!\!\!\!\!
+4{{\xi}^2}\left(
G{}\!\,_{;}{}_{c'}{}_{d'}\,G{}\!\,_{;}{}_{a}{}_{b} +
    G\,G{}\!\,_{;}{}_{a}{}_{b}{}_{c'}{}_{d'} \right)  \cr
&& \!\!\!\!\!-\!2\xi\!\left(1\!-\!2\xi\right)\!
  \left( G{}\!\,_{;}{}_{b}G{}\!\,_{;}{}_{c'}{}_{a}{}_{d'} \!+\!
    G{}\!\,_{;}{}_{a}G{}\!\,_{;}{}_{c'}{}_{b}{}_{d'} \!+\!
    G{}\!\,_{;}{}_{d'}G{}\!\,_{;}{}_{a}{}_{b}{}_{c'} \!+\!
    G{}\!\,_{;}{}_{c'}G{}\!\,_{;}{}_{a}{}_{b}{}_{d'}\! \right)  \cr
&&\!\!\!\!\!+\!2\xi\!\left(1\!-\!2\xi\right)\!\left(
G{}\!\,_{;}{}_{a}\,G{}\!\,_{;}{}_{b}\,
     {R{}_{c'}{}_{d'}} + G{}\!\,_{;}{}_{c'}\,G{}\!\,_{;}{}_{d'}\,
     {R{}_{a}{}_{b}} \right)  \cr
&& \!\!\!\!\! -4\!{{\xi}^2}\!\left(
G{}\!\,_{;}{}_{a}{}_{b}\,{R{}_{c'}{}_{d'}} +
    G{}\!\,_{;}{}_{c'}{}_{d'}\,{R{}_{a}{}_{b}} \right)  G
 +  2\,{{\xi}^2}\,{R{}_{c'}{}_{d'}}\,{R{}_{a}{}_{b}} {G^2},
\end{eqnarray}
\begin{eqnarray}
8 \tilde N'_{ab} \left[ G \right]\!\!\!&=&\!\!\!\! 2(1\!-\!2\xi)
\!\left[
   \left(\!2\xi\!-\!{\frac{1}{2}\!}\right)G{}\!\,_{;}{}_{p'}{}_{b}\,
  G{}\!\,_{;}{}^{p'}{}_{a}
 \!+\! \xi\!\left( G{}\!\,_{;}{}_{b}\,G{}\!\,_{;}{}_{p'}{}_{a}{}^{p'} \!+\!
    G{}\!\,_{;}{}_{a}\,G{}\!\,_{;}{}_{p'}{}_{b}{}^{p'} \right)
\right]\cr &&
%
-4\xi \left[
    \left(\!2\xi\!-\!{\frac{1}{2}}\!\right)\,G{}\!\,_{;}{}^{p'}\,
  G{}\!\,_{;}{}_{a}{}_{b}{}_{p'}
  + \xi\,\left( G{}\!\,_{;}{}_{p'}{}^{p'}\,G{}\!\,_{;}{}_{a}{}_{b} +
    G\,G{}\!\,_{;}{}_{a}{}_{b}{}_{p'}{}^{p'} \right)
\right] \cr
%
&& -({m^2}+\xi R')\,\left[(1-2\,\xi)\,G{}\!\,_{;}{}_{a}\,
     G{}\!\,_{;}{}_{b} - 2\,G\,\xi\,G{}\!\,_{;}{}_{a}{}_{b} \right]  \cr
%
&& + 2\xi\,\left[
\left(\!2\xi\!-\!{\frac{1}{2}}\!\right)\,G{}\!\,_{;}{}_{p'}\,
     G{}\!\,_{;}{}^{p'} + 2\,G\,\xi\,G{}\!\,_{;}{}_{p'}{}^{p'} \right] \,
  {R{}_{a}{}_{b}} \cr
%
&& - ({m^2}+\xi R')\,\xi\,{R{}_{a}{}_{b}} {G^2},
\end{eqnarray}
\begin{eqnarray}
8 \tilde N \left[ G \right]\!\!\!\!&=&\!\!\!\!
2{{\left(2\xi\!-\!{\frac{1}{2}}\right)}^2}G{}\!\,_{;}{}_{p'}{}_{q}\,
  G{}\!\,_{;}{}^{p'}{}^{q}
+ 4{{\xi}^2}\left(
G{}\!\,_{;}{}_{p'}{}^{p'}\,G{}\!\,_{;}{}_{q}{}^{q} +
    G\,G{}\!\,_{;}{}_{p}{}^{p}{}_{q'}{}^{q'} \right)  \cr
&&\!\!\!\! + 4\xi\left(2\xi\!-\!{\frac{1}{2}}\right)\,
  \left( G{}\!\,_{;}{}_{p}\,G{}\!\,_{;}{}_{q'}{}^{p}{}^{q'} +
    G{}\!\,_{;}{}^{p'}\,G{}\!\,_{;}{}_{q}{}^{q}{}_{p'} \right)  \cr
&&\!\!\!\! - \left(2\xi\!-\!{\frac{1}{2}}\right)\,
  \left[ \left({m^2}+\xi R\right)\,G{}\!\,_{;}{}_{p'}\,G{}\!\,_{;}{}^{p'} +
    \left({m^2}+\xi R'\right)\,G{}\!\,_{;}{}_{p}\,G{}\!\,^{;}{}^{p} \right]  \cr
&&\!\!\!\! - 2\xi\left[ \left({m^2}+\xi
R\right)\,G{}\!\,_{;}{}_{p'}{}^{p'} +
    \left({m^2}+\xi R'\right)\,G{}\!\,_{;}{}_{p}{}^{p} \right]  G \cr
&& \!\!\!\! +{\frac{1}{2}} \left({m^2}+\xi R\right)\left({m^2}+\xi
R'\right) {G^2}.
\end{eqnarray}

\subsubsection{Trace of the Noise Kernel}

One of the most interesting and surprising results to come  out of
the investigations undertaken in the 1970's of the quantum stress
tensor was the discovery of the trace anomaly
\cite{CapDuf74,Duf75}. When the trace of the stress tensor
$T=g^{ab}T_{ab}$ is evaluated for a field configuration that
satisties the field equation (\ref{2.2}) the trace is seen to
vanish for  massless conformally coupled fields. When this
analysis is carried over to the renormalized expectation value of
the quantum stress tensor, the trace no longer vanishes. Wald
\cite{Wal78} showed this was due to the failure of the
renormalized Hadamard function $G_{\rm ren}(x,x')$ to be symmetric
in $x$ and $x'$, implying it does not necessarily satisfy the
field equation (\ref{2.2}) in the variable $x'$. The definition of
$G_{\rm ren}(x,x')$ in the context of point separation will come
next.)

With this in  mind, we can now determine the noise associated
with the trace. Taking the trace at both points $x$ and $y$ of
the noise kernel functional (\ref{noiker}):
\begin{eqnarray}
N\left[ G \right]\!\!\!\!&=&\!\!\!\!g^{ab}\,g^{c'd'}\, N_{abc'd'}\left[ G
\right] \cr \!\!\!\!&=&\!\!\!\! - 3\,G\,\xi
    \left[
        \left({m^2} + {1\over 2} \xi R \right) \,{G{}_;{}_{p'}{}^{p'}}
      + \left({m^2} + {1\over 2} \xi R'\right) \,{G{}_;{}_{p}{}^{p}}
    \right] \cr
&&
  \!\!\!\! + {{9\,{{\xi }^2}}\over 2}
       \left[
           {G{}_;{}_{p'}{}^{p'}}\,{G{}_;{}_{p}{}^{p}}
          + G\,{G{}_;{}_{p}{}^{p}{}_{p'}{}^{p'}}
       \right]
\nonumber  \\
&&
  \!\!\!\!  +\left({m^2}\! + \!{1\over 2} \xi R \right)  \,
     \left({m^2} + {1\over 2} \xi R'\right)   G^2
\cr &&\!\!\!\!+ 3\! \left( {1\over 6} \!-\! \xi  \right)
     \! \left[
        3 \!{{\left( {1\over 6} \!-\! \xi  \right) }}
              {G{}_;{}_{p'}{}_{p}}\,{G{}_;{}^{p'}{}^{p}}
        \!-\!3\xi\!
           \left(\!
                {G{}_;{}_{p}}\,{G{}_;{}_{p'}{}^{p}{}^{p'}}
             \!+ \! {G{}_;{}_{p'}}\,{G{}_;{}_{p}{}^{p}{}^{p'}}
          \! \right)
      \right.\cr
&&\!\!\!\!\left.\hspace{10mm}+
          \left({m^2}\! +\! {1\over 2} \xi R \right)
{G{}_;{}_{p'}}{G{}_;{}^{p'}}
       \! +\! \left({m^2} \!+ \!{1\over 2} \xi R'\right) {G{}_;{}_{p}}{G{}^;{}^{p}}
\right]\! .
\end{eqnarray}
For the massless conformal case, this reduces to
\begin{equation}
N\left[ G \right] = \frac{1}{144}\left \{ R R' G^2 - 6G\left(R
\Box' + R' \Box\right) G
  + 18\left[ \left(\Box G\right)\left(\Box' G\right)+ \Box' \Box
  G\right] \right\}\! ,
\end{equation}
which holds for any function $G(x,y)$. For  $G$ being the Green
function, it satisfies the field equation (\ref{2.2}):
\begin{equation}
\Box G = (m^2 + \xi R) G .
\end{equation}
We will only assume the Green function satisfies the field
equation in its first variable. Using the fact $\Box' R=0$
(because the covariant derivatives act at a different point than
at which $R$ is supported), it follows that
\begin{equation}
\Box' \Box  G = (m^2 + \xi R)\Box' G.
\end{equation}
With these results, the noise kernel trace becomes
\begin{eqnarray}
N\left[ G \right]\!\!\! &=& \!\!\!\frac{1}{2} \left[
      {m^2}\,\left( 1 - 3\,\xi  \right)
    + 3\,R\,\left( {1\over 6} - \xi  \right) \,\xi
\right] \cr &&\hspace{18mm}\times
  \left[
       {G^2}\,\left( 2\,{m^2} + {R'} \,\xi  \right)
            + \left( 1 - 6\,\xi  \right) \,{G{}_;{}_{p'}}\,{G{}_;{}^{p'}}
            - 6\,G\,\xi \,{G{}_;{}_{p'}{}^{p'}}
  \right] \cr
&&+ \frac{1}{2} \left( {1\over 6} - \xi  \right)  \left[
3\,\left( 2\,{m^2} + {R'}\,\xi  \right)
\,{G{}_;{}_{p}}\,{G{}^;{}^{p}}
  - 18\,\xi \,{G{}_;{}_{p}}\,{G{}_;{}_{p'}{}^{p}{}^{p'}}
\right.\cr&&\hspace{18mm}\left.
  + 18\,\left( {1\over 6} - \xi  \right) \,
         {G{}_;{}_{p'}{}_{p}}\,{G{}_;{}^{p'}{}^{p}} \right],
\end{eqnarray}
which vanishes for the massless conformal case.  We have thus
shown, based solely on the definition of the point separated noise
kernel, there is no noise associated with the trace anomaly. This
result obtained in Ref. \cite{PhiHu03} is completely general since
it is assumed that the Green function is only satisfying the field
equations in its first variable; an alternative proof of this
result was given in Ref. \cite{MarVer99}. This condition holds not
just for the classical field case, but also for the regularized
quantum case, where one does not expect the Green function to
satisfy the field equation in both variables. One can see this
result from the simple observation used in section \ref{sec2}:
since the trace anomaly is known to be locally determined and
quantum state independent, whereas the noise present in the
quantum field is non-local, it is hard to find a noise associated
with it. This general result is in agreement with previous
findings \cite{CalHu94,HuSin95,CamVer96}, derived from the
Feynman-Vernon influence functional formalism
\cite{FeyVer63,FeyHib65} for some particular cases.

\section{Metric fluctuations in Minkowski spacetime}
\label{sec:flucminspa}

Although the Minkowski vacuum is an eigenstate of the total
four-momentum operator of a field in Minkowski spacetime, it is
not an eigenstate of the stress-energy operator. Hence, even for
those solutions of semiclassical gravity such as the Minkowski
metric, for which the expectation value of the stress-energy
operator can always be chosen to be zero, the fluctuations of this
operator are non-vanishing. This fact leads to consider the
stochastic metric perturbations induced by these fluctuations.

Here we derive the Einstein-Langevin equation for the metric
perturbations in a Minkowski background. We solve this equation
for the linearized Einstein tensor and compute the associated
two-point correlation functions, as well as, the
two-point correlation functions for the metric
perturbations. Even though, in
this case, we expect to have negligibly small values for these
correlation functions for points separated by lengths larger than
the Planck length, there are several reasons why it is worth
carrying out this calculation.

On the one hand, these are the first
back-reaction solutions of
the full Einstein-Langevin
equation.
There are analogous solutions to a ``reduced'' version
of this equation inspired in a ``mini-superspace'' model
\cite{CamVer97,CalCamVer97}, and there is also a previous
attempt to obtain a solution to
the Einstein-Langevin equation in Ref.~\cite{CamVer96},
but, there, the non-local terms in the
Einstein-Langevin equation were neglected.

On the other hand, the results of this calculation, which confirm our
expectations that gravitational fluctuations are negligible at length
scales larger than the Planck length,
but also predict that the fluctuations are strongly
suppressed on small scales, can be considered a first test
of stochastic semiclassical gravity.
These results reveal also an important connection between
stochastic gravity and the large $N$ expansion of quantum gravity.
In addition, they are used in \ref{s6.5} to
study the stability of the Minkowski metric as a solution of semiclassical gravity, which  constitutes an application of the validity criterion introduced in section \ref{s3.3}. This calculation requires also a discussion of the problems posed by the so called \emph{runaway} solutions, which arise in the back-reaction equations of semiclassical and stochastic gravity, and some of the methods to deal with them. As a result we conclude that Minkowski spacetime is a stable and valid solution of semiclassical gravity.

We advise the reader that his section is rather technical since it
deals with an explicit non trivial back-reaction computation in
stochastic gravity. We tried to make it reasonable self-contained
and detailed,
however a more detailed exposition can be found in
Ref.~\cite{MarVer00}.


\subsection{Perturbations around Minkowski spacetime}
\label{s6.1}


The Minkowski metric $\eta_{ab}$, in a manifold ${\cal M}$ which
is topologically ${\rm I\hspace{-0.4 ex}R}^{4}$, and the usual
Minkowski vacuum, denoted as $|0 \rangle$, are the class of
simplest solutions to the semiclassical Einstein equation
(\ref{2.5}), the so called trivial solutions of semiclassical
gravity \cite{FlaWal96}. They constitute the ground state of
semiclassical gravity. In fact, we can always choose a
renormalization scheme in which the renormalized expectation value
$\langle 0|\, \hat{T}_{R}^{ab}\,[\eta] |0 \rangle =0$. Thus,
Minkowski spacetime $({\rm I\hspace{-0.4 ex}R}^{4},\eta_{ab})$ and
the vacuum state $|0 \rangle$ are a solution to the semiclassical
Einstein equation with renormalized cosmological constant
$\Lambda\!=\!0$. The fact that the vacuum expectation value of the
renormalized stress-energy operator in Minkowski spacetime should
vanish was originally proposed by Wald \cite{Wal77} and it may be
understood as a renormalization convention
\cite{Ful89,GriMamMos94}. Note that other possible solutions of
semiclassical gravity with zero vacuum expectation value of the
stress-energy tensor are the exact gravitational plane waves,
since they are known to be vacuum solutions of Einstein equations
which induce neither particle creation nor vacuum polarization
\cite{Gib75,Des75,GarVer91}.

As we have already mentioned the
vacuum $|0 \rangle$ is an eigenstate of the total four-momentum
operator in Minkowski spacetime, but
not an eigenstate of $\hat{T}^{R}_{ab}[\eta]$. Hence, even in
the Minkowski background, there are quantum
fluctuations in the stress-energy tensor and, as a result,
the noise kernel does not vanish.
This fact leads to consider the stochastic corrections
to this class of trivial solutions of semiclassical
gravity.
Since, in this case, the Wightman and Feynman functions
(\ref{Wightman and Feynman functions}), their values in the two-point
coincidence limit, and the products of derivatives of two of such
functions appearing in expressions (\ref{Wightman expression 2}) and
(\ref{Feynman expression 2})
are known in dimensional regularization,
we can compute the Einstein-Langevin
equation using the methods outlined in sections \ref{sec2}
and \ref{sec3}.

To perform explicit calculations it is convenient to work in a global
inertial coordinate system $\{ x^\mu\}$ and in the associated basis,
in which the components of the flat metric are simply
$\eta_{\mu\nu}={\rm diag}(-1,1,\dots,1)$.
In Minkowski spacetime, the components of the classical stress-energy
tensor (\ref{2.3}) reduce to
\be
T^{\mu\nu}[\eta,\phi]=\partial^{\mu}\phi
\partial^{\nu} \phi - {1\over 2}\, \eta^{\mu\nu} \hspace{0.2ex}
\partial^{\rho}\phi \partial_{\rho} \phi
-{1\over 2}\, \eta^{\mu\nu}\hspace{0.2ex} m^2 \phi^2
+\xi \left( \eta^{\mu\nu} \Box
-\partial^{\mu} \partial^{\nu} \right) \phi^2,
\label{flat class s-t}
\ee
where $\Box \!\equiv\! \partial_{\mu}\partial^{\mu}$, and the formal
expression for the
components of the corresponding ``operator''
in dimensional regularization, see Eq. (\ref{regul s-t 2}), is
\be
\hat{T}_{n}^{\mu\nu}[\eta] = {1\over 2} \{
     \partial^{\mu}\hat{\phi}_{n} ,
     \partial^{\nu}\hat{\phi}_{n} \}
     + {\cal D}^{\mu\nu} \hat{\phi}_{n}^2,
\label{flat regul s-t}
\ee
where ${\cal D}^{\mu\nu}$ is the differential operator
(\ref{diff operator}), with $g_{\mu\nu}=\eta_{\mu\nu}$,
$R_{\mu\nu}=0$, and $\nabla_\mu=\partial_\mu$.
The field
$\hat{\phi}_{n}(x)$ is the field operator in the Heisenberg
representation in
a $n$-dimensional Minkowski spacetime, which satisfies the
Klein-Gordon equation (\ref{2.2}).
We use here a stress-energy tensor which differs from the
canonical one, which corresponds to $\xi=0$, both tensors,
however, define the same total momentum.

The Wightman and Feynman functions
(\ref{Wightman and Feynman functions}) when
$g_{\mu\nu}=\eta_{\mu\nu}$, are well
known:
\be
G_n^+(x,y) = i \hspace{0.2ex}\Delta_n^+(x-y),
\ \ \ \
G\!_{\scriptscriptstyle F_{\scriptstyle \hspace{0.1ex}  n}}
 \hspace{-0.2ex}(x,y)
  =
  \Delta_{\scriptscriptstyle F_{\scriptstyle \hspace{0.1ex} n}}
  \hspace{-0.2ex}(x-y),
\label{flat Wightman and Feynman functions}
\ee
with
\bea
&&\Delta_n^+(x)=-2 \pi i \int \! {d^n k \over (2\pi)^n} \,
e^{i kx}\, \delta (k^2+m^2) \,\theta (k^0),
\nn   \\
&&\Delta_{\scriptscriptstyle F_{\scriptstyle \hspace{0.1ex} n}}
  \hspace{-0.2ex}(x)=- \int \! {d^n k \over (2\pi)^n} \,
{e^{i kx}  \over k^2+m^2-i \epsilon} ,
\hspace{5ex} \epsilon \!\rightarrow \! 0^+,
\label{flat propagators}
\eea
where
$k^2 \equiv \eta_{\mu\nu} k^{\mu} k^{\nu}$ and
$k x \equiv \eta_{\mu\nu} k^{\mu} x^{\nu}$.
Note that the derivatives of these functions satisfy
$\partial_{\mu}^{x}\Delta_n^+(x-y)
= \partial_{\mu}\Delta_n^+(x-y)$ and
$\partial_{\mu}^{y}\Delta_n^+(x-y)=
 - \partial_{\mu}\Delta_n^+(x-y)$,
and similarly for the Feynman propagator
$\Delta_{\scriptscriptstyle F_{\scriptstyle \hspace{0.1ex} n}}
 \hspace{-0.2ex}(x-y)$.

To write down the semiclassical Einstein equation
(\ref{2.5}) in $n$-dimensions for this case, we need to
compute the vacuum expectation value of the
stress-energy operator components
(\ref{flat regul s-t}). Since, from
(\ref{flat Wightman and Feynman functions}), we have that
$\langle 0 |\hat{\phi}_{n}^2(x)|0 \rangle=
i\Delta_{\scriptscriptstyle F_{\scriptstyle \hspace{0.1ex} n}}
\hspace{-0.2ex}(0)
=i \Delta_n^+(0)$, which is a constant (independent
of $x$), we have simply
\begin{equation}
\langle 0 | \hat{T}_{n}^{\mu\nu} [\eta]|0 \rangle =
-i \int {d^n k \over (2\pi)^n} \,
{k^{\mu} k^{\nu} \over k^2+m^2-i \epsilon}
= {\eta^{\mu\nu} \over 2} \left( {m^2 \over 4 \pi} \right)^{\! n/2}
\! \Gamma \!\left(- {n \over 2}\right),
\label{vev}
\end{equation}
where the integrals in dimensional regularization have been computed
in the standard way (see Ref. \cite{MarVer00})
and where $\Gamma (z)$ is the
Euler's gamma function. The semiclassical Einstein equation
(\ref{2.5}) in $n$-dimensions before renormalization
reduces now to
\be
{\Lambda_{B} \over 8 \pi G_{B}}\, \eta^{\mu\nu}
= \mu^{-(n-4)}
\langle 0 | \hat{T}_{n}^{\mu\nu}[\eta]|0 \rangle  .
\label{flat semiclassical eq}
\ee
This equation, thus, simply sets the value
of the bare coupling constant
$\Lambda_{B}/G_{B}$.
Note, from (\ref{vev}), that in order to have
$\langle 0|\, \hat{T}_{R}^{\mu\nu}\, |0 \rangle [\eta]\!=\! 0$,
the renormalized and regularized stress-energy tensor
``operator'' for a scalar field in Minkowski spacetime,
see Eq. (\ref{2.4}),
has to be defined as
\be
\hat{T}_{R}^{\mu\nu}[\eta] =
\mu^{-(n-4)}\, \hat{T}_{n}^{\mu\nu}[\eta]
-{ \eta^{\mu\nu} \over 2} \, {m^4 \over (4\pi)^2}
\left( {m^2 \over 4 \pi \mu^2}
\right)^{\!_{\scriptstyle n-4 \over 2}}
\! \Gamma \!\left(- {n \over 2}\right),
\label{flat renorm s-t operator}
\ee
which corresponds to a renormalization of the cosmological constant
\be
{\Lambda_{B} \over G_{B}}={\Lambda \over G}
-{2 \over \pi} \, {m^4 \over n \hspace{0.2ex}(n\!-\!2)}
\: \kappa_n
+O(n-4),
\label{cosmological ct renorm 2}
\ee
where
\be
\kappa_n \equiv {1 \over (n\!-\!4)}
\left({e^\gamma m^2 \over 4 \pi \mu^2} \right)
^{\!_{\scriptstyle n-4 \over 2}}=
{1 \over n\!-\!4}
+{1\over 2}\,
\ln \!\left({e^\gamma m^2 \over 4 \pi \mu^2} \right)+O (n-4),
\label{kappa}
\ee
being $\gamma$ the Euler's constant. In the case of a
massless scalar field, $m^2\!=\!0$, one simply has
$\Lambda_{B} / G_{B}=\Lambda / G$. Introducing this renormalized
coupling constant into Eq.~(\ref{flat semiclassical eq}), we can
take the limit $n \!\rightarrow \! 4$.
We find that,
for $({\rm I\hspace{-0.4 ex}R}^{4}, \eta_{ab},|0 \rangle )$ to
satisfy the semiclassical Einstein equation,
we must take $\Lambda\!=\!0$.

We can now write down the
Einstein-Langevin equations for the components
$h_{\mu\nu}$ of the stochastic metric perturbation
in dimensional regularization.
In our case, using $\langle 0 |\hat{\phi}_{n}^2(x)|0 \rangle=
i\Delta_{\scriptscriptstyle F_{\scriptstyle \hspace{0.1ex} n}}
\hspace{-0.2ex}(0)$
and the explicit expression of
Eq. (\ref{Einstein-Langevin eq 3})
we obtain
\bea
&&\!\!\!\!\!{1\over 8 \pi G_{B}}\Biggl[
G^{{\scriptscriptstyle (1)}\hspace{0.1ex} \mu\nu} +
\Lambda_{B} \left( h^{\mu\nu}
\!-\!{1\over 2}\, \eta^{\mu\nu} h \right)
\Biggr](x) -
{4\over 3}\, \alpha_{B} D^{{\scriptscriptstyle
(1)}\hspace{0.1ex} \mu\nu}(x)
-2\beta_{B}B^{{\scriptscriptstyle (1)}\hspace{0.1ex} \mu\nu}(x)
\nn   \\
&&\!\!\!\!\!-\! \xi G^{{\scriptscriptstyle (1)}\hspace{0.1ex} \mu\nu}\!(x)
\mu^{-(n-4)}\! i\Delta_{\scriptscriptstyle F_{\scriptstyle
\hspace{0.1ex} n}} \hspace{-0.2ex}(0)\! +\! {1\over 2} \!\int\!\! d^ny
 \mu^{-(n-4)}\! H_n^{\mu\nu\alpha\beta}\!(x,y)h_{\alpha\beta}(y)
\!=\! \xi^{\mu\nu}\!(x)\!. \label{flat Einstein-Langevin eq}
\eea
The
indices in $h_{\mu\nu}$ are raised with the Minkowski metric and
$h \equiv h_{\rho}^{\rho}$, and here a superindex ${\scriptstyle
(1)}$  denotes the components of a tensor linearized around the
flat metric. Note that in $n$-dimensions the
two-point correlation functions for the
field $\xi^{\mu\nu}$ is written as \be \langle
\xi^{\mu\nu}(x)\xi^{\alpha\beta}(y) \rangle_s =\mu^{-2
\hspace{0.2ex} (n-4)}  N_n^{\mu\nu\alpha\beta}(x,y),
\label{correlator} \ee

Explicit expressions for $D^{{\scriptscriptstyle
(1)}\hspace{0.1ex} \mu\nu}$ and $B^{{\scriptscriptstyle
(1)}\hspace{0.1ex} \mu\nu}$ are given by
\be
D^{{\scriptscriptstyle (1)}\hspace{0.1ex} \mu\nu}(x)= {1 \over
2}\, {\cal F}^{\mu\nu\alpha\beta}_{x} \, h_{\alpha\beta}(x),
\hspace{7.2ex} B^{{\scriptscriptstyle (1)}\hspace{0.1ex}
\mu\nu}(x)= 2  {\cal F}^{\mu\nu}_{x} {\cal F}^{\alpha\beta}_{x}
h_{\alpha\beta}(x),
\label{D, B tensors}
\ee
with the differential operators ${\cal F}^{\mu\nu}_{x} \equiv
\eta^{\mu\nu} \Box_x -\partial^\mu_{x} \partial^\nu_{x}$ and
${\cal F}^{\mu\nu\alpha\beta}_{x} \equiv 3 {\cal
F}^{\mu(\alpha}_{x} {\cal F}^{\beta)\nu}_{x} -{\cal
F}^{\mu\nu}_{x} {\cal F}^{\alpha\beta}_{x}$.

\subsection{The kernels in the Minkowski background}

Since the two kernels (\ref{finite kernels}) are free of
ultraviolet divergences in the limit $n\!\rightarrow \! 4$, we
can deal directly with the $F^{\mu\nu\alpha\beta}(x-y)\equiv
\lim_{n \rightarrow 4} \mu^{-2 \hspace{0.2ex} (n-4)} \,
F^{\mu\nu\alpha\beta}_n$ in Eq. (\ref{bitensor F}). The kernels $
N^{\mu\nu\alpha\beta}(x,y) ={\rm Re}\, F^{\mu\nu\alpha\beta}(x-y)$
and $ H_{\scriptscriptstyle \!{\rm A}}^{\mu\nu\alpha\beta}(x,y) =
{\rm Im}\, F^{\mu\nu\alpha\beta}(x-y)$ are actually the components
of the ``physical'' noise and dissipation kernels that will appear
in the Einstein-Langevin equations once the renormalization
procedure has been carried out. The bi-tensor
$F^{\mu\nu\alpha\beta}$ can be expressed in terms of the Wightman
function in four spacetime dimensions, according to (\ref{Wightman
expression 2}). The different terms in this kernel can be easily
computed using the integrals \be I(p) \equiv \int\! {d^4 k \over
(2\pi)^4} \: \delta (k^2+m^2) \,\theta (-k^0)  \, \delta
[(k-p)^2+m^2]\,\theta (k^0-p^0), \label{integrals} \ee and
$I^{\mu_1 \dots \mu_r}(p)$ which are defined as the previous one
by inserting the momenta $k^{\mu_1}\dots k^{\mu_r}$ with $r \!=\!
1, 2, 3 ,4$ in the integrand. All these integral can be expressed
in terms of $I(p)$; see Ref. \cite{MarVer00} for the explicit
expressions. It is convenient to separate  $I(p)$ in its even and
odd parts with respect to the variables $p^{\mu}$ as \be
I(p)=I_{\scriptscriptstyle {\rm S}}(p) +I_{\scriptscriptstyle {\rm
A}}(p), \label{I} \ee where $I_{\scriptscriptstyle {\rm S}}(-p)=
I_{\scriptscriptstyle {\rm S}}(p)$ and $I_{\scriptscriptstyle {\rm
A}}(-p)= -I_{\scriptscriptstyle {\rm A}}(p)$. These two functions
are explicitly given by \bea &&I_{\scriptscriptstyle {\rm
S}}(p)={1 \over 8 \, (2 \pi)^3} \; \theta (-p^2-4m^2) \, \sqrt{1+4
\,{m^2 \over p^2} },
\nn  \\
&&I_{\scriptscriptstyle {\rm A}}(p)={-1 \over 8 \, (2 \pi)^3} \;
{\rm sign}\,p^0 \;
\theta (-p^2-4m^2) \, \sqrt{1+4 \,{m^2 \over p^2} }.
\label{S and A parts of I}
\eea
After some manipulations, we find
\bea
F^{\mu\nu\alpha\beta}(x)&\!\!\!\!=\!\!\!\!& {\pi^2 \over 45}\,
 {\cal F}^{\mu\nu\alpha\beta }_{x}
\int\! {d^4 p \over (2\pi)^4} \,
e^{-i px}\hspace{0.1ex}
\left(1+4 \,{m^2 \over p^2} \right)^2 I(p)
\nn   \\
&&+\,{8 \pi^2 \over 9 } \, {\cal F}^{\mu\nu}_{x}{\cal
F}^{\alpha\beta}_{x} \int\! {d^4 p \over (2\pi)^4} \, e^{-i
px}\hspace{0.1ex} \left(3 \hspace{0.3ex}\Delta \xi+{m^2 \over p^2}
\right)^2 I(p), \label{M 3} \eea where $\Delta \xi \equiv \xi -
1/6$. The real and imaginary parts of the last expression, which
yield the noise and dissipation kernels, are easily recognized as
the terms containing $I_{\scriptscriptstyle {\rm S}}(p)$ and
$I_{\scriptscriptstyle {\rm A}}(p)$, respectively. To write them
explicitly, it is useful to introduce the new kernels
\bea
&&N_{\rm A}(x;m^2) \equiv {1 \over 480 \pi} \int\! {d^4 p \over
(2\pi)^4} \, e^{i px}\, \theta (-p^2-4m^2) \, \sqrt{1+4 \,{m^2
\over p^2} } \left(1+4 \,{m^2 \over p^2} \right)^2,
\nn \\
&&N_{\rm B}(x;m^2,\Delta \xi) \equiv {1 \over 72 \pi} \int\! {d^4
p \over (2\pi)^4} \, e^{i px}\, \theta (-p^2-4m^2)
\nonumber\\
&&\hspace{55mm}\times
 \sqrt{1+4
\,{m^2 \over p^2} } \left(3 \hspace{0.3ex}\Delta \xi+{m^2 \over
p^2} \right)^2,
\nn \\
&&D_{\rm A}(x;m^2) \equiv {-i \over 480 \pi} \int\! {d^4 p \over
(2\pi)^4} \, e^{i px}\, {\rm sign}\,p^0 \; \theta (-p^2-4m^2)
\nonumber\\
&&\hspace{55mm}\times \sqrt{1+4 \,{m^2 \over p^2} } \left(1+4
\,{m^2 \over p^2} \right)^2,
\nn \\
&&D_{\rm B}(x;m^2,\Delta \xi) \equiv {-i \over 72 \pi} \int\! {d^4
p \over (2\pi)^4} \, e^{i px}\, {\rm sign}\,p^0 \; \theta
(-p^2-4m^2)
\nonumber\\
&&\hspace{55mm}\times \sqrt{1+4 \,{m^2 \over p^2} } \left(3
\hspace{0.3ex}\Delta \xi+{m^2 \over p^2} \right)^{\!2}\!\!\!\!,
\label{N and D kernels} \eea and we finally get \bea
\!\!\!\!\!&&\!\!\!\!\!N^{\mu\nu\alpha\beta}(x,y)= {1 \over 6}
{\cal F}^{\mu\nu\alpha\beta}_{x} N_{\rm A}(x\!-\!y;m^2) +{\cal
F}^{\mu\nu}_{x}{\cal F}^{\alpha\beta}_{x} N_{\rm
B}(x\!-\!y;m^2,\Delta \xi),
\nn \\
\!\!\!\!\!&&\!\!\!\!\!H_{\scriptscriptstyle \!{\rm
A}}^{\mu\nu\alpha\beta}(x,y)= {1 \over 6}
 {\cal F}^{\mu\nu\alpha\beta}_{x}
D_{\rm A}(x\!-\!y;m^2) +{\cal F}^{\mu\nu}_{x}{\cal
F}^{\alpha\beta}_{x} D_{\rm B}(x\!-\!y;m^2,\Delta \xi) .
\label{noise and dissipation kernels 2} \eea Notice that the noise
and dissipation kernels defined in (\ref{N and D kernels}) are
actually real because, for the noise kernels, only the $\cos px$
terms of the exponentials $e^{i px}$ contribute to the integrals,
and, for the dissipation kernels, the only contribution of such
exponentials comes from the $i \sin px$ terms.

The evaluation of the kernel $H_{\scriptscriptstyle \!{\rm
S}_{\scriptstyle n}} ^{\mu\nu\alpha\beta}(x,y)$ is a more involved
task. Since this kernel contains divergences in the limit
$n\!\rightarrow \! 4$, we use dimensional regularization. Using
Eq.~(\ref{Feynman expression 2}), this kernel can be written in
terms of the Feynman propagator (\ref{flat propagators}) as \be
\mu^{-(n-4)} H_{\scriptscriptstyle \!{\rm S}_{\scriptstyle n}}
^{\mu\nu\alpha\beta}(x,y)= {\rm Im}\, K^{\mu\nu\alpha\beta}(x-y),
\label{kernel H_S} \ee where
\bea \!\!\!\!\!\!\! &&\!\!\!\!\!\!\!
K^{\mu\nu\alpha\beta}(x) \equiv - \mu^{-(n-4)} \biggl\{ 2
\partial^\mu \partial^{( \alpha} \Delta_{\scriptscriptstyle
F_{\scriptstyle \hspace{0.1ex} n}}
   \hspace{-0.2ex}(x) \,
\partial^{\beta )} \partial^\nu
\Delta_{\scriptscriptstyle F_{\scriptstyle \hspace{0.1ex} n}}
   \hspace{-0.2ex}(x)
\nonumber\\
\!\!\!&&\!\!\!\!\!\! +2 {\cal D}^{\mu\nu} \!\left( \partial^\alpha
\Delta_{\scriptscriptstyle F_{\scriptstyle \hspace{0.1ex} n}}
   \hspace{-0.2ex}(x)
\partial^\beta
\Delta_{\scriptscriptstyle F_{\scriptstyle \hspace{0.1ex} n}}
   \hspace{-0.2ex}(x)  \right)
+ 2 {\cal D}^{\alpha\beta}  \Bigl( \partial^\mu
\Delta_{\scriptscriptstyle F_{\scriptstyle \hspace{0.1ex} n}}
   \hspace{-0.2ex}(x) \,
\partial^\nu
\Delta_{\scriptscriptstyle F_{\scriptstyle \hspace{0.1ex} n}}
   \hspace{-0.2ex}(x) \Bigr)
\nonumber\\
\!\!\!&&\!\!\!\!\!\! +2 {\cal D}^{\mu\nu} {\cal D}^{\alpha\beta}
\!\left( \Delta_{\scriptscriptstyle F_{\scriptstyle \hspace{0.1ex}
n}}^2
   \hspace{-0.2ex}(x) \right)
\!+\!\biggl[ \eta^{\mu\nu} \partial^{( \alpha}
\Delta_{\scriptscriptstyle F_{\scriptstyle \hspace{0.1ex} n}}
   \hspace{-0.2ex}(x)
\partial^{\beta )}
\!+ \!\eta^{\alpha\beta} \partial^{( \mu}
 \Delta_{\scriptscriptstyle F_{\scriptstyle \hspace{0.1ex} n}}
   \hspace{-0.2ex}(x)
\partial^{\nu )}
\nn   \\
\!\!\!&&\!\!\!\!\!\! +\! \Delta_{\scriptscriptstyle
F_{\scriptstyle \hspace{0.1ex} n}}
   \hspace{-0.2ex}(0) \! \left( \eta^{\mu\nu}
{\cal D}^{\alpha\beta}\!+\! \eta^{\alpha\beta} {\cal D}^{\mu\nu}
\right) \!+\!{1 \over 4}\eta^{\mu\nu}\! \eta^{\alpha\beta}
\!\left( \Delta_{\scriptscriptstyle F_{\scriptstyle \hspace{0.1ex}
n}}
   \hspace{-0.2ex}(x) \Box
\!-\!m^2\!
\Delta_{\scriptscriptstyle F_{\scriptstyle \hspace{0.1ex} n}}
   \hspace{-0.2ex}(0)\!  \right) \!\biggr] \!\delta^n (x)\!
\biggr\}\!. \label{K} \eea Let us define the integrals \be J_n(p)
\equiv \mu^{-(n-4)} \!\int\! {d^n k \over (2\pi)^n} \: {1 \over
(k^2+m^2-i \epsilon) \, [(k-p)^2+m^2-i \epsilon] },
\label{integrals in n dim} \ee and $J_n^{\mu_1 \dots \mu_r}(p)$
obtained by inserting the momenta $k^{\mu_1}\dots k^{\mu_r}$ into
the previous integral, together with \be I_{0_{\scriptstyle n}}
\equiv \mu^{-(n-4)} \!\int\! {d^n k \over (2\pi)^n} \: {1 \over
(k^2+m^2-i \epsilon) }, \label{constant integrals in n dim} \ee
and $I_{0_{\scriptstyle n}}^{\mu_1 \dots \mu_r}$ which are also
obtained by inserting momenta in the integrand. Then, the
different terms in Eq.~(\ref{K}) can be computed. These integrals
are explicitly given in Ref. \cite{MarVer00}. It is found that
$I_{0_{\scriptstyle n}}^{\mu}=0$ and the remaining integrals can
be written in terms of $I_{0_{\scriptstyle n}}$ and $J_n(p)$. It
is useful to introduce the projector $P^{\mu\nu}$ orthogonal to
$p^\mu$ and the tensor $P^{\mu\nu\alpha\beta}$ as \be p^2
P^{\mu\nu} \!\equiv \! \eta^{\mu\nu} p^2- p^\mu p^\nu, \ \ \ \ \ \
P^{\mu\nu\alpha\beta}\equiv
3P^{\mu(\alpha}P^{\beta)\nu}-P^{\mu\nu}P^{\alpha\beta},
\label{projector} \ee then the action of the operator ${\cal
F}^{\mu\nu}_{x}$ is simply written as ${\cal F}^{\mu\nu}_{x}
\int\! d^n p \, e^{i p x}\, f(p) = - \!\int\! d^n p \, e^{i p x}\,
f(p) \, p^2 P^{\mu\nu}$ where $f(p)$ is an arbitrary function of
$p^\mu$.

Finally after a rather long but straightforward calculation,
and after expanding around $n\!=\!4$, we get,
\bea
&&\!\!\!K^{\mu\nu\alpha\beta}(x)={i \over (4\pi)^2}
\Biggl\{ \kappa_n \left[ {1 \over 90}
{\cal F}^{\mu\nu\alpha\beta}_{x} \delta^n (x)
+4  \Delta \xi^2
{\cal F}^{\mu\nu}_{x}{\cal F}^{\alpha\beta}_{x}
 \delta^n (x)
\right.
\nn  \\
&&
+{2 \over 3}{m^2 \over (n\!-\!2)} \:
\bigr( \eta^{\mu\nu} \eta^{\alpha\beta} \Box_x
-\eta^{\mu (\alpha } \eta^{\beta )\nu} \Box_x
+\eta^{\mu (\alpha } \partial^{\beta )}_x \partial^\nu_x
+\eta^{\nu (\alpha } \partial^{\beta )}_x \partial^\mu_x
\nonumber\\
&&
-\eta^{\mu\nu} \partial^\alpha_x \partial^\beta_x
-\eta^{\alpha\beta} \partial^\mu_x \partial^\nu_x \bigl)
\delta^n (x)
+ {4 \hspace{0.2ex} m^4 \over n (n\!-\!2)} \:
(2 \hspace{0.2ex}\eta^{\mu (\alpha } \eta^{\beta )\nu}
\!- \eta^{\mu\nu} \eta^{\alpha\beta}) \, \delta^n (x)
\biggr]
\nonumber\\
&&
+{1 \over 180}
 {\cal F}^{\mu\nu\alpha\beta}_{x}
  \!\int \! {d^n p \over (2\pi)^n}
e^{i p x} \left(1+4 \,{m^2 \over p^2} \right)^2 \! \bar\phi (p^2)
\nn  \\
&&
 +{2 \over 9}  \, {\cal F}^{\mu\nu}_{x}{\cal
F}^{\alpha\beta}_{x} \!\int \! {d^n p \over (2\pi)^n} \, e^{i p x}
\left(3 \hspace{0.2ex}\Delta \xi+{m^2 \over p^2} \right)^2 \!
\bar\phi (p^2)
\nn  \\
&&
- \left[ {4 \over 675} \,
 {\cal F}^{\mu\nu\alpha\beta}_{x}
+{1 \over 270} \, (60 \hspace{0.1ex}\xi \!-\! 11) \,
{\cal F}^{\mu\nu}_{x}{\cal F}^{\alpha\beta}_{x}
\right] \delta^n (x)
\nn  \\
&&
- m^2 \left[ {2 \over 135} \,
 {\cal F}^{\mu\nu\alpha\beta}_{x}
+{1 \over 27} \, {\cal F}^{\mu\nu}_{x}{\cal F}^{\alpha\beta}_{x}
\right] \Delta_n(x) \Biggr\}+ O(n-4), \label{result for K} \eea
where $\kappa_n$ has been defined in (\ref{kappa}), and $\bar\phi
(p^2)$ and $\Delta_n(x)$ are given by
\be
\bar\phi (p^2)
\!\equiv\! \int_0^1 \!d\alpha \: \ln \biggl(\!1\!+\!{p^2
\over m^2} \alpha (1\!-\!\alpha)\!-\!i \epsilon \!\biggr)\! =\! -i \pi
\theta (-p^2\!-\!4m^2)\sqrt{1\!+\!4 {m^2 \over p^2} } \!+\!\varphi
(p^2\!)\!,
\label{phi}
\ee
\be
\Delta_n(x)\!\equiv\!\int\! {d^n p \over (2\pi)^n}
\: e^{i p x}\; {1 \over p^2},
\label{Delta_n}
\ee
where $ \varphi
(p^2) \equiv  \int_0^1 d\alpha \: \ln | 1+{p^2 \over m^2} \,
\alpha (1\!-\!\alpha)| $.
The imaginary part of (\ref{result for K}) gives the kernel components
$\mu^{-(n-4)}
H_{\scriptscriptstyle \!{\rm S}_{\scriptstyle n}}
^{\mu\nu\alpha\beta}(x,y)$, according to  (\ref{kernel H_S}).
It can be easily obtained multiplying
this expression by $-i$ and retaining only the real part,
$\varphi (p^2)$, of the function $\bar\phi (p^2)$.

\subsection{The Einstein-Langevin equation}
\label{s6.3}

With the previous results for the kernels we can
now write the $n$-dimensional Einstein-Langevin equation
(\ref{flat Einstein-Langevin eq}),
previous to the renormalization.
Taking also into account
Eqs.~(\ref{vev}) and (\ref{flat semiclassical eq}),
we may finally write:
\bea
&&{1\over 8 \pi G_{B}}
G^{{\scriptscriptstyle (1)}\hspace{0.1ex} \mu\nu}(x)
-{4\over 3}\, \alpha_{B} D^{{\scriptscriptstyle
(1)}\hspace{0.1ex} \mu\nu}(x)
-2\beta_{B}B^{{\scriptscriptstyle (1)}\hspace{0.1ex} \mu\nu}(x)
\nonumber\\
&&
+{\kappa_n \over (4\pi)^2} \, \Biggl[
-4 \hspace{0.2ex}\Delta \xi \, {m^2 \over (n\!-\!2)} \,
G^{{\scriptscriptstyle (1)}\hspace{0.1ex} \mu\nu}
+{1 \over 90} \,
D^{{\scriptscriptstyle (1)}\hspace{0.1ex} \mu\nu}
 \Delta \xi^2
B^{{\scriptscriptstyle (1)}\hspace{0.1ex} \mu\nu}
\Biggr]
\hspace{-0.2ex} (x)
\nonumber\\
&&
+{1 \over 2880 \pi^2} \, \Biggl\{
-{16 \over 15} \,
D^{{\scriptscriptstyle (1)}\hspace{0.1ex} \mu\nu}(x)
+\left({1 \over 6}-\! 10\hspace{0.2ex} \Delta \xi \right) \!
B^{{\scriptscriptstyle (1)}\hspace{0.1ex} \mu\nu}(x)
\nn  \\
&&
+ \int\! d^n y \!
\int\! {d^n p \over (2\pi)^n} \, e^{i p (x-y)} \,
\varphi (p^2) \,
\Biggl[\left(1+4 {m^2 \over p^2} \right)^2 \!
D^{{\scriptscriptstyle (1)} \mu\nu}(y)
\nonumber\\
&&
\hspace{18mm}+10 \!
\left(3 \hspace{0.2ex}\Delta \xi+{m^2 \over p^2} \right)^2 \!
B^{{\scriptscriptstyle (1)}\hspace{0.1ex} \mu\nu}(y)
\Biggr]
\nn  \\
&& -\, {m^2 \over 3} \!\int\! d^n y \, \Delta_n(x\!-\!y) \, \Bigl(
8 D^{{\scriptscriptstyle (1)}\hspace{0.1ex} \mu\nu} + 5
B^{{\scriptscriptstyle (1)}\hspace{0.1ex} \mu\nu}
\Bigr)\hspace{-0.2ex} (y) \Biggr\}
\nonumber\\
&&
+{1\over 2} \!\int\! d^ny \,
\mu^{-(n-4)} H_{\scriptscriptstyle \!{\rm A}_{\scriptstyle n}}
^{\mu\nu\alpha\beta}(x,y)\, h_{\alpha\beta}(y) +O(n\!-\!4)
= \xi^{\mu\nu}(x). \label{flat Einstein-Langevin eq 2} \eea
Notice that the terms containing the bare cosmological constant
have cancelled. These equations can now be renormalized, that is,
we can now write the bare coupling constants as renormalized
coupling constants plus some suitably chosen counterterms and take
the limit $n\!\rightarrow \! 4$. In order to carry out such a
procedure, it is convenient to distinguish between massive and
massless scalar fields. The details of the calculation can be
found in Ref.~\cite{MarVer00}.

It is convenient to introduce the two new kernels
\bea &&H_{\rm
A}(x;m^2) \equiv {1 \over 480 \pi^2}\! \int\! {d^4 p \over
(2\pi)^4} \, e^{i px}\, \Biggl\{ \! \left(1+4 \,{m^2 \over p^2}
\right)^{\! 2}
\nonumber\\
&&
\ \ \ \times\Biggl[ - i \pi {\rm sign}p^0 \theta
(-p^2\!-\!4m^2)  \sqrt{1+4 {m^2 \over p^2} }
+\varphi(p^2) \Biggr]
-{8 \over 3} {m^2 \over p^2} \Biggr\},
\nn \\
&&H_{\rm B}(x;m^2,\Delta \xi) \equiv {1 \over 72 \pi^2}\! \int\!
{d^4 p \over (2\pi)^4}  e^{i px} \Biggl\{ \! \left(3
\Delta \xi\!+\!{m^2 \over p^2} \right)^{\! 2}
\nonumber\\
&&
\ \ \ \times\Biggl[\! -
i \pi{\rm sign}\,p^0 \theta (-p^2\!-\!4m^2)  \sqrt{1\!+\!4
{m^2 \over p^2} }
\! +\! \varphi(p^2) \Biggr]\! -{1 \over 6} \! {m^2
\over p^2} \!\Biggr\},
\label{H kernels}
\eea
where $\varphi(p^2)$
is given by the restriction to $n=4$ of expression (\ref{phi}).
The renormalized coupling constants $1/G$, $\alpha$ and $\beta$
are easily computed as it was done in Eq. (\ref{cosmological ct
renorm 2}). Substituting  their expressions into Eq.~(\ref{flat
Einstein-Langevin eq 2}), we can take the limit $n\!\rightarrow \!
4$, using the fact that, for $n=4$, $D^{{\scriptscriptstyle
(1)}\hspace{0.1ex} \mu\nu}(x)= (3/ 2) \, A^{{\scriptscriptstyle
(1)}\hspace{0.1ex} \mu\nu}(x)$, we obtain the corresponding
semiclassical Einstein-Langevin equation.

For the massless case one needs the limit $m \! \rightarrow \! 0$
of equation (\ref{flat Einstein-Langevin eq 2}). In this case it is
convenient to separate $\kappa_n$ in (\ref{kappa}) as
$\kappa_n=\tilde{\kappa}_n +{1 \over 2}\ln (m^2/\mu
^2)+O(n\!-\!4)$, where \be \tilde{\kappa}_n \equiv {1 \over
(n\!-\!4)} \left({e^\gamma \over 4 \pi} \right) ^{\!_{\scriptstyle
n-4 \over 2}}= {1 \over n\!-\!4} +{1\over 2}\, \ln
\!\left({e^\gamma \over 4 \pi } \right)+O (n-4), \label{kappa
tilde} \ee and use that, from Eq.~(\ref{phi}), we have \be
\lim_{m^2 \rightarrow 0} \left[ \varphi (p^2)+\ln (m^2/\mu ^2)
\right]=-2+\ln \left| \hspace{0.2ex} {p^2 \over \mu^2}
\hspace{0.2ex}\right|. \label{massless limit} \ee The coupling
constants are then easily renormalized. We note that in the
massless limit, the Newtonian gravitational constant is not
renormalized and, in the conformal coupling case, $\Delta \xi=0$,
we have that $\beta_{B}\!=\! \beta$. Note also that, by making $m
\!=\!0$ in (\ref{N and D kernels}), the noise and dissipation
kernels can be written as \bea &&N_{\rm A}(x;m^2\!=\!0)=N(x),
\hspace{7ex} N_{\rm B}(x;m^2\!=\!0,\Delta \xi) =60 \hspace{0.2ex}
\Delta \xi^2 \hspace{0.2ex}  N(x),
\nn \\
&&D_{\rm A}(x;m^2\!=\!0)=D(x), \hspace{7ex} D_{\rm
B}(x;m^2\!=\!0,\Delta \xi) =60 \hspace{0.2ex} \Delta \xi^2
\hspace{0.2ex} D(x), \label{massless N and D kernels}
\eea
where
\be N(x) \!\equiv\! {1 \over 480 \pi} \!\int \! {d^4 p \over (2\pi)^4}
\! e^{i px}\! \theta (-p^2)\!,\ \
D(x) \!\equiv\! {-i \over
480 \pi} \!\int \! {d^4 p \over (2\pi)^4} \! e^{i px}\! {\rm
sign}\,p^0  \theta (-p^2)\!.
\label{N and D}
\ee
It is also
convenient to introduce the new kernel \bea H(x;\mu^2) &\equiv &
{1 \over 480 \pi^2} \!\int \! {d^4 p \over (2\pi)^4} \, e^{i px}
\left[ \ln \left| \hspace{0.2ex} {p^2 \over \mu^2}
\hspace{0.2ex}\right| - i \pi \, {\rm sign}\,p^0 \; \theta (-p^2)
\right]
\nn  \\
&=& {1 \over 480 \pi^2} \lim_{\epsilon \rightarrow 0^+} \!\int \!
{d^4 p \over (2\pi)^4} \, e^{i px} \, \ln\! \left( {-(p^0+i
\epsilon)^2+p^i p_i \over \mu^2} \right). \label{Hnew} \eea This
kernel is real and can be written as the sum of an even part and
an odd part in the variables $x^\mu$, where the odd part is the
dissipation kernel $D(x)$. The Fourier transforms (\ref{N and D})
and (\ref{Hnew}) can actually be computed and, thus, in this case
we have explicit expressions for the kernels in position space;
see, for instance, Refs.~\cite{Jon66,CamMarVer95,Hor80}.

Finally, the Einstein-Langevin equation for the physical
stochastic perturbations $h_{\mu\nu}$ can be written in both
cases, for $m \!\neq \!0$ and for $m\!=\!0$, as \bea
\!\!\!\!&&\!\!\!\!{1\over 8 \pi G} G^{{\scriptscriptstyle
(1)}\hspace{0.1ex} \mu\nu}(x) \hspace{-0.2ex}-\hspace{-0.2ex} 2
\left(\bar{\alpha} A^{{\scriptscriptstyle (1)}\hspace{0.1ex}
\mu\nu}(x) \hspace{-0.2ex}+\hspace{-0.2ex} \bar{\beta}
B^{{\scriptscriptstyle (1)}\hspace{0.1ex} \mu\nu}(x) \right)\!
\nn \\
&& \ \ \ + {1\over 4}\int\! d^4y \left[ H_{\rm A}(x\!-\!y)
A^{{\scriptscriptstyle (1)}\hspace{0.1ex} \mu\nu}(y)
\hspace{-0.2ex}+\hspace{-0.2ex} H_{\rm B}(x\!-\!y)
B^{{\scriptscriptstyle (1)}\hspace{0.1ex} \mu\nu}(y) \right] =
\xi^{\mu\nu}(x), \label{unified Einstein-Langevin eq} \eea where
in terms of the renormalized constants $\alpha$ and $\beta$ the
new constants are $\bar{\alpha}=\alpha+(3600\pi^2)^{-1}$ and
$\bar{\beta}=\beta-(1/12-5\Delta\xi)(2880\pi^2)^{-1}$. The kernels
$H_{\rm A}(x)$ and $H_{\rm B}(x)$ are given by Eqs. (\ref{H
kernels}) when $m\neq 0$, and $H_{\rm A}(x)=H(x;\mu^2)$, $H_{\rm
B}(x)=60 \hspace{0.2ex} \Delta \xi^2 H(x;\mu^2)$ when $m \!=\! 0$.
In the massless case, we can use the arbitrariness of the mass
scale $\mu$ to eliminate one of the parameters $\bar{\alpha}$ or
$\bar{\beta}$. The components of the Gaussian stochastic source
$\xi^{\mu\nu}$ have zero mean value and their two-point
correlation functions are given by $
\langle\xi^{\mu\nu}(x)\xi^{\alpha\beta}(y) \rangle_s
=N^{\mu\nu\alpha\beta}(x,y)$, where the noise kernel is given in
(\ref{noise and dissipation kernels 2}), which in the massless
case reduces to (\ref{massless N and D kernels}).

It is interesting to consider the massless conformally coupled
scalar field, {\it i.e.}, the case $\Delta \xi\!=\!0$, of
particular interest because of its similarities with the
electromagnetic field, and also because of its interest in
cosmology: massive fields become conformally invariant when their
masses are negligible compared to the spacetime curvature. We have
already mentioned that for a conformally coupled, field, the
stochastic source tensor must be traceless (up to first order in
perturbation theory around semiclassical gravity), in the sense
that the stochastic variable $\xi^\mu_\mu \!\equiv
\!\eta_{\mu\nu}\xi^{\mu\nu}$ behaves deterministically as a
vanishing scalar field. This can be directly checked by noticing,
from Eqs.~(\ref{noise and dissipation kernels 2}) and
(\ref{massless N and D kernels}), that, when $\Delta \xi\!=\!0$,
one has $\langle\xi^\mu_\mu(x)\xi^{\alpha\beta}(y) \rangle_s =0$,
since ${\cal F}^\mu_\mu\!=\! 3 \hspace{0.2ex}\Box $ and ${\cal
F}^{\mu \alpha}{\cal F}^\beta_\mu \!=\! \Box {\cal
F}^{\alpha\beta}$. The Einstein-Langevin equations for this
particular case (and generalized to a spatially flat
Robertson-Walker background) were first obtained in
Ref.~\cite{CamVer96}, where the coupling constant $\beta$ was
fixed to be  zero. See also Ref.~\cite{HuVer03a} for a discussion
of this result and its connection to the problem of structure
formation in the trace anomaly driven inflation
\cite{Sta80,Vil85,HawHerRea01}.

Note that the expectation value of the renormalized stress-energy
tensor for a scalar field can be obtained by comparing
Eq.~(\ref{unified Einstein-Langevin eq}) with the
Einstein-Langevin equation (\ref{2.11}), its explicit expression
is given in Ref.~\cite{MarVer00}. The results agree with the
general form found by Horowitz \cite{Hor80,Hor81} using an
axiomatic approach and coincides with that given in
Ref.~\cite{FlaWal96}. The particular cases of conformal coupling,
$\Delta \xi \!=\!0$, and minimal coupling, $\Delta \xi \!=\!-1/6$,
are also in agreement with the results for this cases given in
Refs.~\cite{Hor80,Hor81,Sta81,CamVer94,Jor87}, modulo local terms
proportional to $A^{{\scriptscriptstyle (1)}\hspace{0.1ex}
\mu\nu}$ and $B^{{\scriptscriptstyle (1)}\hspace{0.1ex} \mu\nu}$
due to different choices of the renormalization scheme. For the
case of a massive minimally coupled scalar field, $\Delta \xi
\!=\!-1/6$, our result is equivalent to that of
Ref.~\cite{TicFla98}.


\subsection{Correlation functions for gravitational
perturbations}
\label{s6.4}

Here we solve the Einstein-Langevin equations (\ref{unified
Einstein-Langevin eq}) for the components $G^{{\scriptscriptstyle
(1)}\hspace{0.1ex} \mu\nu}$ of the linearized Einstein tensor.
Then we use these solutions to compute the corresponding two-point
correlation functions, which give a measure of the gravitational
fluctuations predicted by the stochastic semiclassical theory of
gravity in the present case. Since the linearized Einstein tensor
is invariant under gauge transformations of the metric
perturbations, these two-point correlation functions are also
gauge invariant. Once we have computed the two-point correlation
functions for the linearized Einstein tensor, we find the
solutions for the metric perturbations and compute the associated
two-point correlation functions. The procedure used to solve the
Einstein-Langevin equation is similar to the one used by Horowitz
\cite{Hor80}, see also Ref.~\cite{FlaWal96}, to analyze the
stability of Minkowski spacetime in semiclassical gravity.

We first note that the tensors $A^{{\scriptscriptstyle
(1)}\hspace{0.1ex} \mu\nu}$ and $B^{{\scriptscriptstyle
(1)}\hspace{0.1ex} \mu\nu}$ can be written in terms of
$G^{{\scriptscriptstyle (1)}\hspace{0.1ex} \mu\nu}$ as \be
A^{{\scriptscriptstyle (1)}\hspace{0.1ex} \mu\nu} = {2 \over 3} \,
({\cal F}^{\mu\nu} G^{{\scriptscriptstyle
(1)}}\mbox{}^{\alpha}_\alpha -{\cal F}^{\alpha}_\alpha
G^{{\scriptscriptstyle (1)}\hspace{0.1ex} \mu\nu}), \hspace{10ex}
B^{{\scriptscriptstyle (1)}\hspace{0.1ex} \mu\nu} = 2
\hspace{0.2ex} {\cal F}^{\mu\nu} G^{{\scriptscriptstyle
(1)}}\mbox{}^{\alpha}_\alpha, \label{A and B} \ee where we have
used that $3 \hspace{0.2ex}\Box={\cal F}^{\alpha}_\alpha$.
Therefore, the Einstein-Langevin equation (\ref{unified
Einstein-Langevin eq})  can be seen as a linear
integro-differential stochastic equation for the components
$G^{{\scriptscriptstyle (1)}\hspace{0.1ex} \mu\nu}$. In order to
find solutions to Eq. (\ref{unified Einstein-Langevin eq}), it is
convenient to Fourier transform. With the convention $\tilde
f(p)=\int d^4x e^{-ipx}f(x)$ for a given field $f(x)$, one finds,
from (\ref{A and B}),
\begin{eqnarray}
&&
\tilde{A}^{{\scriptscriptstyle
(1)}\mu\nu}(p)\!=\! 2 p^2
\tilde{G}^{{\scriptscriptstyle (1)}\mu\nu}(p)\! -\! {2
\over 3} \! p^2\! P^{\mu\nu} \tilde{G}^{{\scriptscriptstyle
(1)}}\mbox{}^{\alpha}_\alpha(p),
\nonumber\\
&&
\tilde{B}^{{\scriptscriptstyle (1)}\mu\nu}(p)\!=\! -2
p^2 \!P^{\mu\nu}\! \tilde{G}^{{\scriptscriptstyle
(1)}}\mbox{}^{\alpha}_\alpha(p).
\end{eqnarray}

The Fourier transform of the
Einstein-Langevin Eq.~(\ref{unified Einstein-Langevin eq}) now reads
\be F^{\mu\nu}_{\hspace{2ex}\alpha\beta}(p) \,
\tilde{G}^{{\scriptscriptstyle (1)}\hspace{0.1ex} \alpha\beta}(p)=
8 \pi G \, \tilde{\xi}^{\mu\nu}(p),
\label{Fourier transf of E-L
eq}
\ee
where
\be F^{\mu\nu}_{\hspace{2ex} \alpha\beta}(p) \equiv
F_1(p) \, \delta^\mu_{( \alpha} \delta^\nu_{\beta )}+ F_2(p) \,
p^2 P^{\mu\nu} \eta_{\alpha\beta},
\label{F def}
\ee
with
\begin{eqnarray}
&&F_1(p) \equiv 1+16 \pi G \, p^2 \left[ {1\over 4}\tilde{H}_{\rm A}(p)-2
\bar{\alpha}\right],
\nonumber\\
&&
F_2(p) \equiv -{16 \over 3} \,
\pi G \left[ {1\over 4}\tilde{H}_{\rm A}(p)+{3\over 4} \tilde{H}_{\rm B}(p) -2
\bar{\alpha}-6 \bar{\beta}\right].
\label{F_1 and F_2}
\end{eqnarray}
In the
Fourier transformed Einstein-Langevin Eq.~(\ref{Fourier transf of
E-L eq}), $\tilde{\xi}^{\mu\nu}(p)$, the Fourier transform of
$\xi^{\mu\nu}(x)$, is a Gaussian stochastic source of zero average
and
\be \langle \tilde{\xi}^{\mu\nu}(p)
\tilde{\xi}^{\alpha\beta}(p^\prime) \rangle_s = (2 \pi)^4 \,
\delta^4(p+p^\prime) \, \tilde{N}^{\mu\nu\alpha\beta}(p),
\label{Fourier transf of corr funct}
\ee
where we have introduced
the Fourier transform of the noise kernel. The explicit expression
for $\tilde{N}^{\mu\nu\alpha\beta}(p)$ is found from (\ref{noise
and dissipation kernels 2}) and (\ref{N and D kernels}) to be
\bea
\tilde{N}^{\mu\nu\alpha\beta}(p)
\!\!\!\! &=& \!\!\!\! {1 \over 720 \pi}\theta
(-p^2\!-\!4m^2) \, \sqrt{1+4 \,{m^2 \over p^2} }
\left[ {1 \over 4} \left(1+4 \,{m^2 \over p^2} \right)^{\!2}
(p^2)^2  P^{\mu\nu\alpha\beta} \right.
\nn  \\
&&\hspace{25mm} \left. + 10 \! \left(3 \hspace{0.2ex}\Delta
\xi+{m^2 \over p^2} \right)^{\!2}\! (p^2)^2 P^{\mu\nu}
P^{\alpha\beta} \right], \label{Fourier transf of noise 2} \eea
which in the massless case reduces to \be \lim_{m \rightarrow 0}\!
\tilde{N}^{\mu\nu\alpha\beta}(p)= {1 \over 480 \pi} \: \theta
(-p^2) \left[ {1 \over 6} \, (p^2)^2 \, P^{\mu\nu\alpha\beta} +60
\hspace{0.2ex} \Delta \xi^2 (p^2)^2 P^{\mu\nu} P^{\alpha\beta}
\right]. \label{Fourier transf of massless noise} \ee

\subsubsection{Correlation functions for the linearized
Einstein tensor}


In general, we can write
$G^{{\scriptscriptstyle (1)}\hspace{0.1ex} \mu\nu}=
\langle G^{{\scriptscriptstyle (1)}\hspace{0.1ex} \mu\nu} \rangle_s
+G_{\rm f}^{{\scriptscriptstyle (1)}\hspace{0.1ex} \mu\nu}$,
where
$G_{\rm f}^{{\scriptscriptstyle (1)}\hspace{0.1ex} \mu\nu}$
is a solution to Eqs.~(\ref{unified Einstein-Langevin eq})
with zero average,
or (\ref{Fourier transf of E-L eq}) in the Fourier transformed version.
The averages
$\langle G^{{\scriptscriptstyle (1)}\hspace{0.1ex} \mu\nu} \rangle_s$
must be a solution of the linearized semiclassical Einstein equations
obtained by averaging Eqs.~(\ref{unified Einstein-Langevin eq}),
or (\ref{Fourier transf of E-L eq}).
Solutions to these equations (specially in the massless case,
$m \!=\! 0$) have been studied by several authors
\cite{Hor80,Hor81,HorWal78,Ran81,Ran82,Sue89a,Sue89b,HarHor81,Sim91,%
Jor87,FlaWal96}, particularly in connection with the problem of
the stability of the ground state of semiclassical gravity. The
two-point correlation functions for the linearized Einstein tensor
are defined by
\begin{eqnarray}
{\cal G}^{\mu\nu\alpha\beta}(x,x^{\prime}) \!\!\! &\equiv &\!\!\!
\langle G^{{\scriptscriptstyle (1)}\hspace{0.1ex} \mu\nu}(x)
G^{{\scriptscriptstyle (1)}\hspace{0.1ex} \alpha\beta}(x^{\prime})
\rangle_s
-\langle G^{{\scriptscriptstyle (1)}\hspace{0.1ex} \mu\nu}(x)
\rangle_s
\langle G^{{\scriptscriptstyle (1)}\hspace{0.1ex} \alpha\beta}
(x^{\prime})\rangle_s
\nonumber\\
 &= &\!\!\!
\langle G_{\rm f}^{{\scriptscriptstyle (1)}\hspace{0.1ex} \mu\nu}(x)
G_{\rm f}^{{\scriptscriptstyle (1)}\hspace{0.1ex} \alpha\beta}
(x^{\prime})\rangle_s.
\label{two-p corr funct}
\end{eqnarray}

Now we shall seek the family of solutions to the Einstein-Langevin
equation which can be written as a linear functional of the
stochastic source and whose Fourier transform,
$\tilde{G}^{{\scriptscriptstyle (1)}\hspace{0.1ex} \mu\nu}(p)$,
depends locally on $\tilde{\xi}^{\alpha\beta}(p)$. Each of such
solutions is a Gaussian stochastic field and, thus, it can be
completely characterized by the averages $\langle
G^{{\scriptscriptstyle (1)}\hspace{0.1ex} \mu\nu} \rangle_s$ and
the two-point correlation functions (\ref{two-p corr funct}). For
such a family of solutions, $\tilde{G}_{\rm
f}^{{\scriptscriptstyle (1)}\hspace{0.1ex} \mu\nu}(p)$ is the most
general solution to Eq.~(\ref{Fourier transf of E-L eq}) which is
linear, homogeneous and local in $\tilde{\xi}^{\alpha\beta}(p)$.
It can be written as \be \tilde{G}_{\rm f}^{{\scriptscriptstyle
(1)}\hspace{0.1ex} \mu\nu}(p) = 8 \pi G \,
D^{\mu\nu}_{\hspace{2ex} \alpha\beta}(p) \,
\tilde{\xi}^{\alpha\beta}(p), \label{G_f} \ee where
$D^{\mu\nu}_{\hspace{2ex} \alpha\beta}(p)$ are the components of a
Lorentz invariant tensor field distribution in Minkowski spacetime
(by ``Lorentz invariant'' we mean invariant under the
transformations of the orthochronous Lorentz subgroup; see
Ref.~\cite{Hor80} for more details on the definition and
properties of these tensor distributions), symmetric under the
interchanges $\alpha \! \leftrightarrow \!\beta$ and $\mu \!
\leftrightarrow \!\nu$, which is the most general solution of \be
F^{\mu\nu}_{\hspace{2ex} \rho\sigma}(p) \,
D^{\rho\sigma}_{\hspace{2ex} \alpha\beta}(p)= \delta^\mu_{(
\alpha} \delta^\nu_{\beta )}. \label{eq for D} \ee In addition, we
must impose the conservation condition: $p_\nu \tilde{G}_{\rm
f}^{{\scriptscriptstyle (1)}\hspace{0.1ex} \mu\nu}(p) = 0$, where
this zero must be understood as a stochastic variable which
behaves deterministically as a zero vector field. We can write
$D^{\mu\nu}_{\hspace{2ex} \alpha\beta}(p)= D^{\mu\nu}_{p
\hspace{1.2ex} \alpha\beta}(p)+ D^{\mu\nu}_{h \hspace{1.2ex}
\alpha\beta}(p)$, where $D^{\mu\nu}_{p \hspace{1.2ex}
\alpha\beta}(p)$ is a particular solution to Eq.~(\ref{eq for D})
and $D^{\mu\nu}_{h \hspace{1.2ex} \alpha\beta}(p)$ is the most
general solution to the homogeneous equation. Consequently, see
Eq.~(\ref{G_f}), we can write $\tilde{G}_{\rm
f}^{{\scriptscriptstyle (1)}\hspace{0.1ex} \mu\nu}(p)
=\tilde{G}_p^{{\scriptscriptstyle (1)}\hspace{0.1ex} \mu\nu}(p)+
\tilde{G}_h^{{\scriptscriptstyle (1)}\hspace{0.1ex} \mu\nu}(p)$.
To find the particular solution, we try an ansatz of the form \be
D^{\mu\nu}_{p \hspace{1.2ex} \alpha\beta}(p)= d_1(p) \,
\delta^\mu_{( \alpha} \delta^\nu_{\beta )} + d_2(p) \, p^2
P^{\mu\nu} \eta_{\alpha\beta}. \label{ansatz for D} \ee
Substituting this ansatz into Eqs.~(\ref{eq for D}), it is easy to
see that it solves these equations if we take \be d_1(p)=\left[ {1
\over F_1(p)} \right]_r, \hspace{7ex} d_2(p)= - \left[
{F_2(p)\over F_1(p) F_3(p)} \right]_r, \label{d's} \ee with \be
F_3(p) \equiv F_1(p) + 3 p^2 F_2(p)= 1-48 \pi G \, p^2 \left[
{1\over 4}\tilde{H}_{\rm B}(p)-2 \bar{\beta}\right], \label{F_3}
\ee and where the notation $[ \;\; ]_r$ means that the zeros of
the denominators are regulated with appropriate prescriptions in
such a way that $d_1(p)$ and $d_2(p)$ are well defined Lorentz
invariant scalar distributions. This yields a particular solution
to the Einstein-Langevin equations: \be
\tilde{G}_p^{{\scriptscriptstyle (1)}\hspace{0.1ex} \mu\nu}(p) = 8
\pi G \, D^{\mu\nu}_{p \hspace{1.2ex} \alpha\beta}(p) \,
\tilde{\xi}^{\alpha\beta}(p), \label{solution} \ee which, since
the stochastic source is conserved, satisfies the conservation
condition. Note that, in the case of a massless scalar field,
$m\!=\!0$, the above solution has a functional form analogous to
that of the solutions of linearized semiclassical gravity found in
the Appendix of Ref.~\cite{FlaWal96}. Notice also that, for a
massless conformally coupled field, $m\!=\!0$ and $\Delta
\xi\!=\!0$, the second term on the right hand side of
Eq.~(\ref{ansatz for D}) will not contribute in the correlation
functions (\ref{two-p corr funct}), since in this case the
stochastic source is traceless.

A detailed analysis given in Ref.~\cite{MarVer00}
concludes that the homogeneous solution
$\tilde{G}_h^{{\scriptscriptstyle (1)}\hspace{0.1ex} \mu\nu}(p)$ gives
no contribution to the correlation functions
(\ref{two-p corr funct}). Consequently
${\cal G}^{\mu\nu\alpha\beta}(x,x^{\prime}) \!=\!
\langle G_p^{{\scriptscriptstyle (1)}\hspace{0.1ex} \mu\nu}(x)
G_p^{{\scriptscriptstyle (1)}\hspace{0.1ex} \alpha\beta}
(x^{\prime})\rangle_s$, where
$G_p^{{\scriptscriptstyle (1)}\hspace{0.1ex} \mu\nu}(x)$ is the
inverse Fourier transform of (\ref{solution}),
and the correlation functions (\ref{two-p corr funct}) are
\be
\langle \tilde{G}_p^{{\scriptscriptstyle
(1)}\hspace{0.1ex} \mu\nu}\!(p)  \tilde{G}_p^{{\scriptscriptstyle
(1)}\hspace{0.1ex} \alpha\beta}\! (p^\prime) \rangle_s\! = \! 64 \!(2
\pi)^6 \! G^2 \! \delta^4\!(p\!+\!p^\prime) \! D^{\mu\nu}_{p
\hspace{1.2ex}\rho\sigma}(p) \! D^{\alpha\beta}_{p \hspace{1.2ex}
\lambda\gamma}(-p) \! \tilde{N}^{\rho\sigma\lambda\gamma}\!(p).
\ee
It is easy to see from the above analysis that the prescriptions
$[ \;\; ]_r$ in the factors $D_p$ are irrelevant in the last
expression and, thus, they can be suppressed. Taking into account
that $F_l(-p) \!=\! F^{\displaystyle \ast}_l(p)$, with $l \!=\!
1,2,3$, we get from Eqs.~(\ref{ansatz for D}) and (\ref{d's}) \bea
\!\!\!\! &&\!\!\!\!\langle \tilde{G}_p^{{\scriptscriptstyle (1)}
\hspace{0.1ex}
\mu\nu}(p) \, \tilde{G}_p^{{\scriptscriptstyle (1)}\hspace{0.1ex}
\alpha\beta} (p^\prime) \rangle_s =
\nonumber\\
\!\!\!\!\!\! && \!\!\!\!\!\!
64
(2 \pi)^6 \, G^2 \: {\delta^4(p+p^\prime) \over \left|
\hspace{0.1ex} F_1(p) \hspace{0.1ex}\right|^2 } \left[
\tilde{N}^{\mu\nu\alpha\beta}(p) - {F_2(p) \over F_3(p)} \: p^2
P^{\mu\nu} \hspace{0.2ex}
\tilde{N}^{\alpha\beta\rho}_{\hspace{3.3ex} \rho}(p)
\right.     \nn  \\
\!\! \!\!\! && \!\!\!\!\!
\left. -
{F_2^{\displaystyle \ast}(p) \over F_3^{\displaystyle \ast}(p)}
\: p^2 P^{\alpha\beta} \hspace{0.2ex}
\tilde{N}^{\mu\nu\rho}_{\hspace{3.3ex} \rho}(p)
\! +\! { \left| F_2(p) \right|^2
\over \left| F_3(p) \right|^2 }
p^2 P^{\mu\nu}  p^2 P^{\alpha\beta}
\tilde{N}
^{\rho \hspace{0.9ex} \sigma}_{\hspace{1ex} \rho \hspace{1.1ex}
\sigma} (p) \right]\!.   \nn  \\
\mbox{} \eea This last expression is well defined as a
bi-distribution and can be easily evaluated using
Eq.~(\ref{Fourier transf of noise 2}). The final explicit result
for the Fourier transformed correlation function for the Einstein
tensor is thus \bea
 \!\!\!\!\! && \!\!\!\!\!
\langle \tilde{G}_p^{{\scriptscriptstyle
(1)}\hspace{0.1ex} \mu\nu}(p) \, \tilde{G}_p^{{\scriptscriptstyle
(1)}\hspace{0.1ex} \alpha\beta} (p^\prime) \rangle_s=
\nonumber\\
 \!\!\!\!\! &=& \!\!\!\!\!{2 \over
45} \!(2 \pi)^5 \, G^2 \:
{\delta^4(p+p^\prime) \over \left| \hspace{0.1ex} F_1(p)
\hspace{0.1ex}\right|^2 } \: \theta (-p^2\!-\!4m^2) \, \sqrt{1+4
\,{m^2 \over p^2} }
\nn   \\
&&  \!\!\!\!  \times \!
\left[{1 \over 4} \left(1+4 \,{m^2 \over p^2} \right)^{\!2} \!
(p^2)^2 \,
 P^{\mu\nu\alpha\beta }  \right.
\nn   \\
&&  \left.
+\! 10 \!
\left(3 \hspace{0.2ex}\Delta \xi\!+\!{m^2 \over p^2} \right)^{\!2} \!
(p^2)^2 P^{\mu\nu} P^{\alpha\beta}
\left| 1\!-\!3 p^2  {F_2(p) \over F_3(p)} \right|^2
\right]\!\!.
\label{Fourier tr corr funct}
\eea

To obtain the correlation functions in coordinate space,
Eq.~(\ref{two-p corr funct}), we take the inverse Fourier
transform. The final result is: \be {\cal
G}^{\mu\nu\alpha\beta}(x,x^{\prime})= {\pi \over 45} \, G^2
\,{\cal F}^{\mu\nu\alpha\beta}_{x} \, {\cal G}_{\rm A}
(x-x^{\prime})+ {8 \pi \over 9} \, G^2 \, {\cal F}^{\mu\nu}_{x}
{\cal F}^{\alpha\beta}_{x} \, {\cal G}_{\rm B} (x-x^{\prime}),
\label{corr funct} \ee with \bea
\!\!\!\!\!\!&&\!\!\!\!\!\!\tilde{{\cal G}}_{\rm A}(p) \!\equiv\!
\theta (-p^2-4m^2) \, \sqrt{1+4 \,{m^2 \over p^2} } \left(1+4
\,{m^2 \over p^2} \right)^{\!2} \! {1 \over \left| \hspace{0.1ex}
F_1(p) \hspace{0.1ex}\right|^2 }  ,
\nn   \\
\!\!\!\!\!\!&&\!\!\!\!\!\!\tilde{{\cal G}}_{\rm B}(p)\! \equiv\!
\theta (-p^2\!-\!4m^2)  \sqrt{1\!+\!4{m^2 \over p^2} }\! \left(\!3
\Delta \xi\!+\!{m^2 \over p^2}\!\right)^{\!2} \!\! {1 \over \left|
F_1(p) \right|^2 }\! \left| 1\!-\! 3 p^2 {F_2(p) \over F_3(p)}
\right|^2\!\!, \label{distri} \eea where $F_l(p)$, $l=1,2,3$, are
given in (\ref{F_1 and F_2}) and (\ref{F_3}). Notice that, for a
massless field ($m \!=\! 0$), we have \bea &&F_1(p)= 1+ 4 \pi G
\hspace{0.2ex} p^2 \hspace{0.2ex}
          \tilde{H}(p;\bar{\mu}^2),
\nn  \\
&&F_2(p)= - {16 \over 3} \, \pi G \left[
(1 +180 \hspace{0.2ex}\Delta \xi^2 ) \, {1\over 4}\tilde{H}(p;\bar{\mu}^2)
-6 \Upsilon \right],
\nn  \\
&&F_3(p)= 1- 48 \pi G \hspace{0.2ex} p^2
\left[ 15 \hspace{0.2ex}\Delta \xi^2 \hspace{0.2ex}
\tilde{H}(p;\bar{\mu}^2) -2 \Upsilon \right],
\eea
with $\bar{\mu} \equiv \mu\, \exp (1920 \pi^2 \bar{\alpha})$
and
$\Upsilon \equiv \bar{\beta}
 -60 \hspace{0.2ex}\Delta \xi^2 \hspace{0.2ex}\bar{\alpha}$, and where
$\tilde{H}(p;\mu^2)$ is the Fourier transform of
$H(x;\mu^2)$ given in (\ref{Hnew}).


\subsubsection{Correlation functions for the metric
perturbations}
\label{s6.4.2}


Starting from the solutions found for the linearized Einstein tensor,
which are characterized by the two-point correlation functions
(\ref{corr funct}) [or, in terms of Fourier transforms,
(\ref{Fourier tr corr funct})], we can now solve the equations for the
metric perturbations. Working in the harmonic gauge,
$\partial_{\nu} \bar{h}^{\mu\nu} \!=\! 0$ (this zero must be
understood in a statistical sense) where
$\bar{h}_{\mu\nu} \!\equiv \! h_{\mu\nu}
\!-\! (1/2)\hspace{0.2ex} \eta_{\mu\nu} \hspace{0.2ex}h_\alpha^\alpha$,
the equations for the metric perturbations in terms of
the Einstein tensor are
\be
\Box \bar{h}^{\mu\nu}(x) \!=\! -2
G^{{\scriptscriptstyle (1)}\hspace{0.1ex} \mu\nu}(x),
\label{metric and G}
\ee
or, in terms of
Fourier transforms,
$p^2
\tilde{\bar{h}}^{\mbox{}_{\mbox{}_{\mbox{}_{\mbox{}_{\mbox{}
_{\scriptstyle \mu\nu}}}}}}\hspace{-0.5ex} (p)
\!=\! 2 \tilde{G}^{{\scriptscriptstyle (1)}\hspace{0.1ex} \mu\nu}(p)$.
Similarly to the analysis of  the equation
for the Einstein tensor, we can write
$\bar{h}^{\mu\nu} \!=\! \langle \bar{h}^{\mu\nu} \rangle_s
\!+\! \bar{h}^{\mu\nu}_{\rm f}$, where $\bar{h}^{\mu\nu}_{\rm f}$ is a
solution to these equations with zero average, and the two-point
correlation functions are defined by
\begin{eqnarray}
{\cal H}^{\mu\nu\alpha\beta}(x,x^{\prime})\!\!\!\! &\equiv &\!\!\!\!
\langle \bar{h}^{\mu\nu}(x) \bar{h}^{\alpha\beta}(x^{\prime})
\rangle_s - \langle \bar{h}^{\mu\nu}(x) \rangle_s
\langle \bar{h}^{\alpha\beta}(x^{\prime}) \rangle_s
\nonumber\\
\!\!\!\! &=&\!\!\!\!
\langle \bar{h}^{\mu\nu}_{\rm f}(x)
\bar{h}^{\alpha\beta}_{\rm f}(x^{\prime}) \rangle_s.
\label{co fu}
\end{eqnarray}

We seek solutions of
the Fourier transform of Eq.~(\ref{metric and G})
of the form
$\tilde{\bar{h}}^{\mbox{}_{\mbox{}_{\mbox{}_{\mbox{}_{\mbox{}
_{\scriptstyle \mu\nu}}}}}}_{\rm f}\hspace{-0.5ex} (p)
\!=\! 2 D(p)
\tilde{G}^{{\scriptscriptstyle (1)}\hspace{0.1ex} \mu\nu}_{\rm f}(p)$,
where $D(p)$ is a Lorentz invariant scalar distribution in Minkowski
spacetime, which is the most general solution of $p^2 D(p) \!=\! 1$.
Note that, since the linearized Einstein tensor is conserved,
solutions of this form automatically satisfy the harmonic
gauge condition. As in the previous subsection,
we can write $D(p) \!=\! [ 1/p^2 ]_r
\!+\! D_h(p)$, where $D_h(p)$ is the most general solution to the
associated homogeneous equation and, correspondingly, we have
$\tilde{\bar{h}}^{\mbox{}_{\mbox{}_{\mbox{}_{\mbox{}_{\mbox{}
_{\scriptstyle \mu\nu}}}}}}_{\rm f}\hspace{-0.5ex} (p)
\!=\!
\tilde{\bar{h}}^{\mbox{}_{\mbox{}_{\mbox{}_{\mbox{}_{\mbox{}
_{\scriptstyle \mu\nu}}}}}}_p\hspace{-0.5ex} (p) +
 \tilde{\bar{h}}^{\mbox{}_{\mbox{}_{\mbox{}_{\mbox{}_{\mbox{}
_{\scriptstyle \mu\nu}}}}}}_h \hspace{-0.5ex} (p)$. However, since
$D_h(p)$ has support on the set of points for which $p^2 \!=\! 0$,
it is easy to see from Eq.~(\ref{Fourier tr corr funct}) (from the
factor $\theta (-p^2-4 m^2)$) that $\langle
\tilde{\bar{h}}^{\mbox{}_{\mbox{}_{\mbox{}_{\mbox{}_{\mbox{}
_{\scriptstyle \mu\nu}}}}}}_h \hspace{-0.5ex} (p)
\tilde{G}^{{\scriptscriptstyle (1)}\hspace{0.1ex} \alpha\beta}
_{\rm f}(p^{\prime}) \rangle_s \!=\! 0$ and, thus, the two-point
correlation functions (\ref{co fu}) can be computed from $\langle
\tilde{\bar{h}}^{\mbox{}_{\mbox{}_{\mbox{}_{\mbox{}_{\mbox{}
_{\scriptstyle \mu\nu}}}}}}_{\rm f}\hspace{-0.5ex} (p)
\tilde{\bar{h}}^{\mbox{}_{\mbox{}_{\mbox{}_{\mbox{}_{\mbox{}
_{\scriptstyle \alpha\beta}}}}}}_{\rm
f}\hspace{-0.5ex}(p^{\prime}) \rangle_s \!=\! \langle
\tilde{\bar{h}}^{\mbox{}_{\mbox{}_{\mbox{}_{\mbox{}_{\mbox{}
_{\scriptstyle \mu\nu}}}}}}_p\hspace{-0.5ex} (p)
\tilde{\bar{h}}^{\mbox{}_{\mbox{}_{\mbox{}_{\mbox{}_{\mbox{}
_{\scriptstyle \alpha\beta}}}}}}_p\hspace{-0.5ex}(p^{\prime})
\rangle_s$. {}From Eq.~(\ref{Fourier tr corr funct}) and due to
the factor $\theta (-p^2-4 m^2)$, it is also easy to see that the
prescription $[ \;\; ]_r$ is irrelevant in this correlation
function and we obtain \be \langle
\tilde{\bar{h}}^{\mbox{}_{\mbox{}_{\mbox{}_{\mbox{}_{\mbox{}
_{\scriptstyle \mu\nu}}}}}}_p\hspace{-0.5ex} (p)
\tilde{\bar{h}}^{\mbox{}_{\mbox{}_{\mbox{}_{\mbox{}_{\mbox{}
_{\scriptstyle \alpha\beta}}}}}}_p\hspace{-0.5ex}(p^{\prime})
\rangle_s = {4 \over (p^2)^2} \, \langle
\tilde{G}_p^{{\scriptscriptstyle (1)}\hspace{0.1ex} \mu\nu}(p) \,
\tilde{G}_p^{{\scriptscriptstyle (1)}\hspace{0.1ex} \alpha\beta}
(p^\prime) \rangle_s, \ee where $\langle
\tilde{G}_p^{{\scriptscriptstyle (1)}\hspace{0.1ex} \mu\nu}(p) \,
\tilde{G}_p^{{\scriptscriptstyle (1)}\hspace{0.1ex} \alpha\beta}
(p^\prime) \rangle_s$ is given by  Eq.~(\ref{Fourier tr corr
funct}). The right hand side of this equation is a well defined
bi-distribution, at least for $m \!\neq \! 0$ (the $\theta$
function provides the suitable cutoff). In the massless field
case, since the noise kernel is obtained as the limit $m
\!\rightarrow \!0$ of the noise kernel for a massive field, it
seems that the natural prescription to avoid the divergences on
the lightcone $p^2 \!=\! 0$ is a Hadamard finite part, see
Refs.~\cite{Sch57,Zem87} for its definition. Taking this
prescription, we also get a well defined bi-distribution for the
massless limit of the last expression.

The final result for the two-point correlation function
for the field $\bar h^{\mu\nu}$ is:
\be
{\cal H}^{\mu\nu\alpha\beta}(x,x^{\prime})=
{4 \pi \over 45} \, G^2 \,{\cal F}^{\mu\nu\alpha\beta}_{x} \,
{\cal H}_{\rm A} (x-x^{\prime})+
{32 \pi \over 9} \, G^2 \,
{\cal F}^{\mu\nu}_{x} {\cal F}^{\alpha\beta}_{x} \,
{\cal H}_{\rm B} (x-x^{\prime}),
\label{corr funct 2}
\ee
where $\tilde{{\cal H}}_{\rm A}(p) \!\equiv \!
[1/(p^2)^2]\, \tilde{{\cal G}}_{\rm A}(p)$ and
$\tilde{{\cal H}}_{\rm B}(p) \!\equiv \!
[1/(p^2)^2]\, \tilde{{\cal G}}_{\rm B}(p)$, with
$\tilde{{\cal G}}_{\rm A}(p)$ and
$\tilde{{\cal G}}_{\rm B}(p)$ given by (\ref{distri}).
The two-point correlation functions for the metric perturbations
can be easily obtained using $h_{\mu\nu} \!=\!
\bar{h}_{\mu\nu}
\!-\! (1/2) \hspace{0.2ex}\eta_{\mu\nu}
\hspace{0.2ex}\bar{h}^{\alpha}_{\alpha}$.


\subsubsection{Conformally coupled field}
\label{s6.4.3}


For a conformally coupled field, {\it i.e.}, when $m = 0$ and
$\Delta \xi=0$, the previous correlation functions are greatly
simplified and can be approximated explicitly in terms of analytic
functions. The detailed results are given in Ref. \cite{MarVer00},
here we outline the main features.

When $m=0$ and $\Delta \xi=0$ we have
${\cal G}_{\rm B} (x) \!=\!0$ and
$
\tilde{{\cal G}}_{\rm A}(p)= \theta(-p^2)
\left|\hspace{0.2ex} F_1(p) \hspace{0.2ex} \right|^{-2}$.
Thus the two-point correlations functions for
the Einstein tensor is
\be
{\cal G}^{\mu\nu\alpha\beta}(x,x^{\prime})=
{\pi \over 45} \, G^2 \,{\cal F}^{\mu\nu\alpha\beta}_{x} \,
\int {d^4p\over (2\pi)^4}\frac
{e^{ip(x-x^\prime)}\,\theta(-p^2)}
{| 1+4\pi Gp^2\tilde H(p;\bar\mu^2)|^2},
\label{corr funct conf}
\ee
where $\tilde H(p,\mu^2)=(480\pi^2)^{-1}\ln
[-((p^0+i\epsilon)^2+p^ip_i)/\mu^2]$, see Eq. (\ref{Hnew}).

To estimate this integral
let us consider spacelike separated points
$(x-x^{\prime})^\mu=(0,\!{\bf x}-{\bf x}^\prime)$, and define
${\bf y}={\bf x}-{\bf x}^\prime$. We may now
formally change the momentum variable $p^\mu$
by the dimensionless vector $s^\mu$: $p^\mu=s^\mu/|{\bf y}|$, then
the previous integral denominator is
$|1+16\pi (l_p/|{\bf y}|)^2s^2\tilde H(s)|^2$,
where we have introduced the Planck length $l_p=\sqrt{G}$.
It is clear that we can consider two regimes: (a) when
$l_p \ll |{\bf y}|$, and (b) when $|{\bf y}|\sim l_p$.
In  case (a) the correlation function, for the $0000$
component, say,
will be of the order
$$
{\cal G}^{0000}({\bf y})\sim {l_p^4\over|{\bf y}|^8}.
$$
In case (b) when the denominator has zeros
a detailed calculation  carried out in Ref. \cite{MarVer00}
shows that:
$$
{\cal G}^{0000}({\bf y})\sim e^{-|{\bf y}|/l_p}\left(
{l_p\over |{\bf y}|^5}+\dots +{1\over l_p^2|{\bf y}|^2}\right)
$$
which indicates an exponential decay at distances around
the Planck scale. Thus short scale fluctuations are
strongly suppressed.

For the two-point metric correlation the results
are similar. In this case we have
\be
{\cal H}^{\mu\nu\alpha\beta}(x,x^{\prime})=
{4\pi \over 45} \, G^2 \,{\cal F}^{\mu\nu\alpha\beta}_{x} \,
\int {d^4p\over (2\pi)^4}\frac
{e^{ip(x-x^\prime )}\theta(-p^2)}
{(p^2)^2| 1+4\pi Gp^2\tilde H(p;\bar\mu^2)|^2}.
\label{corr funct conf metric}
\ee
The integrand has the same behavior of the correlation function
of Eq. (\ref{corr funct conf})
thus matter fields tends to suppress the short scale metric
perturbations.
In this case we find, as for the correlation of
the Einstein tensor,
that for case (a) above we have,
$$
{\cal H}^{0000}({\bf y})\sim {l_p^4\over|{\bf y}|^4},
$$
and for case (b) we have
$$
{\cal H}^{0000}({\bf y})\sim e^{-|{\bf y}|/l_p}\left(
{l_p\over |{\bf y}|}+\dots \right).
$$

It is interesting to write expression (\ref{corr funct conf
metric}) in an alternative way. If we use the dimensionless tensor
$P^{\mu\nu\alpha\beta}$ introduced in Eq.~(\ref{projector}), which
accounts for the effect of the operator ${\cal
F}^{\mu\nu\alpha\beta}_{\,x}$, we can write \be {\cal
H}^{\mu\nu\alpha\beta}(x,x^{\prime})= {4\pi \over 45} \, G^2 \,
\int {d^4p\over (2\pi)^4}\,\frac
{e^{ip(x-x^\prime)}\,P^{\mu\nu\alpha\beta}\,\theta(-p^2)} {|
1+4\pi Gp^2\tilde H(p;\bar\mu^2)|^2}. \label{corr funct conf
metric2} \ee This expression allows a direct comparison with the
graviton propagator for linearized quantum gravity in the $1/N$
expansion found by Tomboulis \cite{Tom77}. One can see that the
imaginary part of the graviton propagator leads, in fact, to
Eq.~(\ref{corr funct conf metric2}). In Ref.~\cite{RouVer03b} it
is shown that, in fact, the two-point correlation functions for
the metric perturbations derived from the Einstein-Langevin
equation are equivalent to the symmetrized quantum two-point
correlation functions for the metric fluctuations in the large $N$
expansion of quantum gravity interacting with $N$ matter fields.

The main results of this section are the
correlation functions  (\ref{corr funct})
and (\ref{corr funct 2}). In the case of a conformal field, the
correlation functions of the linearized Einstein
tensor have been explicitly estimated.
{}From the exponential factors $e^{-|{\bf y}|/l_p}$ in these results
for scales near the Planck length,
we see that the correlation functions of the linearized Einstein
tensor have the  Planck length as the correlation length.
A similar behavior is found for the
correlation functions of the metric perturbations.
Since these fluctuations are induced by the matter fluctuations
we infer that the effect of the matter fields is to suppress the
fluctuations of the metric at very small scales.
On the other hand,
at scales much larger than the Planck length
the induced metric fluctuations are small
compared with the free graviton propagator which goes like
$l_p^2/|{\bf y}|^2$, since the action for the free
graviton goes like $S_h\sim\int d^4 x\,l_p^{-2}h\Box h$

For background solutions of semiclassical gravity with other
scales present apart from the Planck scales (for instance, for
matter fields in a thermal state), stress-energy fluctuations may
be important at larger scales. For such backgrounds, stochastic
semiclassical gravity might predict correlation functions with
characteristic correlation lengths larger than the Planck scales.
It seems quite plausible, nevertheless, that these correlation
functions would remain non-analytic in their characteristic
correlation lengths. This would imply that these correlation
functions could not be obtained from a calculation involving a
perturbative expansion in the characteristic correlation lengths.
In particular, if these correlation lengths are proportional to
the Planck constant $\hbar$, the gravitational correlation
functions could not be obtained from an expansion in $\hbar$.
Hence, stochastic semiclassical gravity might predict a behavior
for gravitational correlation functions different from that of the
analogous functions in perturbative quantum gravity
\cite{Don94a,Don94b,Don96a,Don96b}. This is not necessarily
inconsistent with having neglected action terms of higher order in
$\hbar$ when considering semiclassical gravity as an effective
theory \cite{FlaWal96}. It is, in fact, consistent with the closed
connection of stochastic gravity with the large $N$ expansion of
quantum gravity interacting with $N$ matter fields.

\subsection{Stability of Minkowski spacetime}
\label{s6.5}

In this section we apply the validity criterion
for semiclassical gravity introduced in section
\ref{s3.3} to flat spacetime. The Minkowski metric is a
particularly simple and interesting solution of semiclassical
gravity. In fact, as we have seen in section \ref{s6.1},
when the quantum fields are in the Minkowski vacuum state one
may take the renormalized expectation value of the stress tensor as
$\langle\hat T^R_{ab}[\eta]\rangle=0$
(this is equivalent to assuming that the cosmological constant is
zero), then the Minkowski metric $\eta_{ab}$ is a solution of the
semiclassical Einstein equation (\ref{2.5}). Thus, we can look for
the stability of Minkowski spacetime against quantum matter fields.
According to the criteria we have established we have to look for
the behavior of the two-point quantum correlations for the metric
perturbations $h_{ab}(x)$ over the Minkowski background which are given by
Eqs.~(\ref{1.5b}) and (\ref{1.5c}). As we have emphasized before
these metric fluctuations separate in two parts: the first term on
the right hand side of Eq.~(\ref{1.5c}) which corresponds to the
\emph{intrinsic} metric fluctuations, and the second term which corresponds
to the \emph{induced} metric fluctuations.

\subsubsection{Intrinsic metric fluctuations}

Let us first consider the intrinsic metric fluctuations,
\begin{equation}
\langle h_{ab}(x)h_{cd}(y)\rangle_{\mathrm{int}}=\langle h^0_{ab}(x) h^0_{cd}(y)\rangle_s, \label{6.1}
\end{equation}
where $h^0_{ab}$ are the homogeneous solutions of the
Einstein-Langevin equation (\ref{2.11}), or equivalently the
linearly perturbed semiclassical equation, and where the statistical
average is taken with respect to the Wigner distribution that
describes the initial quantum state of
the metric perturbations. Since these solutions are
described by the linearized semiclassical equation around flat
spacetime we can make use of the results derived in
Refs.~\cite{Hor80,FlaWal96,AndMolMot03,AndMolMot02}.
The solutions for
the case of a massless scalar field were first discussed in
Ref.~\cite{Hor80} and an exhaustive description can be found
in Appendix~A of Ref.~\cite{FlaWal96}.
It is convenient to decompose the perturbation around Minkowski
spacetime into scalar, vectorial and tensorial parts, as
\begin{equation}
h_{ab} = \bar\phi\,\eta_{ab} + \left( \nabla_{(a}\nabla_{b)} - \eta_{ab}
\nabla_c\nabla^c
\right) \psi + 2 \nabla_{(a} v_{b)} + h_{ab}^{\mathrm{TT}},
\label{metric decomp}
\end{equation}
where $v^a$ is a transverse vector and $h_{ab}^{\mathrm{TT}}$ is a
transverse and traceless symmetric tensor, \emph{i.e.}, $\nabla_a
v^a=0$, $\nabla^a h_{ab}^{\mathrm{TT}}=0$ and $(h^{\mathrm{TT}})^a_a=0$.
A vector field $\zeta^a$ characterizes
the gauge freedom due to
infinitesimal diffeomorphisms
$h_{ab}\to h_{ab}+\nabla_a\zeta_b+ \nabla_b\zeta_a$. We may use this freedom to choose a gauge, a convenient election is the so called Lorentz or
harmonic gauge defined as
\begin{equation}
\nabla^a\left(h_{ab}-\frac{1}{2}\eta_{ab}h_c^c\right)=0.
\label{Lorentz gauge}
\end{equation}
When this gauge is imposed we have the following conditions on the metric
perturbations $\nabla_a\nabla^a v^b=0$ and $\nabla_b\bar \phi=0$, which implies
$\bar\phi=\mathrm{const}$. A remaining gauge freedom compatible with the Lorentz gauge is still possible provided the vector field $\zeta^a$ satisfies the condition $\nabla_a\nabla^a \zeta^b=0$. One can easily see
\cite{HuRouVer04a} that the vectorial and scalar part $\bar\phi$ can be eliminated,
as well as the contribution of the scalar part $\psi$
which corresponds to Fourier modes  $\tilde\psi(p)$ with $p^2=0$.
Thus we will assume that we impose the Lorentz gauge with additional gauge
transformations which leave only the tensorial component and the modes of the scalar component $\psi$ with $p^2\not=0$ in Fourier space.

Using the metric decomposition (\ref{metric decomp}) we may compute
the linearized Einstein tensor
$G^{(1)}_{ab}$. It is found that the vectorial part of the metric
perturbation gives no contribution to this tensor,
and the scalar and tensorial components give rise, respectively, to scalar and tensorial components: $G^{(1)\,(S)}_{ab}$ and
$G^{(1)\,(T)}_{ab}$, respectively. Thus let us now write the Fourier transform of
the homogeneous
Einstein-Langevin equation (\ref{Fourier transf of E-L eq}), which
is equivalent to the linearized semiclassical Einstein equation,
\begin{equation}
F^{\mu\nu}_{\hspace{2ex}\alpha\beta}(p) \,
\tilde{G}^{{\scriptscriptstyle (1)}\hspace{0.1ex} \alpha\beta}(p)=0.
\label{Fourier transf of homog E-L eq}
\end{equation}
Using the previous decomposition of the Einstein tensor this equation can be re-written in terms of its scalar and tensorial parts as
\begin{eqnarray}
\left[ F_1(p) + 3 p^2 F_2(p) \right] \tilde{G}^{(1)\, \mathrm{(S)}}
_{\mu\nu}(p) &=& 0 , \\
F_1(p) \tilde{G}^{(1)\, \mathrm{(T)} }_{\mu\nu}(p) &=& 0 .
\end{eqnarray}
where $F_1(p)$ and $F_2(p)$ are given by Eqs.~(\ref{F_1 and F_2}),
and $\tilde{G}^{(1) \, \mathrm{(S)}}_{\mu\nu}$ and $\tilde{G}^{(1)\,
\mathrm{(T)}}_{\mu\nu}$ denote, respectively,
the Fourier transformed scalar and tensorial
parts of the linearized Einstein tensor.
To simplify the problem and to
illustrate, in particular, how the runaway solutions arise we will consider the case of a massless and conformally coupled field; see Ref.~\cite{FlaWal96} for the massless case with arbitrary coupling and Refs.~\cite{MarVer00,AndMolMot03} for
the general massive case.
Thus substituting $m=0$ and $\xi=1/6$
into the functions $F_1(p)$ and $F_2(p)$, and using Eq.~(\ref{Hnew}),
the above equations become:
\begin{eqnarray}
\left( 1 + 12 \kappa \bar{\beta} p^2 \right) \tilde{G}^{(1)\,
\mathrm{(S)}}_{\mu\nu}(p) &=& 0, \label{scalar} \\
\lim\limits_{\;\epsilon \rightarrow 0^+} \left[ 1 + \frac{\kappa p^2}{960 \pi^2}
\ln \left( \frac{-(p^0 + i \epsilon)^2 + \vec{p}^{\,2} }{\mu^2} \right) \right]
\tilde{G}^{(1)\, \mathrm{(T)} }_{\mu\nu}(p) &=& 0, \label{tensorial}
\end{eqnarray}
where $\kappa=8\pi G$. Let us consider these two equations separately.

For the scalar component when $\bar{\beta} = 0$ the only solution
is $\tilde{G}^{(1)\, \mathrm{(S)} }_{\mu\nu}(p) = 0$. When $\bar{\beta}
> 0$ the solutions for the scalar component exhibit an oscillatory
behavior in spacetime coordinates which corresponds to a massive
scalar field with $m^2 = (12 \kappa |\bar{\beta}|)^{-1}$; for
$\bar{\beta} < 0$ the solutions correspond to a tachyonic field
with $m^2 = - (12 \kappa |\bar{\beta}|)^{-1}$: in spacetime
coordinates they exhibit an exponential behavior in time, growing
or decreasing, for wavelengths larger than $4 \pi (3 \kappa
|\bar{\beta}|)^{1/2}$, and an oscillatory behavior for wavelengths
smaller than $4 \pi (3 \kappa |\bar{\beta}|)^{1/2}$. On the other
hand, the solution $\tilde{G}^{(1)\, \mathrm{(S)} }_{\mu\nu}(p) = 0$ is
completely trivial since any scalar metric perturbation
$\tilde{h}_{\mu\nu}(p)$ giving rise to a vanishing linearized Einstein
tensor can be eliminated by a gauge transformation.

For the tensorial component, when $\mu \le \mu_{\mathrm{crit}} =
l_p^{-1} (120\pi)^{1/2} e^{\gamma}$, where $l_p$ is the Planck length
($l_p^2\equiv \kappa/8\pi$) the first factor in
Eq.~(\ref{tensorial}) vanishes for four complex values of $p^0$ of
the form $\pm \omega$ and $\pm \omega^*$, where $\omega$ is some
complex value.
This means that in the corresponding propagator, there are two poles on the upper half plane of the complex $p^0$ plane and
two poles in the lower half plane.
We will consider here the case in which $\mu <
\mu_{\mathrm{crit}}$; a detailed description of the situation for
$\mu \ge \mu_{\mathrm{crit}}$ can be found in Appendix~A of
Ref.~\cite{FlaWal96}. The two zeros on the upper half of the
complex plane correspond to solutions in spacetime coordinates which
exponentially grow in time, whereas the two on the lower half
correspond to solutions exponentially decreasing in time. Strictly
speaking, these solutions only exist in spacetime coordinates,
since their Fourier transform is not well defined. They are
commonly referred to as \emph{runaway} solutions and for $\mu \sim
l_p^{-1}$ they grow exponentially in time scales comparable to the
Planck time.

Consequently, in addition to the solutions with $G_{ab}^{(1)}(x)=0$,
there are other solutions that in Fourier space take the form
$ \tilde G_{ab}^{(1)}(p)\propto \delta(p^2-p^2_0)$ for some particular values of $p_0$, but all of them exhibit exponential instabilities with characteristic Planckian time scales.
In order to deal with those unstable solutions, one possibility is
to make use of the \textit{order reduction} prescription
\cite{ParSim93}, which we will briefly summarize in the last
subsection. Note that the $p^2$ terms in
Eqs.~(\ref{scalar}) and  (\ref{tensorial})
come from two spacetime derivatives of the Einstein tensor, moreover,
the $p^2\ln p^2$ term comes from the nonlocal term of the expectation value of the stress tensor. The order reduction prescription amounts here to neglecting these
higher derivative terms. Thus, neglecting the terms proportional to $p^2$
in Eqs.~(\ref{scalar}) and  (\ref{tensorial}), we are left only with the
solutions which satisfy $\tilde{G}^{(1)}_{ab}(p)=0$.
The result for the metric perturbation in the gauge introduced above can
be obtained by solving for the Einstein tensor, which in the Lorentz gauge
of Eq.~(\ref{Lorentz gauge}) reads:
\begin{equation}
\tilde{G}^{(1)}_{ab}(p)=\frac{1}{2}p^2\left(\tilde h_{ab}(p)-\frac{1}{2}\eta_{ab}\tilde h_c^c (p)\right).
\label{lin Einstein in Lorentz}
\end{equation}
These solutions for
$\tilde{h}_{ab}(p)$ simply correspond to free linear gravitational
waves propagating in Minkowski spacetime expressed in the
transverse and traceless (TT) gauge. When substituting back into
Eq.~(\ref{6.1}) and averaging over the initial conditions we
simply get the symmetrized quantum correlation function for free
gravitons in the TT gauge for the state given by the
Wigner distribution. As far as the intrinsic fluctuations are
concerned, it seems that the order reduction prescription is too
drastic, at least in the case of Minkowski spacetime, since no
effects due to the interaction with the quantum matter fields are
left.

A second possibility, proposed by Hawking \emph{et al.}
\cite{HawHerRea01,HawHer02}, is to impose boundary conditions which
discard the runaway solutions that grow unbounded in time.
These boundary conditions
correspond to a special prescription for the integration contour
when Fourier transforming back to spacetime coordinates.
As we will discuss in some more detail in the next subsection,
this prescription reduces here to integrating along the real axis in the $p^0$ complex plane. Following that procedure we get, for example, that for a massless
conformally coupled matter field with  $\bar{\beta} > 0$ the
intrinsic contribution to the symmetrized quantum correlation
function coincides with that of free gravitons plus an extra
contribution for the scalar part of the metric perturbations.
This extra massive scalar
renders Minkowski spacetime stable, but also
plays a crucial role in
providing a graceful exit in inflationary models driven by the
vacuum polarization of a large number of conformal fields. Such a
massive scalar field would not be in conflict with present
observations because, for the range of parameters
considered, the mass would be far too large to have observational
consequences \cite{HawHerRea01}.

\subsubsection{Induced metric fluctuations}

The induced metric fluctuations are described by the second term in
Eq.~(\ref{1.5c}). They are dependent on the noise kernel that describes
the stress tensor fluctuations of the matter fields,
\begin{equation}
\langle h_{ab}(x)h_{cd}(y)\rangle_{\mathrm{ind}} =
\frac{\bar\kappa^2}{N}\int d^4x^\prime
d^4y^\prime \sqrt{g(x^\prime)g(y^\prime)}
G_{abef}^{\mathrm{ret}}(x,x^\prime) N^{efgh}(x^\prime,y^\prime)
G_{cdgh}^{\mathrm{ret}}(y,y^\prime),
\label{6.2}
\end{equation}
where here we have written the expression in the large $N$ limit, so that $\bar\kappa=N\kappa$ ($\kappa=8\pi G$) and $N$ is the number of
independent free scalar fields. The
contribution corresponding to the induced quantum fluctuations is
equivalent to the stochastic correlation function obtained by
considering just the inhomogeneous part of the solution to the
Einstein-Langevin equation. We can make use of the results for the metric correlations obtained in sections
\ref{s6.3} and \ref{s6.4} for solving the Einstein-Langevin
equation. In fact, one should simply take $N=1$ to transform our
expressions here to those of sections
\ref{s6.3} and \ref{s6.4} or, more precisely,
one should multiply the noise kernel in the expressions of sections
\ref{s6.3} and \ref{s6.4} by
$N$ in order to use those expressions here, as follows
{}from the fact that now we have $N$ independent matter fields.

As we have seen in section \ref{s6.4},
following Ref.~\cite{MarVer00}, the Einstein-Langevin equation can
be entirely written in terms of the linearized Einstein tensor.
The equation involves second spacetime derivatives of
that tensor and in terms of its Fourier components is given
in Eq.~(\ref{Fourier transf of E-L eq}) as
\begin{equation}
F^{\mu\nu}_{\hspace{2ex}\alpha\beta}(p) \,
\tilde{G}^{{\scriptscriptstyle (1)}\hspace{0.1ex} \alpha\beta}(p)=
\bar\kappa \, \tilde{\xi}^{\mu\nu}(p),
\label{6.3}
\end{equation}
where we have used now the
rescaled coupling $\bar\kappa$. The solution for the linearized
Einstein tensor is given in Eq.~(\ref{solution})
in terms of the retarded propagator
$D_{\mu\nu\rho\sigma}(p)$ defined in Eq.~(\ref{eq for D}).
Now this propagator, which is written in
Eq.~(\ref{ansatz for D}), exhibits two poles in the upper half complex $p^0$ plane and two poles in the lower half plane, as we have seen analyzing the zeros in
Eqs.~(\ref{scalar}) and (\ref{tensorial}) for the massless and conformally coupled case.
The retarded propagator in spacetime coordinates is obtained as usual by taking
the appropriate integration contour in the $p^0$ plane.
It is convenient in this case to deform the integration path along the real $p^0$ axis so as to leave the two poles of the upper half plane below that path. In this way when closing the contour by un upper half circle, in order to compute the anti-causal part of the propagator, there will be no contribution.
The problem now is that when closing the contour on the lower half plane, in order to compute the causal part, the contribution of the upper half plane poles gives an unbounded solution, a runaway instability. If we adopt Hawking \emph{et al.}
\cite{HawHerRea01,HawHer02} criterion of imposing final boundary conditions which discard solutions growing unboundedly in time this implies that we just need to take the integral along the real axis, as was done in section \ref{s6.4.2}. But now the propagator is no longer strictly retarded, there are causality violations in time scales of the order of $\sqrt{N}l_p$, which should have no observable consequences. This propagator, however, has well defined Fourier transform.

Following the steps after Eq.~(\ref{solution}) the Fourier transform of the two-point correlation for the linearized Einstein tensor can be written in our case as,
\begin{equation}
\langle \tilde{G}^{(1)}_{ \mu\nu}(p)  \tilde{G}^{
(1)}_{\alpha\beta} (p^\prime) \rangle_{\mathrm{ind}} =
\frac{\bar\kappa^2}{N}(2\pi)^4
\delta^4\!(p+p^\prime)
D_{\mu\nu\rho\sigma}(p) D_{\alpha\beta\lambda\gamma}(-p)  \tilde{N}^{\rho\sigma\lambda\gamma}(p),
\label{6.3a}
\end{equation}
where
the noise kernel $\tilde{N}^{\rho\sigma\lambda\gamma}(p)$ is given by Eq. (\ref{Fourier transf of noise 2}).
Note that these correlation functions are invariant
under gauge transformations of the metric perturbations because the linearized Einstein tensor is  invariant under those transformations.

We may also take the order reduction prescription which amounts in this case to neglecting  terms in the propagator which are proportional to $p^2$, corresponding to two spacetime derivatives of the Einstein tensor. The propagator then becomes a constant, and we have
\begin{equation}
\langle \tilde{G}^{(1)}_{ \mu\nu}(p)  \tilde{G}^{
(1)}_{\alpha\beta} (p^\prime) \rangle_{\mathrm{ind}} =
\frac{\bar\kappa^2}{N}(2\pi)^4
 \delta^4\!(p\!+\!p^\prime)   \tilde{N}_{\mu\nu\alpha\beta}(p).
\label{6.3b}
\end{equation}

Finally we may derive the correlations for the metric perturbations from
these equations (\ref{6.3a}) or (\ref{6.3b}).
In the Lorentz or harmonic gauge the linearized Einstein tensor takes the particularly simple form  of Eq. (\ref{lin Einstein in Lorentz}) in terms of the metric perturbation. One may derive the correlation functions for
$\tilde{h}_{\mu\nu}(p)$ as it was done in section \ref{s6.4.2} to get
\begin{equation}
\langle
\tilde{\bar{h}}_{\mu\nu}(p)
\tilde{\bar{h}}_{\alpha\beta}(p^{\prime})
\rangle_{\mathrm{ind}} = {4 \over (p^2)^2} \, \langle
\tilde{G}_{\mu\nu}(p)
\tilde{G}_{ \alpha\beta}
(p^\prime) \rangle_{\mathrm{ind}}.
\label{6.3c}
\end{equation}
There will be one possible expression for the two-point metric correlation  which corresponds to the Einstein tensor correlation of Eq.~(\ref{6.3a}) and another expression corresponding to Eq.~(\ref{6.3a}), when the order reduction prescription is used. We should note that contrary to the correlation functions for the Einstein tensor, the two-point metric correlation is not gauge invariant (it is given in the Lorentz gauge). Moreover, when taking the Fourier transform to get the correlations in spacetime coordinates there is an apparent infrared divergence when $p^2=0$
in the massless case. This can be seen from the expression for the noise kernel $\tilde{N}_{\mu\nu\alpha\beta}(p)$ defined in
Eq.~(\ref{Fourier transf of noise 2}). For the massive case no such divergence due to the factor $\theta(-p^2-4m^2)$ exists, but as one takes the limit $m\to 0$ it will show up. This infrared divergence, however, is a gauge artifact that has been enforced by the use of the Lorentz gauge. A gauge different from the Lorentz gauge should be used in the massless case; see Ref.~\cite{HuRouVer04a} for a more detailed discussion of this point.

Let us now write
the two-point metric correlation function in spacetime coordinates
for the massless and conformally coupled fields.
In order to avoid runaway solutions we use the prescription that the propagator should have a well defined Fourier transform, by integrating along the real axis in the complex $p^0$ plane. This was, in fact, done in section \ref{s6.4.3}, see
Eq. (\ref{corr funct conf metric2}), which we may write now as
\begin{equation}
\langle
\tilde{\bar{h}}_{\mu\nu}(x)
\tilde{\bar{h}}_{\alpha\beta}(y)
\rangle_{\mathrm{ind}}=\frac{\bar\kappa^2}{720\pi N}
\int\frac{d^4p}{(2\pi)^4}\frac
{e^{ip(x-y)}\,P_{\mu\nu\alpha\beta}\,\theta(-p^2)} {|
1+(\bar\kappa/2)p^2\tilde H(p;\bar\mu^2)|^2},
\label{6.4}
\end{equation}
where the projector $P_{\mu\nu\alpha\beta}$ is defined in Eq. (\ref{projector}).
This correlation
function for the metric perturbations is in agreement with the
real part of the graviton propagator obtained by Tomboulis in
Ref.~\cite{Tom77} using a large $N$ expansion with Fermion fields.
Note that when the
order reduction prescription is used the terms in the denominator of Eq.~(\ref{6.4}) which are proportional to $p^2$ are
neglected. Thus, in contrast to
the intrinsic metric fluctuations, there is still a nontrivial
contribution to the induced metric fluctuations due to the quantum matter
fields in this case.

To estimate the above integral let us follow section \ref{s6.4.3}
and consider space like separated
points $x-y=(0,{\mathbf{r}})$ and introduce the Planck length
$l_p$. For space
separations $|{\mathbf{r}}|\gg l_p$ we have that the two-point correlation
(\ref{6.4}) goes like
$
 \sim{N l_p^4}/{|{\mathbf{r}}|^4},
$
and for $|{\mathbf{r}}|\sim \sqrt{N} l_p$ we have that they go like
$
\sim
\exp(-|{\mathbf{r}}|/\sqrt{N}l_p){l_p}/{|{\mathbf{r}}|}.
$
Since these metric fluctuations are induced by the matter stress fluctuations we
infer that the effect of the matter fields is to suppress metric
fluctuations at small scales. On the other hand, at large scales
the induced metric fluctuations are small compared to the free
graviton propagator which goes like $l_p^2/|{\mathbf{r}}|^2$.

We thus conclude that, once the instabilities giving rise to the
unphysical runaway solutions have been discarded, the fluctuations
of the metric perturbations around the Minkowski spacetime induced
by the interaction with quantum scalar fields are indeed stable
(instabilities lead to divergent results when Fourier
transforming back to spacetime coordinates).
We have found that, indeed, both the intrinsic and the induced
contributions to the quantum correlation functions of metric perturbations
are stable, and consequently Minkowski spacetime is stable.

\subsubsection{Order reduction prescription and large $N$}

Runaway solutions are a typical feature of equations describing
back-reaction effects, such is in classical electrodynamics, and
are due to higher than two time derivatives in the dynamical
equations. Here we will give a qualitative analysis of this problem
in semiclassical gravity. In a very schematic way the semiclassical Einstein
equations have the form
\begin{equation}
G_h+l_p^2\ddot G_h=0,
\label{6.7}
\end{equation}
where $G_h$ stands for the linearized Einstein tensor
over the Minkowski background, say,
and we have simplified the equation as much as
possible. The second term of the equation is due to the vacuum
polarization of matter fields and contains four time derivatives
of the metric perturbation. Some specific examples of such an equation
are, in momentum space, Eqs.~(\ref{scalar}) and (\ref{tensorial}).
The order reduction
procedure is based on treating perturbatively the terms involving
higher order derivatives, differentiating the equation under
consideration and substituting back the higher derivative terms in
the original equation keeping only terms up to the required order
in the perturbative parameter. In the case of the semiclassical
Einstein equation, the perturbative parameter is $l_p^2$. If we
differentiate twice Eq.~(\ref{6.7}) with respect to time it is
clear that the second order derivatives of the Einstein tensor are
of order $l_p^2$. Substituting back into the original equation, we
get the following equation up to order $l_p^4$: $ G_h=0+ O(l_p^4). $
Now, there are certainly no runaway solutions but also no effect
due to the vacuum polarization of matter fields. Note that the
result is not so trivial when there is an inhomogeneous term on
the right hand side of Eq.~(\ref{6.7}), this is what happens with the
induced fluctuations predicted by the Einstein-Langevin equation.

Semiclassical gravity is expected to provide reliable results as
long as the characteristic length scales under consideration, say
$L$, satisfy that $L\gg l_p$ \cite{FlaWal96}. This can be
qualitatively argued by estimating the magnitude of the different
contributions to the effective action for the gravitational field,
considering the relevant Feynman diagrams and using dimensional
arguments. Let us write the effective gravitational action, again
in a very schematic way, as
\begin{equation}
S_{\mathrm{eff}}=\int d^4x \sqrt{-g}\left( \frac{1}{l_p^2}R
+\alpha R^2+l_p^2 R^3+\dots \right),
\label{6.8}
\end{equation}
where $R$ is the Ricci scalar. The first term is the usual
classical Einstein-Hilbert term, the second stands for terms
quadratic in the curvature (square of Ricci and Weyl tensors) these
terms appear as radiative corrections due to vacuum polarization
of matter fields, here $\alpha$ is a dimensionless parameter
presumably of order 1, the $R^3$ terms are higher order
corrections which appear for instance when one considers internal
graviton propagators inside matter loops. Let us assume that
$R\sim L^{-2}$ then the different terms in the action are of the
order of $R^2\sim L^{-4}$ and $l_p^2R^3\sim l_p^2L^{-6}$.
Consequently when $L\gg l_p^2$, the term due to matter loops is a
small correction to the Einstein-Hilbert term $(1/l_p^2)R\gg R^2$,
and this term can be treated as a perturbation. The justification
of the order reduction prescription is actually based on this
fact. Therefore, significant effects from the vacuum polarization
of the matter fields are only expected when their small
corrections accumulate in time, as would be the case, for
instance, for an evaporating macroscopic black hole all the way
before reaching Planckian scales (see section \ref{sec8.3}).

However if we have a large number $N$ of matter fields the
estimates for the different terms change in a remarkable way. This
is interesting because the large $N$ expansion seems,
as we have argued in section \ref{largeN}, the best
justification for semiclassical gravity. In fact, now the $N$ vacuum
polarization terms involving loops of matter are of order
$NR^2\sim NL^{-4}$. For this reason the contribution of the
graviton loops, which is just of order $R^2$
as any other loop of matter, can be neglected in front
of the matter loops; this justifies the semiclassical
limit. Similarly higher order corrections are of order
$Nl_p^2R^3\sim Nl_p^2L^{-6}$. Now there is a regime, when $L\sim
\sqrt{N}l_p$, where the Einstein-Hilbert term is comparable to the
vacuum polarization of matter fields, $(1/l_p^2)R\sim NR^2$, and
yet the higher correction terms can be neglected because we still
have $L\gg l_p$, provided $N\gg 1$. This is the kind of situation
considered in trace anomaly driven inflationary models
\cite{HawHerRea01}, such as that originally proposed by Starobinsky
\cite{Sta80}, see also Ref.~\cite{Vil85},
where exponential inflation is driven by
a large number of massless conformal fields. The order reduction
prescription would completely discard the effect from the vacuum
polarization of the matter fields even though it is comparable to
the Einstein-Hilbert term. In contrast, the procedure proposed by
Hawking \emph{et al.} keeps the contribution from the matter
fields. Note that here the actual physical Planck length $l_p$ is
considered, not the rescaled one, $\bar{l}_p^2 = \bar\kappa/8\pi$,
which is related
to $l_p$ by $l_p^2 = \kappa/8\pi= \bar{l}_p^2/ N$.

\subsubsection{Summary}

An analysis of the stability of any solution of semiclassical
gravity with respect to small quantum perturbations should include
not only the evolution of the expectation value of the metric
perturbations around that solution, but also their fluctuations,
encoded in the quantum correlation functions. Making use of the
equivalence (to leading order in $1/N$, where $N$ is the number of
matter fields) between the stochastic correlation functions
obtained in stochastic semiclassical gravity and the quantum
correlation functions for metric perturbations around a solution
of semiclassical gravity, the symmetrized two-point quantum
correlation function for the metric perturbations can be
decomposed into two different parts: the intrinsic metric fluctuations
due to the fluctuations of the initial state of the metric
perturbations itself, and the fluctuations induced by their
interaction with the matter fields. From the
linearized perturbations of the semiclassical Einstein equation,
information on the intrinsic metric fluctuations can be retrieved.
On the other hand, the information on the induced metric fluctuations
naturally follows from the solutions of the Einstein-Langevin
equation.

We have analyzed the symmetrized two-point
quantum correlation function for the metric perturbations around
the Minkowski spacetime interacting with $N$ scalar fields
initially in the Minkowski vacuum state. Once the
instabilities that arise in semiclassical gravity which
are commonly regarded as unphysical, have been properly dealt with
by using the order reduction prescription or the procedure
proposed by Hawking \emph{et al.} \cite{HawHerRea01,HawHer02}, both
the intrinsic and the induced contributions to the quantum
correlation function for the metric perturbations are found to be
stable \cite{HuRouVer04a}. Thus, we conclude that Minkowski
spacetime is a valid solution of semiclassical gravity.

\section{Structure formation in the early universe}
\label{sec:strfor}

Structure formation in the early universe is a key problem in modern
cosmology. It is believed that galaxies, clusters or any large scale
structure  observed today originated from random sources of
primordial inhomogeneities (density contrast) in the early universe
amplified by the expansion of the universe. Theories of structure
formation based on general relativity theory have been in existence
for over 60 years \cite{Lif46,Lif63,BelKhaLif70} (see e.g.,
\cite{Pee80,Bar80,Pad93}) long before the advent of inflationary
cosmology \cite{KolTur90,Lin90,Muk05} but the inflation paradigm
\cite{Gut81,AlbSte82,Lin82,Lin85} provided at least two major
improvements in the modern theory of cosmological structure formation
\cite{Haw82,GutPi82,BarSteTur82,MukFelBra92}:
\begin{itemize}
\item The sources: instead
of a classical white noise source arbitrarily specified, the seeds of
structures of the new theory are from quantum fluctuations which obey
equations derivable from the dynamics of the inflaton field, the
quantum field which is responsible for driving inflation.
\item The
spectrum: The almost scale- invariant spectrum (masses of galaxies as
a function of their scales) has a more natural explanation from the
almost exponential expansion of the inflationary universe than from
the power law expansion of the FRW universe in the traditional
theory.
\end{itemize}

Stochastic gravity provides a sound and natural formalism for the
derivation of the cosmological perturbations generated during
inflation. In Ref.~\cite{RouVer07} it was shown that the correlation
functions that follow from the Einstein-Langevin equation which
emerges in the framework of stochastic gravity coincide with that
obtained with the usual quantization procedures \cite{MukFelBra92}
when both the metric perturbations and the inflaton fluctuations are
linearized. Stochastic gravity, however, can naturally deal with the
fluctuations of the inflaton field even beyond the linear
approximation. In the last subsection we will enumerate possible
advantages of the stochastic gravity treatment of this problem over
the usual methods based on the quantization of the linear
cosmological and inflaton perturbations.

We should point out that the equivalence at the linearized level is proved
in Ref.~\cite{RouVer07} directly from the field equations of the perturbations and by showing that the stochastic and the quantum correlations are both given by identical expressions. Within the stochastic gravity framework an explicit computation of the curvature perturbation correlations was performed by Urakawa and Maeda \cite{UraMae07}. A convenient approximation for that computation, used by these authors, leads only to a small discrepancy with the usual approach for the observationally relevant part of the spectrum. We think the deviation from the standard result found for superhorizon modes would not arise if an exact calculation were used.

Here, we illustrate the equivalence with the conventional
approach with one of the simplest chaotic inflationary models in
which the background spacetime is a quasi de Sitter universe
\cite{RouVer00,RouVer07}.


\subsection{The model}


In this chaotic inflationary model \cite{Lin90} the inflaton field
$\phi$ of mass $m$ is described by the following Lagrangian
density,
\begin{equation} {\cal
L}(\phi)={1\over 2}g^{ab}\nabla_a\phi \nabla_b\phi + {1\over 2}m^2\phi^2.
\label{1.14}
\end{equation}
The conditions for the existence of an inflationary period, which is
characterized by an accelerated cosmological expansion, is that the
value of the field over a region with the typical size of the Hubble
radius is higher than the Planck mass $m_P$. This is because in order
to solve the cosmological horizon and flatness problem more than 60
e-folds of expansion are needed, to achieve this the scalar field
should begin with a value higher than $3m_P$. The inflaton mass is
small: as we will see, the large scale anisotropies measured in the
cosmic background radiation \cite{Smo92} restrict the inflaton mass
to be of the order of $10^{-6}m_P$. We will not discuss the
naturalness of this inflationary model but simply assume that if one
such region is found (inside a much larger universe) it will inflate
to become our observable universe.

We want to study the metric perturbations produced by the
stress-energy tensor
fluctuations of the inflaton field on the homogeneous background of a flat
Friedmann-Robertson-Walker model, described by the cosmological scale
factor $a(\eta)$, where $\eta$ is the conformal time, which is driven by
the homogeneous inflaton field $\phi(\eta)=\langle\hat\phi\rangle$. Thus we
write the inflaton field in the following form
\begin{equation}
\hat\phi=\phi(\eta)+ \hat\varphi (x),
\label{1.15}
\end{equation}
where $\hat\varphi (x)$ corresponds to a free massive quantum scalar field
with zero expectation value on the homogeneous background metric:
$\langle\hat\varphi\rangle=0$.
We will restrict ourselves to scalar-type metric perturbations because
these are the ones that couple to the inflaton fluctuations in the linear
theory. We note that this is not so if we were to consider inflaton
fluctuations beyond the linear approximation, then tensorial and vectorial
metric perturbations would also arise. The perturbed metric $\tilde
g_{ab}=g_{ab}+h_{ab}$ can be written in the longitudinal gauge as,
\begin{equation}
ds^2=a^2(\eta)[-(1+2\Phi(x))d\eta^2+(1-2\Psi(x))\delta_{ij}dx^idx^j],
\label{1.16} \end{equation} where the scalar metric perturbations
$\Phi(x)$ and $\Psi(x)$ correspond to Bardeen's gauge invariant
variables \cite{Bar80}.


\subsection{Einstein-Langevin equation for scalar metric perturbations}


The Einstein-Langevin equation as described in  Sec. \ref{sec2} is
gauge invariant, and thus we can work in a desired gauge and then
extract the gauge invariant quantities. The Einstein-Langevin
equation (\ref{2.11}) now reads:
\begin{equation}
G^{(0)}_{ab}-8\pi G\langle\hat T^{(0)}_{ab}\rangle+
G^{(1)}_{ab}(h)-8\pi G\langle\hat T^{(1)}_{ab}(h)\rangle=
 8\pi G\xi_{ab},
\label{1.17}
\end{equation}
Note that the two first terms cancel, that is, $G^{(0)}_{ab}-8\pi
G\langle\hat T^{(0)}_{ab}\rangle=0$, if the background metric is a
solution of the semiclassical Einstein equations. Here the subscripts
$(0)$ and $(1)$ refer to functions in the background metric $g_{ab}$
and linear in the metric perturbation $h_{ab}$, respectively. The
stress tensor operator $\hat T_{ab}$ for the minimally coupled
inflaton field in the perturbed metric is:
\begin{equation}
\hat T_{ab}=
\tilde\nabla_{a}\hat\phi \tilde\nabla_{b}\hat\phi+{1\over2}\tilde
g_{ab} (\tilde\nabla_{c} \hat\phi \tilde\nabla^{c}\hat\phi+
m^2\hat\phi^2).
\label{1.18}
\end{equation}

Using the decomposition of the scalar field into its
homogeneous and inhomogeneous part, see  Eq.~(\ref{1.15}), and the metric
$\tilde g_{ab}$ into its homogeneous background $g_{ab}$ and
its perturbation $h_{ab}$,
the renormalized expectation value for the stress-energy
tensor operator can be written as
\begin{equation}
\langle \hat T^R_{ab}[\tilde g]\rangle=
\langle \hat T_{ab}[\tilde g]\rangle_{\phi\phi}+
\langle \hat T_{ab}[\tilde g]\rangle_{\phi\varphi}+
\langle \hat T^R_{ab}[\tilde g]\rangle_{\varphi\varphi},
\label{1.19}
\end{equation}
where the subindices indicate the degree of dependence
on the homogeneous field $\phi$ and its perturbation $\varphi$.
The first term in this equation depends only on the homogeneous field
and it is given by the classical expression.
The second term is proportional to
$\langle\hat\varphi[\tilde g]\rangle$ which is not zero because the field
dynamics is considered on the perturbed spacetime, {\it i.e.}, this term
includes the coupling of the field with $h_{ab}$ and
may be obtained from the expectation value of the linearized
Klein-Gordon equation,
\begin{equation}
\left( \Box_{g+h}-m^2\right)\hat\varphi =0.
\label{1.19a}
\end{equation}
The last term in Eq.~(\ref{1.19}) corresponds to the expectation
value of the stress tensor for a free scalar field on the spacetime
of the perturbed metric.

After using the previous decomposition, the noise kernel
$N_{abcd}[g;x,y)$ defined in Eq.~(\ref{2.8}) can be written as
\begin{eqnarray}
\langle \{\hat t_{ab}[g;x),\hat t_{cd}[g;y)\}\rangle
\!\!\! &=&\!\!\!
\langle \{\hat t_{ab}[g;x),\hat t_{cd}[g;y)\}
\rangle_{(\phi\varphi)^2}
\nonumber\\
&&\!\!\! +
\langle \{\hat t_{ab}[g;x),\hat t_{cd}[g;y)\}
\rangle_{(\varphi\varphi)^2},
\label{1.20}
\end{eqnarray}
where we have used the fact that $\langle\hat\varphi\rangle=0
=\langle\hat\varphi\hat\varphi\hat\varphi\rangle$ for Gaussian states
on the background geometry. We consider the vacuum state to be the
Euclidean vacuum which is preferred in the de Sitter background, and
this state is Gaussian. In the above equation the first term is
quadratic in $\hat\varphi$ whereas  the second one is quartic, both
contributions to the noise kernel are separately conserved since both
$\phi(\eta)$ and $\hat\varphi$ satisfy the Klein-Gordon field
equations on the background spacetime. Consequently, the two terms
can be considered separately. On the other hand,  if one treats $\hat
\varphi$ as a small perturbation the second term in (\ref{1.20}) is
of lower order than the first and may be consistently neglected, this
corresponds to neglecting the last term of Eq.~(\ref{1.19}). The
stress tensor fluctuations due to a term of that kind were considered
in Ref. \cite{RouVer99}.

We can now write down the Einstein-Langevin equations (\ref{1.17})
to linear order in the inflaton fluctuations. It is easy to check
\cite{RouVer07} that the {\it space-space} components coming from
the stress tensor expectation value terms and the stochastic
tensor are diagonal, i.e. $\langle\hat T_{ij}\rangle=0= \xi_{ij}$
for $i\not= j$. This, in turn, implies that the two functions
characterizing the scalar metric perturbations are equal:
$\Phi=\Psi$ in agreement with ref. \cite{MukFelBra92}. The
equation for $\Phi$ can be obtained from the $0i$-component of the
Einstein-Langevin equation which, neglecting a nonlocal term, reads in Fourier space as
\begin{equation} 2ik_i({\cal H}\Phi_k+\Phi'_k)= 8\pi
G(\xi_{0i})_k, \label{1.21}
\end{equation}
where $k_i$ is the comoving momentum component associated to the
comoving coordinate $x^i$, and we have used the definition
$\Phi_k(\eta)= \int d^3 x \exp(-i\vec k\cdot\vec x)\Phi(\eta,\vec
x)$. Here primes denote derivatives with respect to the conformal
time $\eta$ and ${\cal H}=a'/a$. A nonlocal term of dissipative
character which comes from the second term in Eq.~(\ref{1.19}) should
also appear on the left hand side of Eq.~(\ref{1.21}), but we have
neglected it to simplify the forthcoming expressions (the large scale spectrum does not change in a substantial way).
We must emphasize, however, that the proof of the equivalence of the
stochastic approach to linear order in $\hat\varphi$ to the usual
linear cosmological perturbations approach does not assume that simplification \cite{RouVer07}. To solve Eq.~(\ref{1.21}), whose
left-hand side comes from the linearized Einstein tensor for the
perturbed metric \cite{MukFelBra92}, we need the retarded propagator
for the gravitational potential $\Phi_k$,
\begin{equation} G_k(\eta,\eta')= -i {4\pi\over k_i m_P^2}\left(
\theta(\eta-\eta') {a(\eta')\over a(\eta)}+f(\eta,\eta')\right),
\label{1.22} \end{equation} where $f$ is a
homogeneous solution of Eq.~(\ref{1.21}) related to the initial conditions
chosen and $m_P^2=1/G$. For instance, if we take
$f(\eta,\eta')=-\theta(\eta_0-\eta')a(\eta')/a(\eta)$ the solution would
correspond to ``turning on" the stochastic source at $\eta_0$.
With the solution of the Einstein-Langevin equation (\ref{1.21}) for the
scalar metric perturbations we are in the position to compute the
two-point correlation functions for these perturbations.


\subsection{Correlation functions for scalar metric perturbations}


The two-point
correlation function for the scalar metric perturbations induced by the
inflaton fluctuations is thus given by
\begin{eqnarray}
\!\!\! && \!\!\!\langle\Phi_k(\eta)\Phi_{k'}(\eta')\rangle_s=
(2\pi)^2\delta(\vec
k+\vec k')
\nonumber\\
&& \times\int^\eta \!d\eta_1\int^{\eta'}\!d\eta_2 G_k(\eta,\eta_1)
G_{k'}(\eta',\eta_2)
\langle(\xi_{0i})_k(\eta_1)(\xi_{0i})_{k'}(\eta_2)\rangle_s .
\label{1.23}
\end{eqnarray}
Here the two-point correlation function for the stochastic source,
which is connected to the stress-energy tensor fluctuations through
the noise kernel, is given by,
\begin{eqnarray}
\langle (\xi_{0i})_k(\eta_1)(\xi_{0i})_{-k}(\eta_2)\rangle_s
\!\!\! &= &\!\!\!{1\over2}
\langle\{(\hat t_{0i})_k(\eta_1),(\hat
t_{0i})_{-k}(\eta_2)\}\rangle_{\phi\varphi}
\nonumber\\
\!\!\! &=&\!\!\! {1\over2}
k_ik_i\phi'(\eta_1)\phi'(\eta_2)G_k^{(1)}(\eta_1,\eta_2),
\label{1.24}
\end{eqnarray}
where $G_k^{(1)}(\eta_1,\eta_2)=\langle\{\hat\varphi_k(\eta_1),
\hat\varphi_{-k}(\eta_2)\}\rangle$ is the $k$th-mode Hadamard
function for a free minimally coupled scalar field in the appropriate
vacuum state on the Friedmann-Robertson-Walker background.

In practice, to make the explicit computation of the Hadamard
function we will assume that the field state is in the Euclidean
vacuum and the background spacetime is de Sitter. Furthermore we will
compute the Hadamard function for a massless field, and will make a
perturbative expansion in terms of the dimensionless parameter
$m/m_P$.
Thus we consider
\begin{displaymath}
  \bar G_k^{(1)}(\eta_1,\eta_2) =
  \langle 0|\{\hat y_k(\eta_1),\hat y_{-k}(\eta_2)\}|0\rangle =
  2{\cal R}\left(u_k(\eta_1)u_k^*(\eta_2)\right),
\end{displaymath}
with
\begin{displaymath}
  \hat y_k(\eta)= a(\eta)\hat\varphi_k(\eta) =
  \hat a_k u_k(\eta)+\hat a_{-k}^\dagger u_{-k}^*(\eta),
\end{displaymath}
and where
\begin{displaymath}
  u_k = (2k)^{-1/2} \, e^{ik\eta} \, (1-i/\eta)
\end{displaymath}
are the positive
frequency $k$-modes for a massless minimally coupled scalar field
on a de Sitter background, which define the
Euclidean vacuum state, $\hat a_k|0\rangle=0$~\cite{BirDav82}.

The assumption of a massless field for the computation of the
Hadamard function is made because massless modes in de Sitter are
much simpler to deal with than massive modes. We can see that this is
nonetheless a reasonable approximation as follows:  For a given mode
the $m=0$ approximation is reasonable when its wavelength $\lambda$
is shorter than the Compton wavelength, $\lambda_c=1/m$. In our case
we have a very small mass $m$ and the horizon size $H^{-1}$, where
$H$ is the Hubble constant $H=\dot a/a$ (here $a(t)$ with $t$ the
physical time $dt=ad\eta$), satisfies that $H^{-1}<\lambda_c$. Thus, for
modes inside the horizon $\lambda<\lambda_c$ and $m=0$ is a
good approximation. Outside the horizon massive modes decay in
amplitude as $\sim \exp (-m^2 t/3H)$
whereas massless modes remain
constant, thus when modes leave the horizon the approximation will
eventually break down. However, we only need to ensure that the
approximation is still valid after $60$ e-folds, {\it i.e.} $H\Delta t\sim
60$ ($\Delta t$ being the time between horizon exit and the end of inflation).
But this is the case provided $3H^2/m^2>60$,
since the decay factor $\sim \exp[-(m^2/3H^2)H\Delta t]$
will not be too different from unity for those modes that left the horizon
during the last sixty e-folds of inflation. This condition is indeed satisfied
given that $m\ll H$ in most slow-roll inflationary models
\cite{KolTur90,Pad93}, and in particular for the model considered here where $m\sim 10^{-6}m_P$.

We note that the background geometry is not exactly that of de Sitter
spacetime, for which $a(\eta)=-(H\eta)^{-1}$ with $-\infty <\eta< 0$.
One can expand in terms of the ``slow-roll" parameters and assume
that to first order $\dot\phi(t)\simeq m_P^2(m/m_P)$, where $t$ is
the physical time. The correlation function for the metric
perturbation (\ref{1.23}) can then be easily computed; see
Ref.~\cite{RouVer00,RouVer07} for details. The final result,
however, is very weakly dependent on the initial conditions as one
may understand from the fact that the accelerated expansion of the
quasi-de Sitter spacetime during inflation erases the information
about the initial conditions. Thus one may take the initial time to
be $\eta_0=-\infty$ and obtain to lowest order in $m/m_P$ the
expression
\begin{equation}
\langle\Phi_k(\eta)\Phi_{k'}(\eta')\rangle_s\simeq
8\pi^2\left( {m\over m_P}\right)^2 k^{-3}(2\pi)^3\delta(\vec k+\vec k')
\cos k(\eta-\eta').
\label{1.25}
\end{equation}

{}From this result two main conclusions can be derived. First, the
prediction of an almost scale-invariant Harrison-Zel'dovich spectrum
for large scales, i.e. small values of $k$. Second, since the
correlation function is of the order of $(m/m_P)^2$ a severe bound to
the mass $m$ is imposed by the gravitational fluctuations derived
from the small values of the Cosmic Microwave Background (CMB)
anisotropies detected by COBE. This bound is of the order of
$(m/m_P)\sim 10^{-6}$ \cite{Smo92,MukFelBra92}.

We now comment on some differences with
Refs.~\cite{CalHu95,Mat97a,Mat97b,CalGon97} which used a
self-interacting scalar field or a scalar field interacting
nonlinearly with other fields. In those works an important relaxation
of the ratio $m/m_P$ was found. The long wavelength modes of the
inflaton field  were regarded as an open system in an environment
made out of the shorter wavelength modes. Then, Langevin type
equations were used to compute the correlations of the long
wavelength modes driven by the fluctuations of the shorter wavelength
modes. In order to get a significant relaxation on the above ratio,
however, one had to assume that the correlations of the free long
wavelength modes, which correspond to the dispersion of the system's
initial state, had to be very small. Otherwise they dominate by
several orders of magnitude those fluctuations that come from the
noise of the environment.
This would require a great amount of
fine-tuning for the initial quantum state of each mode
\cite{RouVer07}.

We should remark that in the linear model discussed here
there is no environment for the inflaton fluctuations.
When one linearizes with
respect to both the scalar metric perturbations and the inflaton
perturbations, the system cannot be
regarded as a true open quantum system. The reason is that Fourier
modes decouple and the dynamical constraints due to diffeomorphism
invariance link the metric perturbations of scalar type with the
perturbations of the inflaton field so that only one true dynamical
degree of freedom is left for each Fourier mode.
Nevertheless, the inflaton
fluctuations are responsible for the noise that induces the
metric perturbations.


\subsection{Summary and outlook}


Stochastic gravity provides an alternative framework to study the
generation of primordial inhomogeneities in inflationary models.
Besides the interest of the problem in its own right, there are also
other reasons that make this problem worth discussing from the point
of view of stochastic gravity. The Einstein-Langevin equation is not
restricted by the use of linearized perturbations of the inflaton
field. In practice this may not be very important for inflationary models which
are driven by an inflaton field which takes a non-zero expectation
value, because the linear perturbations will give the leading
contribution. However, the importance of considering corrections due to one-loop
contributions from the inflaton field perturbations, beyond the tree level
of the linear cosmological perturbation theory, has recently been
emphasized by Weinberg \cite{Wei05,Wei06} as a means to understand the consequences of the theory even where in practice its predictions may not be observed. Note also
Refs. \cite{KahWoo07a,KahWoo07b} for the effect of
the quantum gravitational loop corrections on the dynamics of the inflaton field.

In the stochastic gravity approach some insights on the exact treatment of the inflaton
scalar field perturbations have been discussed in Refs. \cite{RouVer07,UraMae07,UraMae08}.
The main features that would characterize an exact treatment of the
inflaton perturbations are the following. First, the three types of metric
perturbations (scalar, vectorial and tensorial perturbations)
couple to the perturbations of the inflaton field.
Second, the corresponding Einstein-Langevin
equation for the linear metric perturbations will explicitly couple
the scalar and tensorial metric perturbations. Furthermore, although
the Fourier modes (with respect to the spatial coordinates) for the
metric perturbations will still decouple in the Einstein-Langevin
equation, any given mode of the noise and dissipation kernels will get
contributions from an infinite number of Fourier modes of the inflaton
field perturbations. This fact will imply, in addition, the need to
properly renormalize the ultraviolet divergences arising in the
dissipation kernel, which actually correspond to the divergences
associated with the expectation value of the stress tensor operator of
the quantum matter field evolving on the perturbed geometry.

We should remark that although the gravitational fluctuations are
here assumed to be classical, the correlation functions obtained
correspond to the expectation values of the symmetrized quantum
metric perturbations \cite{CalRouVer03,RouVer07}. This means that
even in the absence of decoherence, the fluctuations predicted by the
Einstein-Langevin equation still
give the correct symmetrized quantum two-point correlation functions.
In Ref.~\cite{CalRouVer03}
it was explained how a stochastic description based on a Langevin-type
equation could be introduced to gain information on fully quantum
properties of simple linear open systems. In a forthcoming paper
\cite{RouVer03b} it will be shown that, by carefully dealing with the
gauge freedom and the consequent dynamical constraints, this
result can be extended to the case of $N$ free quantum matter fields
interacting with the metric perturbations around a given
background. In particular, the correlation functions for the
metric perturbations obtained using the Einstein-Langevin equation are
equivalent to the correlation functions that would follow from a purely quantum
field theory calculation up to the leading order contribution in the large $N$ limit.
This will generalize the results already obtained on a Minkowski background \cite{HuRouVer04a,HuRouVer04b}.

These results
have important implications on the use of the Einstein-Langevin
equation to address situations in which the background configuration
for the scalar field vanishes. This includes not only the case
of a Minkowski background spacetime, but also the remarkably
interesting case of the trace anomaly induced inflation.
That is, inflationary models driven by the vacuum
polarization of a large number of conformal fields \cite{Sta80,Vil85,HawHerRea01}, where the
usual approaches based on the linearization of both the metric
perturbations and the scalar field perturbations and their subsequent
quantization can no longer be applied.
More specifically, the
semiclassical Einstein equations (\ref{2.5}) for massless quantum
fields conformally coupled to the gravitational field admit
an inflationary solution that begins in an almost de Sitter like regime and ends
up in a matter-dominated like regime \cite{Sta80,Vil85}. In these
models the standard approach based on the quantization of the
gravitational and the matter fields to linear order cannot be used
because the calculation of the metric perturbations
correspond to having only the last term in the noise kernel in
Eq.~(\ref{1.20}), since there is no homogeneous field $\phi(\eta)$ as
the expectation value $\langle\hat \phi\rangle=0$, and linearization
becomes trivial.

In the trace anomaly induced inflation model Hawking et al.
\cite{HawHerRea01} were able to compute the two-point quantum
correlation function for scalar and tensorial metric perturbations in
a spatially closed de Sitter universe, making use of the anti-de
Sitter / conformal field theory correspondence. They find that short
scale metric perturbations are strongly suppressed by the conformal
matter fields. This is similar to what we obtained in Sec.
\ref{sec:flucminspa} for the induced metric fluctuations in Minkowski
spacetime. In the stochastic gravity context, the noise kernel in a
spatially closed de Sitter background was derived in
Ref.~\cite{RouVer99}, and in a spatially-flat arbitrary
Friedmann-Robertson-Walker model the Einstein-Langevin equations
describing the metric perturbations were first obtained in
Ref.~\cite{CamVer96}. The computation of the corresponding two-point
correlation functions for the metric perturbations is now work in progress.


\section{Black Hole Backreaction and Fluctuations}
\label{sec:bhbkrn}


As another illustration of the application of stochastic gravity we
now consider the backreaction and fluctuations in black hole
spacetimes. Backreaction refers to the quantum effects of matter
fields such as vacuum polarization, quantum fluctuations and particle
creation on the spacetime structure and dynamics.  Studying the
dynamics of quantum fields in a fixed background spacetime, Hawking found
that black holes emit thermal radiation with a temperature inversely
proportional to their mass \cite{Haw75,Isr75,Par75,Wal75}. When the backreaction of
the quantum fields on the spacetime dynamics is included, one expects
that the mass of the black hole decreases as thermal radiation at
higher and higher temperatures is emitted. The reduction of the mass
of a black hole due to particle creation is often referred to as the
black hole `evaporation' process.  Backreaction of Hawking radiation
\cite{HajIsr80,Bar81,Yor83,Yor85,Yor86,HocKep93,HocKepYor93,AndEtal94}
could alter the evolution of the background spacetime and change the
nature of its end state, more drastically so for Planck size black
holes.

Backreaction is a technically challenging but conceptually rewarding
problem. Progress is slow in this long standing problem but it cannot
be ignored because existing results from test field approximations or
semiclassical analysis are not trustworthy when backreaction becomes
strong enough as to alter the structure and dynamics of the
background spacetime. At the least one needs to know how strong the
backreaction effects are under what circumstances the existing
predictions make sense. Without an exact quantum solution of the
black hole plus quantum field system or at least a full backreaction
consideration including the intrinsic and induced effects of metric
fluctuations, much of the long speculations on the end-state of black
hole collapse -- remnants,  naked singularity, baby universe
formation or complete evaporation (see e.g.,
\cite{Wit91,CalEtal92,RusSusTho92a,RusSusTho92b,HolWil92,Wil93,%
PolStr94,AngEtal95,Hu96,LowTho06,Gid06}.)
-- and the information loss issue \cite{Haw76a,Haw76b,Haw05,Pag80} (see,
e.g., \cite{Pre93,Pag94} for an overview, and recent results
from quantum information \cite{SmoOpp06,BraPat07}) will remain to be
speculations and puzzles. This issue also enters in the extension of
the well-known black hole thermodynamics
\cite{Bek73,Bek94,BekMuk95,Sor98,Jac99,Wal02,Wal01,%
SusUgl94,KabSheStr95,StrVaf96,Hor96,HorPol97,MalStrWit97,Mal98} to
nonequilibrium conditions \cite{EliJac06} and can lead to new
inferences on the microscopic structure of spacetime and the true
nature of Einstein's equations \cite{Jac95} along the viewpoint
regarding general relativity as geometro-hydrodynamics and gravity as
emergent phenomena. (See the nontraditional views of Volovik
\cite{Vol03,Vol07}, Hu \cite{grhydro,Hu07,meso} on spacetime
structure, Wen \cite{Wen04,Wen05a,Wen05,Wen06} on quantum
order, Seiberg \cite{Sei05}, Horowitz and Polchinsky
\cite{HorPol06} on emergent gravity, Herzog on the hydrodynamics
of M-theory \cite{Her02} and the seminal work of Unruh, Jacobson
\cite{Unr80,Jac93} leading to analog gravity
\cite{BarLibVis05,BalEtal05,Sch07}.)

\subsection{General Issues on Backreaction}

Backreaction studies of quantum field processes in cosmological
spacetimes have progressed further than the corresponding black hole
problems, partly because of relative technical simplicity associated
with the higher symmetry of relevant cosmological background
geometries. (For a summary of the cosmological backreaction problem
treated in the stochastic gravity theory, see \cite{HuVer03a}). The
way how the problem is set up and approached (e.g. via effective
action) in these previously studied models can be carried over to
black hole problems. In fact since the interior of a black hole can
be described by a cosmological model (e.g. the Kantowski-Sachs
universe for a spherically symmetric black hole), some aspects even
convey directly. The latest important work on this problem is that of
Hiscock, Larson and Anderson \cite{HisLarAnd97} on backreaction in
the interior of a black hole, where one can find a concise summary of
earlier work.

\subsubsection{Regularized Energy Momentum Tensor}

The first step in a backreaction problem is to find a regularized
energy momentum tensor of the quantum fields using reasonable
techniques since the expectation value of which serves as the source
in the semiclassical Einstein equation. For this,  much work started
in the 80's (and still ongoing sparingly) are concerned with finding
the right approximations for the regularized energy momentum tensor
\cite{JenMcLOtt95,ParPir94,Mas95,AndHisSam93,AndHisSam95,AndHisLor95,HisLarAnd97}.
Even in the simplest spherically symmetric spacetime, including the
important Schwarzschild metric, is technically quite involved. To
name a few of the important landmarks in this endeavor (this is
adopted from \cite{HisLarAnd97}), Howard and Candelas
\cite{HowCan84,How84} have computed the stress-energy of a
conformally invariant scalar field in the Schwarzschild geometry.
Jensen and Ottewill \cite{JenOtt89} have computed the vacuum
stress-energy of a massless vector field in Schwarzschild.
Approximation methods have been developed by Page, Brown, and
Ottewill \cite{Pag82,BroOtt85,BroOttPag86} for conformally invariant
fields in Schwarzschild spacetime, Frolov and Zel'nikov
\cite{FroZel87} for conformally invariant fields in a general static
spacetime, Anderson, Hiscock and Samuel
\cite{AndHisSam93,AndHisSam95} for massless arbitrarily coupled
scalar fields in a general static spherically symmetric spacetime.
Furthermore the DeWitt-Schwinger approximation has been derived by
Frolov and Zel'nikov \cite{FroZel82,FroZel84} for massive fields in
Kerr spacetime, Anderson Hiscock and Samuel
\cite{AndHisSam93,AndHisSam95} for a general (arbitrary curvature
coupling and mass) scalar field in a general static spherically
symmetric spacetime and have applied their method to the
Reissner-Nordstr\"{o}m geometry \cite{AndHisLor95}. Though arduous
and demanding, the effort continues on because of its importance in
finding the backreaction effects of Hawking radiation on the
evolution of black holes and the quantum structure of spacetime.

Here we wish to address the black hole backreaction problem with new
insights and methods provided by stochastic  gravity. (For the latest
developments see reviews, e.g.,
\cite{Banff,stogra,HVErice,HuVer03a}). It is not our intention to
seek better approximations for the regularized energy momentum
tensor, but to point out new ingredients lacking in the existing
semiclassical gravity framework. In particular one needs to consider
both the dissipation and the fluctuations aspects in the back
reaction of particle creation or vacuum polarization.

In a short note  Hu, Raval and Sinha \cite{Vishu} first used the
stochastic gravity formalism to address the backreaction of
evaporating black holes. A more detailed analysis is given by the
recent work of Hu and Roura \cite{HuRou06a,HuRou07}  For the class of
quasi-static black holes, the formulation of the problem in this new
light was sketched out by Sinha Raval and Hu \cite{SinRavHu03}. We
follow these two latter works in the stochastic gravity theory
approach to the black hole fluctuations and backreaction problems.

\subsubsection{Backreaction and Fluctuation-Dissipation relation}

{}From the statistical field theory perspective provided by
stochastic gravity one can understand that backreaction effect is the
manifestation of a fluctuation-dissipation relation
\cite{Ein05,Ein06,Nyq28,CalWel51,CalGre52,Web56}. This was first
conjectured by Candelas and Sciama \cite{CanSci77,Sci79,SciCanDeu81}
for a dynamic Kerr black hole emitting Hawking radiation, and Mottola
\cite{Mottola} for a static black hole (in a box) in
quasi-equilibrium with its radiation via linear response theory
\cite{Kub57,BerCal59,Kub66,LanLifPit80,KubTodHas85}. This postulate
was shown to hold for fully dynamical spacetimes. From cosmological
backreaction problem  Hu and Sinha \cite{HuSin95} derived a
generalized fluctuation-dissipation relation relating dissipation (of
anisotropy in Bianchi Type I universes) and fluctuations (measured by
particle numbers created in neighboring histories).

While the fluctuation-dissipation relation in a linear response
theory captures the response of the system (e.g., dissipation of the
black hole) to the environment (in these cases the quantum matter
field) linear response theory (in the way it is commonly presented in
statistical thermodynamics)  cannot provide a full description of
self-consistent backreaction on at least two counts:

First, because it is usually based on the assumption of a specified
background spacetime (static in this case) and state (thermal) of the
matter field(s) (e.g., \cite{Mottola}). The spacetime and the state
of matter should be determined in a self-consistent manner by their
dynamics and mutual influence. Second, the fluctuation part
represented by the noise kernel is amiss (e.g.,
\cite{AndMolMot02,AndMolMot03}) This is also a problem in the
fluctuation-dissipation relation proposed by Candelas and Sciama
\cite{CanSci77,Sci79,SciCanDeu81} (see below). As demonstrated by
many authors \cite{HuSin95,CamVer96} backreaction is intrinsically a
dynamic process. The Einstein-Langevin equation in stochastic gravity
overcomes both of these deficiencies.

For Candelas and Sciama \cite{CanSci77,Sci79,SciCanDeu81}, the
classical formula they showed relating the dissipation in area
linearly to the squared absolute value of the shear amplitude is
suggestive of  a fluctuation-dissipation relation. When the
gravitational perturbations are quantized (they choose the quantum
state to be the Unruh vacuum) they argue that it approximates a flux
of radiation from the hole at large radii. Thus the dissipation in
area due to the Hawking flux of gravitational radiation is allegedly
related to the quantum fluctuations of gravitons.
The criticism in Ref. \cite{Vishu} is that their's is not a
fluctuation-dissipation relation in the truly statistical mechanical
sense because it does not relate dissipation of a certain quantity
(in this case, horizon area) to the fluctuations of {\it the same
quantity}. To do so would require one to compute the two point
function of the area, which, being a four-point function of the
graviton field, is related to a two-point function of the stress
tensor. The stress tensor is the true ``generalized force'' acting on
the spacetime via the equations of motion, and the dissipation in the
metric must eventually be related to the fluctuations of this
generalized force for the relation to qualify as a
fluctuation-dissipation relation.

\subsubsection{Noise and Fluctuations -- the missing ingredient in older treatment}

{}From this reasoning, we see that the vacuum expectation value of
the stress energy bi-tensor,  known as the noise kernel, is the
necessary new ingredient in addition to the dissipation kernel, and
stochastic gravity as an extension of semiclasical gravity is the
appropriate framework for backreaction considerations. The noise
kernel for quantum fields in Minkowski and de Sitter spacetime has
been carried out by Martin, Roura and Verdaguer
\cite{MarVer99a,MarVer00,RouVer07}, for thermal fields in black hole
spacetime and scalar fields in general spacetimes by Campos, Hu and
Phillips \cite{CamHu98,CamHu99,PhiHu01,PhiHu03}.

\subsection{Backreaction on black holes under quasi-static conditions}

As an illustration of the application of stochastic gravity theory we
outline the essential steps in a black hole backreaction calculation,
focusing on a more manageable quasi-static class. We adopt the
Hartle-Hawking picture \cite{HarHaw76} where the black hole is bathed
eternally -- actually in quasi-thermal equilibrium -- in the Hawking
radiance it emits. It is described here by a massless scalar quantum
field at the Hawking temperature. As is well-known, this
quasi-equilibrium condition is possible only if the black hole is
enclosed in a box of size suitably larger than
the event horizon. 
We can divide our consideration into the far field case and the near
horizon case. Campos and Hu \cite{CamHu98,CamHu99} have treated a
relativistic thermal plasma in a weak gravitational field.  Since the
far field limit of a Schwarzschild metric is just the perturbed
Minkowski spacetime, one can perform a perturbation expansion off hot
flat space using the thermal Green functions \cite{GibPer78}.
Strictly speaking the location of the box holding the black hole in
equilibrium with its thermal radiation is as far as one can go, thus
the metric may not reach the perturbed Minkowski form. But one can
also put the black hole and its radiation in an anti-de Sitter space
\cite{HawPag83}, which contains such a region. Hot flat space has
been studied before for various purposes. See e.g.,
\cite{GroPerYaf82,Reb91,Reb92,AlmBraFre94,BraFre98}. Campos and Hu
derived a stochastic CTP effective action and from it an equation of
motion, the Einstein Langevin equation, for the dynamical effect of a
scalar quantum field on a background spacetime. To perform
calculations leading to the Einstein-Langevin equation one needs to
begin with a self-consistent solution of the semiclassical Einstein
equation for the thermal field and the perturbed background
spacetime. For a black hole background, a semiclassical gravity
solution is provided by York \cite{Yor83,Yor85,Yor86}. For a
Robertson-Walker background with thermal fields it is given by Hu
\cite{Hu81}.

We follow the strategy outlined by Sinha, Raval and Hu
\cite{SinRavHu03} for treating the near horizon case, following the
same scheme of Campos and Hu. In both cases two new terms appear
which are absent in semiclassical gravity considerations: a nonlocal
dissipation and a (generally colored) noise kernel. When one takes
the noise average one recovers York's \cite{Yor83,Yor85,Yor86}
semiclassical equations for radially perturbed quasi-static black
holes. For the near horizon case one cannot obtain the full details
yet, because the Green function for a scalar field in the
Schwarzschild metric comes only in an approximate form (e.g. Page
approximation \cite{Pag82}), which, though reasonably accurate for
the stress tensor, fails poorly for the noise kernel \cite{PhiHu03}.
In addition a formula is derived in \cite{SinRavHu03} expressing the
CTP effective action in terms of the Bogoliubov coefficients. Since
it measures not only the number of particles created, but also the
difference of particle creation in alternative histories, this
provides a useful avenue to explore the wider set of issues in black
hole physics related to noise and fluctuations.

Since backreaction calculations in semiclassical gravity has been
under study for a much longer time than in stochastic gravity we will
concentrate on explaining how the new stochastic features arise from
the framework of semiclassical gravity, i.e., noise and fluctuations
and their consequences. Technically the goal is to obtain an
influence action for this model of a black hole coupled to a scalar
field and to derive an Einstein-Langevin equation from it. As a
by-product, from the fluctuation-dissipation relation, one can derive
the vacuum susceptibility function and the isothermal compressibility
function for black holes, two quantities of fundamental interest in
characterizing the nonequilibrium thermodynamic properties of black
holes.

\subsubsection{The model}

In this model the black hole spacetime is described by a
spherically symmetric static metric with line element of the
following general form written in advanced time
Eddington-Finkelstein coordinates \be ds^2 =
g_{\mu\nu}dx^{\mu}dx^{\nu} = -e^{2\psi}\left(1 - {2m\over
r}\right)dv^2 + 2 e^{2\psi}dvdr + r^2~d{\Omega}^2,
\label{ssmetric} \te where $\psi = \psi(r)$ and $m = m(r)$ , $ v =
t + r + 2Mln\left({r\over 2M} -1 \right)$ and $d{\Omega}^2$ is the
line element on the two sphere. Hawking radiation is described by
a massless, conformally coupled quantum scalar field $\phi$ with
the classical action \be S_m[\phi, g_{\mu\nu}] = -{1\over 2}\int
d^n x \sqrt{-g}[g^{\mu\nu}\partial_{\mu}\phi
\partial_{\nu}\phi + \xi(n) R{\phi}^2], \label{phiact} \te where
$\xi(n) = {(n-2)\over 4(n-1)}$ ($n$ is the dimension of
spacetime) and $R$ is the curvature scalar of the spacetime it
lives in.

Let us consider linear perturbations $h_{\mu\nu}$ off a background
Schwarzschild metric $g^{(0)}_{\mu\nu}$ \be g_{\mu\nu} =
g^{(0)}_{\mu\nu} + h_{\mu\nu}, \label{linearize} \te with standard
line element \be (ds^2)^0 = \left( 1 - {2M\over r}\right)dv^2 +
2dvdr + r^2d{\Omega}^2. \label{schwarz} \te We look for this class
of perturbed metrics in the form given by (\ref{ssmetric}), (thus
restricting our consideration only to spherically symmetric
perturbations): \be e^\psi \simeq  1+ \epsilon \rho(r)
,\label{rho} \te and \be m \simeq M[ 1 + \epsilon \mu (r)],
\label{mu} \te where ${\epsilon\over \lambda M^2} = {1\over 3}a
T_H^4 ;$ $ a ={{\pi}^2\over 30} ; \lambda = 90(8^4)\pi^2$. $T_H$
is the Hawking temperature. This particular parametrization of the
perturbation is chosen following York's \cite{Yor83,Yor85,Yor86} notation. Thus
the only non-zero components of $h_{\mu\nu}$ are \be h_{vv} =
-\left((1 - {2M\over r})2\epsilon \rho(r) + {2M\epsilon \mu
(r)\over r}\right), \label{hvv} \te and \be h_{vr} = \epsilon\rho
(r) \label{hvr}. \te So this represents a metric with small static
and radial perturbations about a Schwarzschild black hole. The
initial quantum state of the scalar field is taken to be the
Hartle Hawking vacuum, which is essentially a thermal state at the
Hawking temperature and it represents a black hole in (unstable)
thermal equilibrium with its own Hawking radiation. In the far
field limit, the gravitational field is described by a linear
perturbation from Minkowski spacetime. In equilibrium  the thermal
bath can be characterized by a relativistic fluid with a
four-velocity (time-like normalized vector field) $u^\mu$, and
temperature in its own rest frame $\beta^{-1}$.

To facilitate later comparisons with our program we briefly recall
York's work \cite{Yor83,Yor85,Yor86}. See also work by Hochberg
and Kephart \cite{HocKep93} for a massless vector field, Hochberg,
Kephart and York \cite{HocKepYor93} for a massless spinor field,
and Anderson, Hiscock, Whitesell, and York \cite{AndEtal94} for a
quantized massless scalar field with arbitrary coupling to
spacetime curvature. York considered the semiclassical Einstein
equation \be G_{\m\n} (g_{\alpha \beta}) = \k \langle
T_{\m\n}\rangle ,\te with $G_{\mu\nu} \simeq G^{(0)}_{\mu\nu} +
\delta G_{\mu\nu}$ where $G^{(0)}_{\mu\nu}$ is the Einstein tensor
for the background spacetime. The zeroth order solution gives a
background metric in empty space, i.e, the Schwarzschild metric.
$\delta G_{\mu\nu}$ is the linear correction to the Einstein
tensor in the perturbed metric. The semiclassical Einstein
equation  in this approximation therefore reduces to \be \delta
G_{\mu\nu}(g^{(0)}, h) = \kappa \langle T_{\mu\nu} \rangle
.\label{pertscee} \te York solved this equation to first order by
using the expectation value of the energy momentum tensor for a
conformally coupled scalar field in the Hartle-Hawking vacuum in
the unperturbed (Schwarzschild) spacetime on the right hand side
and using (\ref{hvv}) and (\ref{hvr}) to calculate $\delta
G_{\mu\nu}$ on the left hand side. Unfortunately, no exact
analytical expression is available for the $\langle
T_{\mu\nu}\rangle$ in a Schwarzschild metric with the quantum
field in the Hartle-Hawking vacuum that goes on the right hand
side. York therefore uses the approximate expression given by Page
\cite{Pag82} which is known to give excellent agreement with
numerical results. Page's approximate expression for $\langle
T_{\mu\nu}\rangle$ was constructed using a thermal Feynman Green's
function obtained  by a conformal transformation of a WKB
approximated Green's function for an optical Schwarzschild metric.
York then solves the semiclassical Einstein equation
(\ref{pertscee}) self consistently to obtain the corrections to
the background metric induced by the backreaction encoded in the
functions $\mu(r)$ and $\rho(r)$. There was no mention of
fluctuations or its effects. As we shall see, in the language of
the previous section, the semiclassical gravity procedure which
York followed working at the equation of motion level is
equivalent to looking at the noise-averaged backreaction effects.

\subsubsection{CTP Effective Action for the Black Hole}

We first derive the CTP effective action for the model described
in the previous section. Using the  metric (\ref{schwarz}) (and
neglecting the surface terms that appear in an integration by
parts) we have  the action for the scalar field  written
perturbatively as
\begin{equation}
   S_m[\phi,h_{\mu\nu}]
        \ = \  {1\over 2}\int d^nx{\sqrt{-g^{(0)}}}\ \phi
               \left[ \Box^{(0)} + V^{(1)} + V^{(2)} + \cdots
              \right] \phi,
\label{phipert}
\end{equation}
where the first and second order perturbative operators $V^{(1)}$
and $V^{(2)}$ are given by
\begin{eqnarray}
V^{(1)}  \!\!\!\!\!\! & \equiv  &\!\!\!\!\!\! - {1\over
\sqrt{\!-g^{(0)}}} \left\{
\!\partial_\mu\left(\!\sqrt{\!-g^{(0)}}\bar h^{\mu\nu}\right)
                                \!\partial_\nu
                              \!+\!\bar h^{\mu\nu}\partial_\mu
                              \partial_\nu
                            \!+\!\xi(n) R^{(1)}
                     \right\}\!,
               \nonumber \\
V^{(2)}
  \!\!\!\!\!\!  &  \equiv & \!\!\!\!\!\!- \!{1\over \sqrt{\!-g^{(0)}}}
\left\{ \!\partial_\mu \!\left(\!\sqrt{\!-g^{(0)}} \hat
h^{\mu\nu}\right)
                              \!\partial_\nu
                            +\hat h^{\mu\nu}\partial_\mu
                            \partial_\nu
                          \!-\!\xi(n) ( R^{(2)}
                               \!+\!{1\over 2}hR^{(1)})\!\right\}\!\!.
\end{eqnarray}
In the above expressions, $R^{(k)}$ is the $k$-order term in the
perturbation $h_{\mu\nu}(x)$ of the scalar curvature $R$ and $\bar
h_{\mu\nu}$ and $\hat h_{\mu\nu}$ denote a linear and a quadratic
combination of the perturbation, respectively,
\begin{eqnarray}
   \bar h_{\mu\nu}
        &  \equiv  & h_{\mu\nu} - {1\over 2} h g^{(0)}_{\mu\nu},
                     \nonumber \\
   \hat h_{\mu\nu}
        &  \equiv  & h^{\,\, \alpha}_\mu h_{\alpha\nu}
                      -{1\over 2} h h_{\mu\nu}
                      +{1\over 8} h^2 g^{(0)}_{\mu\nu}
                      -{1\over 4} h_{\alpha\beta}h^{\alpha\beta} g^{(0)}_{\mu\nu}.
   \label{eq:def bar h}
\end{eqnarray}
{}From quantum field theory in curved spacetime considerations
discussed above we take the following action for the gravitational
field:
\begin{eqnarray}
  \!\!\!\! S_g[g_{\mu\nu}]
       \!\!\!\!\! &=& \!\!\!\!\!{1\over {(16 \pi G)^{\frac{n-2}{2}}}}\int d^nx\ \sqrt{-g(x)}R(x)
                +{\alpha\bar\mu^{n-4}\over4(n-4)}
                   \int d^nx\ \sqrt{-g(x)} \nonumber \\
      \!\!\!\!\!  &&\!\!\!\!\!
                   \times\left\{ 3R_{\mu\nu\alpha\beta}(x)
                           R^{\mu\nu\alpha\beta}(x)
                         \!-\!\left[ 1\!-\!360 \left(\xi(n)\! - \!{1\over6}\right)^2
                          \right]R^2(x)
                   \right\}\!.
\end{eqnarray}
The first term is the classical Einstein-Hilbert action and the
second term is the counterterm in four dimensions used  to
renormalize the divergent effective action. In this action
$\ell^2_P = 16\pi G_N$, $\alpha = (2880\pi^2)^{-1}$ and $\bar\mu$
is an arbitrary mass scale.

We are interested in computing the CTP effective action
(\ref{phipert}) for the matter action and when the field $\phi$ is
initially in the Hartle-Hawking vacuum. This is equivalent to
saying that the initial state of the field is described by a
thermal density matrix at a finite temperature $T = T_H$. The CTP
effective action at finite temperature $T \equiv 1/\beta$ for
this model is given by (for details see \cite{CamHu98,CamHu99})
\begin{equation}
   S_{rm eff}^\beta [h^\pm_{\mu\nu}]
        \ = \ S_g[h^+_{\mu\nu}]
             -S_g[h^-_{\mu\nu}]
             -{i\over2}Tr\{ \ln\bar G^\beta_{ab}[h^\pm_{\mu\nu}]\},
   \label{eq:eff act two fields}
\end{equation}
where $\pm$ denote the forward and backward time path of the CTP
formalism and $\bar G^\beta_{ab}[h^\pm_{\mu\nu}]$ is the complete
$2\times 2$ matrix propagator ($a$ and $b$ take $\pm$ values:
$G_{++},G_{+-}$ and $G_{--}$ correspond to the Feynman, Wightman
and Schwinger Greens functions respectively) with thermal boundary
conditions for the differential operator $\sqrt{-g^{(0)}}(\Box +
V^{(1)} + V^{(2)} + \cdots)$. 
The actual form of $\bar G^\beta_{ab}$ cannot be explicitly
given. However, it is easy to obtain a perturbative expansion in
terms of $V^{(k)}_{ab}$, the $k$-order matrix version of the
complete differential operator defined by $V^{(k)}_{\pm\pm}
\equiv \pm V^{(k)}_{\pm}$ and $V^{(k)}_{\pm\mp} \equiv 0$, and
$G^\beta_{ab}$, the thermal matrix propagator for a massless
scalar field in Schwarzschild spacetime . To second order $\bar
G^\beta_{ab}$ reads,
\begin{eqnarray}
   \bar G^\beta_{ab}
        \ = \  G^\beta_{ab}
              -G^\beta_{ac}V^{(1)}_{cd}G^\beta_{db}
              -G^\beta_{ac}V^{(2)}_{cd}G^\beta_{db}
              +G^\beta_{ac}V^{(1)}_{cd}G^\beta_{de}
               V^{(1)}_{ef}G^\beta_{fb}
              +\cdots
\end{eqnarray}
Expanding the logarithm and dropping one term independent of the
perturbation $h^\pm_{\mu\nu}(x)$, the CTP effective action may be
perturbatively written as,
\begin{eqnarray}
  \!\!\!\! S_{\rm eff}^\beta [h^\pm_{\mu\nu}]
      \!\! \!\!\!\!\! & = & \!\!\!\!\!\!\! S_g[h^+_{\mu\nu}] \!- \!S_g[h^-_{\mu\nu}]
                \nonumber \\
       \!\! \!\!\!\!\!& &\!\!\!\!\!\!\! +\!{i\over2}Tr[ V^{(1)}_{+}G^\beta_{++}
                               \!-\!V^{(1)}_{-}G^\beta_{--}
                               \!+\!V^{(2)}_{+}G^\beta_{++}
                               \!-\!V^{(2)}_{-}G^\beta_{--}
                              ]
                \nonumber \\
       \!\! \!\!\!\!\!& & \!\!\!\!\!\!\!-{i\over4}Tr[  V^{(1)}_{+}G^\beta_{++}
                                 V^{(1)}_{+}G^\beta_{++}
                               \!+\! V^{(1)}_{-}G^\beta_{--}
                                 V^{(1)}_{-}G^\beta_{--}
                                 \nonumber\\
      \!\!\!\!\!\!\!&&\ \ \ \ \ \ \ \
                               \!-\!2V^{(1)}_{+}G^\beta_{+-}
                                 V^{(1)}_{-}G^\beta_{-+}
                              ]\!.
   \label{eq:effective action}
\end{eqnarray}
In computing the traces, some terms containing divergences are
canceled using counterterms introduced in the classical
gravitational action after dimensional regularization.

\subsubsection{Near Flat Case}

At this point we divide our considerations into two cases. In the
far field limit  $h_{\mu\nu}$ represent perturbations about flat
space, i.e., $g^{(0)}_{\mu\nu}= \eta_{\m\n}$. The exact
``unperturbed" thermal propagators for scalar fields are known,
i.e., the Euclidean propagator with periodicity $\beta$. Using
the Fourier transformed  form (those quantities are denoted with a
tilde) of the thermal propagators $\tilde G^\beta_{ab}(k)$ , the
trace terms of the form
$Tr[V^{(1)}_{a}G^\beta_{mn}V^{(1)}_{b}G^\beta_{rs}]$ can be
written as \cite{CamHu98,CamHu99},
\begin{eqnarray}
   \!\!\!\!\!\!&&\!\!\!\!\!Tr[V^{(1)}_{a}G^\beta_{mn}V^{(1)}_{b}G^\beta_{rs}]
       = \int d^nxd^nx'\
               h^a_{\mu\nu}(x)h^b_{\alpha\beta}(x')
               \nonumber\\
               &&\ \ \ \ \
               \times\int {d^nk\over(2\pi)^n}{d^nq\over(2\pi)^n}
               e^{ik\cdot (x-x')}
               \tilde G^\beta_{mn}(k+q)\tilde G^\beta_{rs}(q)
               \kl{T}{}{q,k},
   \label{eq:trace}
\end{eqnarray}
where the tensor $\kl{T}{}{q,k}$ is defined in
\cite{CamHu98,CamHu99} after an  expansion in terms of a basis of
14 tensors \cite{Reb91,Reb92}. In particular, the last trace of
(\ref{eq:effective action}) may be split in two different kernels
$\kl{N}{}{x-x'}$ and $\kl{D}{}{x-x'}$,
\begin{eqnarray}
   \!\!\!\!\!&&\!\!\!\!\!{i\over2}
   Tr[V^{(1)}_{+}G^\beta_{+-}V^{(1)}_{-}G^\beta_{-+}]=
   \nonumber\\
   &&\ \ \
   -\int d^4xd^4x'\
               h^+_{\mu\nu}(x)h^-_{\alpha\beta}(x')
               [   \kl{D}{}{x-x'}
                +i \kl{N}{}{x-x'}
               ].
\end{eqnarray}
One can express the Fourier transforms of these kernels,
respectively, as
\begin{eqnarray}
   \kl{\tilde N}{}{k}
       \!\!\!\!\! & = &\!\!\!\!\! \pi^2\int {d^4q\over(2\pi)^4}\
                  \left\{ \theta(k^o+q^o)\theta(-q^o)
                         +\theta(-k^o-q^o)\theta(q^o)
                         \right.
                         \nonumber\\
                         &&\!\!\hspace{2mm}
                         \left.
                         +n_\beta(|q^o|)+n_\beta(|k^o+q^o|)
                  \right.
                \nonumber \\
        & & \!\!\hspace{2mm}
                  \left. +2n_\beta(|q^o|)n_\beta(|k^o+q^o|)
                  \right\}\delta(q^2)\delta[(k+q)^2]\kl{T}{}{q,k},
   \label{eq:N}
\end{eqnarray}
\begin{eqnarray}
   \kl{\tilde D}{}{k}
     \!\!\!\!\!\!   & = &\!\!\!\!\!\! -i\pi^2\int {d^4q\over(2\pi)^4}\
                  \left\{ \theta(k^o+q^o)\theta(-q^o)
                         -\theta(-k^o-q^o)\theta(q^o)
                         \right.
                         \nonumber\\
                         &&\!\!\hspace{2mm}
                         \left.
                         +sg(k^o+q^o) n_\beta(|q^o|)
                  \right.
                \nonumber \\
        &  & \!\!\hspace{2mm}
                  \left. -sg(q^o)n_\beta(|k^o+q^o|)
                  \right\}\delta(q^2)\delta[(k+q)^2]\kl{T}{}{q,k}.
   \label{eq:D}
\end{eqnarray}
Using the property $\kl{T}{}{q,k} = \kl{T}{}{-q,-k}$, it is easy
to see that the kernel $\kl{N}{}{x-x'}$ is symmetric
and $\kl{D}{}{x-x'}$
antisymmetric in their arguments; that is, $\kl{N}{}{x} =
\kl{N}{}{-x}$ and $\kl{D}{}{x} = -\kl{D}{}{-x}$.

The physical meanings of these kernels can be extracted if we
write the renormalized CTP effective action at finite temperature
(\ref{eq:effective action}) in an influence functional form
\cite{CalLeg83,GraSchIng88,HuPazZha92,HuPazZha93}. N, the
imaginary part of the CTP effective action can be identified with
the noise kernel and D, the antisymmetric piece of the real part,
with the dissipation kernel. Campos and Hu \cite{CamHu98,CamHu99}
have shown that these kernels identified as such indeed satisfy a
thermal fluctuation-dissipation relation.

If we denote the difference and the sum of the perturbations
$h^\pm_{\mu\nu}$ defined along each branch $C_\pm$ of the complex
time path of integration $C$ by $[h_{\mu\nu}] \equiv h^+_{\mu\nu}
- h^-_{\mu\nu}$ and $\{h_{\mu\nu}\} \equiv h^+_{\mu\nu} +
h^-_{\mu\nu}$, respectively, the influence functional form of the
thermal CTP effective action may be written to second order in
$h_{\mu\nu}$ as,
\begin{eqnarray}
   S_{\rm eff}^\beta [h^\pm_{\mu\nu}]
        & \ \simeq \ & {1\over 2(16 \pi G_N)} \int d^4x\ d^4x'\
                       [h_{\mu\nu}](x)\kl{L}{(o)}{x-x'}
                       \{h_{\alpha\beta}\}(x')
                     \nonumber \\
        &            &+{1\over2}\int d^4x\
                       [h_{\mu\nu}](x)T^{\mu\nu}_{(\beta)}
                     \nonumber \\
        &            &+{1\over2}\int d^4x\ d^4x'\
                       [h_{\mu\nu}](x)\kl{H}{}{x-x'}
                       \{h_{\alpha\beta}\}(x')
                     \nonumber \\
        &            &-{1\over2}\int d^4x\ d^4x'\
                       [h_{\mu\nu}](x)\kl{D}{}{x-x'}
                       \{h_{\alpha\beta}\}(x')
                     \nonumber \\
        &            &+{i\over2}\int d^4x\ d^4x'\
                       [h_{\mu\nu}](x)\kl{N}{}{x-x'}
                       [h_{\alpha\beta}](x').
\label{CTPbh}
\end{eqnarray}
The first line is the Einstein-Hilbert action to second order in
the perturbation $h^\pm_{\mu\nu}(x)$. $\kl{L}{(o)}{x}$ is a
symmetric kernel ({\sl i.e.} $\kl{L}{(o)}{x}$ =
$\kl{L}{(o)}{-x}$). In the near flat case its Fourier transform
is given by
\begin{eqnarray}
   \kl{\tilde L}{(o)}{k}
       &= &{1\over4}\left[ - k^2 \kl{T}{1}{q,k}
                              +2k^2 \kl{T}{4}{q,k}
                              \right.
                              \nonumber\\
                              && \ \ \ \
                              \left.
                              + \kl{T}{8}{q,k}
                              -2\kl{T}{13}{q,k}
                       \right].
\end{eqnarray}
The fourteen elements of the tensor basis $\kl{T}{i}{q,k}$
($i=1,\cdots,14$) are defined in \cite{Reb91,Reb92}. The second is
a local term linear in $h^\pm_{\mu\nu}(x)$.  Only when far away
from the hole that it takes the form of the stress tensor of
massless scalar particles at temperature $\beta^{-1}$, which has
the form of a perfect fluid stress-energy tensor
\begin{equation}
   T^{\mu\nu}_{(\beta)}
        \ = \ {\pi^2\over30\beta^4}
              \left[ u^\mu u^\nu + {1\over3}(\eta^{\mu\nu}+u^\mu u^\nu)
              \right],
\end{equation}
where $u^\mu$ is the four-velocity of the plasma
and the factor ${\pi^2\over30\beta^4}$ is the
familiar thermal energy density for massless scalar particles at
temperature $\beta^{-1}$. In the
far field limit, taking into account the four-velocity $u^\mu$ of
the fluid, a manifestly Lorentz-covariant approach to thermal
field theory may be used \cite{Wel82}. However, in order to
simplify the involved tensorial structure we work in the
co-moving coordinate system of the fluid where $u^\mu =
(1,0,0,0)$. In the third line, the Fourier
transform of the symmetric kernel $\kl{H}{}{x}$ can be expressed
as
\begin{eqnarray}
   \kl{\tilde H}{}{k}
     \!\!\!\!   & =  & \!\!\!\!\! -{\alpha k^4\over4}
                   \left\{ {1\over2}\ln {|k^2|\over\mu^2}\kl{Q}{}{k}
                          +{1\over3}\kl{\bar Q}{}{k}
                   \right\}
                \nonumber \\
        &       & \!\!\!\! +{\pi^2\over180\beta^4}
                   \left\{ - \kl{T}{1}{u,k}
                           -2\kl{T}{2}{u,k}
                           \right.
                           \nonumber\\
                           &&\ \ \ \ \ \ \ \ \ \
                           \left.
                           + \kl{T}{4}{u,k}
                           +2\kl{T}{5}{u,k}
                   \right\}
                \nonumber \\
        &       & \!\!\!\! +{\xi\over96\beta^2}
                   \left\{    k^2 \kl{T}{1}{u,k}
                           -2 k^2 \kl{T}{4}{u,k}
                           \right.
                           \nonumber\\
                           &&\ \ \ \ \ \ \ \ \ \
                           \left.
                           -      \kl{T}{8}{u,k}
                           +2     \kl{T}{13}{u,k}
                   \right\}
                \nonumber \\
        &       & \!\!\!\! +\pi\int {d^4q\over(2\pi)^4}\
                   \left\{ \delta(q^2)n_\beta(|q^o|)
                           {\cal P}\left[ {1\over(k+q)^2}
                                   \right]
                                   \right.
                           \nonumber\\
                           &&\ \
                           \left.
                          +\delta[(k\!+\!q)^2]n_\beta(|k^o\!+\!q^o|)
                           {\cal P}\!\left[ {1\over q^2}
                                   \right]\!\right\}
        \!\times\!\kl{T}{}{q,k}\!,
   \label{eq:grav pol tensor}
\end{eqnarray}
where $\mu$ is a simple redefinition of the renormalization
parameter $\bar\mu$ given by $\mu \equiv \bar\mu \exp
({23\over15} + {1\over2}\ln 4\pi - {1\over2}\gamma)$, and the
tensors $\kl{Q}{}{k}$ and $\kl{\bar Q}{}{k}$ are defined,
respectively, by
\begin{eqnarray}
   \kl{Q}{}{k}
       \!\!\!\!\! &=& \!\!\!\!\!{3\over2} \left\{               \kl{T}{1}{q,k}
                                    -{1\over k^2} \kl{T}{8}{q,k}
                                    +{2\over k^4} \kl{T}{12}{q,k}
                            \right\}
                \nonumber \\
       & &\hspace{-20mm}-[1\!-\!360(\xi\!-\!{1\over6})^2]\!
                  \left\{ \!              \kl{T}{4}{q,k}
                          \!+\!{1\over k^4} \kl{T}{12}{q,k}
                          \!-\!{1\over k^2} \kl{T}{13}{q,k}
                  \!\right\}\!,
   \label{eq:Q tensor}
\end{eqnarray}
\begin{eqnarray}
   \kl{\bar Q}{}{k}
    \!\!\!\!\!\!    &=&\!\!\!\!\!  [1+576(\xi-{1\over6})^2-60(\xi-{1\over6})(1-36\xi')]
                  \left\{               \kl{T}{4}{q,k}
                  \right.
                  \nonumber\\
                  &&\ \ \ \ \ \ \ \ \ \ \
                  \left.
                          +{1\over k^4} \kl{T}{12}{q,k}
                          -{1\over k^2} \kl{T}{13}{q,k}
                  \right\}.
\end{eqnarray}
In the above and subsequent equations, we denote the coupling
parameter in four dimensions $\xi(4)$ by $\xi$ and consequently
$\xi'$ means $d\xi(n)/dn$ evaluated at $n=4$. $\kl{\tilde H}{}{k}$
is the complete contribution of a free massless quantum scalar
field to the thermal graviton polarization
tensor\cite{Reb91,Reb92,AlmBraFre94,BraFre98} and it is
responsible for the instabilities found in flat spacetime at
finite temperature
\cite{GroPerYaf82,Reb91,Reb92,AlmBraFre94,BraFre98}. Note that the
addition of the contribution of other kinds of matter fields to
the effective action, even graviton contributions, does not change
the tensor structure of these kernels and only the overall factors
are different to leading order \cite{Reb91,Reb92}.
Eq.~(\ref{eq:grav pol tensor}) reflects the fact that the kernel
$\kl{\tilde H}{}{k}$ has thermal as well as non-thermal
contributions. Note that it reduces to the first term in the zero
temperature limit ($\beta\rightarrow\infty$)
\begin{equation}
   \kl{\tilde H}{}{k}
        \ \simeq \ -{\alpha k^4\over4}
                     \left\{ {1\over2}\ln {|k^2|\over\mu^2}\kl{Q}{}{k}
                            +{1\over3}\kl{\bar Q}{}{k}
                     \right\}.
\end{equation}
and at high temperatures the leading term ($\beta^{-4}$) may be
written as
\begin{equation}
   \kl{\tilde H}{}{k}
        \ \simeq \ {\pi^2\over30\beta^4}
                    \sum^{14}_{i=1}
                    \mbox{\rm H}_i(r) \kl{T}{i}{u,K},
\end{equation}
where we have introduced the dimensionless external momentum
$K^\mu \equiv k^\mu/|\vec{k}| \equiv (r,\hat k)$. The $\mbox{\rm
H}_i(r)$ coefficients were first given in \cite{Reb91,Reb92} and
generalized to the next-to-leading order ($\beta^{-2}$) in
\cite{AlmBraFre94,BraFre98}. (They are given with the MTW sign
convention  in \cite{CamHu98,CamHu99}.)

Finally, as defined above, $\kl{N}{}{x}$ is the noise kernel
representing the random fluctuations of the thermal radiance and
$\kl{D}{}{x}$ is the dissipation kernel, describing the
dissipation of energy of the gravitational field.


\subsubsection{Near Horizon Case}

In this case, since the perturbation is taken around the
Schwarzschild spacetime, exact expressions for the corresponding
unperturbed propagators $G^\beta_{ab}[h^\pm_{\mu\nu}]$ are not
known. Therefore apart from the approximation of computing the CTP
effective action to certain order in perturbation theory, an
appropriate approximation scheme for the unperturbed Green's
functions is also required. This feature manifested itself in
York's calculation of backreaction as well, where, in writing the
$\langle T_{\mu \nu}\rangle$ on the right hand side of the
semiclassical Einstein equation in the unperturbed Schwarzschild
metric, he had to use an approximate expression for $\langle
T_{\mu\nu} \rangle$ in the Schwarzschild metric given by Page
\cite{Pag82}. The additional complication here is that while to
obtain $\langle T_{\mu\nu}\rangle$ as in York's calculation, the
knowledge of only the thermal Feynman Green's function is
required, to calculate the CTP effective action one needs the
knowledge of the full matrix propagator, which involves the
Feynman, Schwinger and Wightman functions.

It is indeed possible to construct the full thermal matrix propagator
$G^\beta_{ab}[h^\pm_{\mu\nu}]$ based on Page's approximate Feynman
Green's function by using identities relating the Feynman Green's
function with the other Green's functions with different boundary
conditions. One can then proceed to explicitly compute a CTP
effective action and hence the influence functional based on this
approximation. However, we desist from delving into such a
calculation for the following reason. Our main interest in performing
such a calculation is to identify and analyze the noise term which is
the new ingredient in the backreaction. We have mentioned that the
noise term gives a stochastic contribution $\xi^{\mu\nu}$ to the
Einstein-Langevin equation (\ref{2.11}). We had also stated that this
term is related to the variance of fluctuations in $T_{\mu\nu}$, i.e,
schematically, to $\langle T^2_{\mu\nu}\rangle$. However, a
calculation of $\langle T^2_{\mu\nu}\rangle$ in the Hartle-Hawking
state in a Schwarzschild background using the Page approximation was
performed by Phillips and Hu \cite{PhiHu01,PhiHu03} and it was shown
that though the approximation is excellent as far as $\langle
T_{\mu\nu}\rangle$ is concerned, it gives unacceptably large errors
for $\langle T^2_{\mu\nu}\rangle$ at the horizon. In fact, similar
errors will be propagated in the non-local dissipation term as well,
because both terms originate from the same source, that is, they come
from the last trace term in (\ref{eq:effective action}) which
contains terms quadratic in the Green's function. However, the
Influence Functional or CTP formalism itself does not depend on the
nature of the approximation, so we will attempt to exhibit the
general structure of the calculation without resorting to a specific
form for the Greens function and conjecture on what is to be
expected. A more accurate computation can be performed using this
formal structure once a better approximation becomes available.

The general structure of the CTP effective action arising from
the calculation of the traces in equation (\ref{eq:effective
action}) remains the same. But to write down explicit expressions
for the non-local kernels one requires the input of the explicit
form of $G^\beta_{ab}[h^\pm_{\mu\nu}]$  in the Schwarzschild
metric, which is not available in closed form. We can make some
general observations about the terms in there. The first line
containing L does not have an explicit Fourier representation as
given in the far field case, neither will $T_{(\beta)}^{\mu\nu}$
in the second line representing the zeroth order contribution to
$\langle T_{\mu\nu} \rangle$ have a perfect fluid form. The third
and fourth terms containing the remaining quadratic component of
the real part of the effective action will not have any simple or
even complicated analytic form. The symmetry properties of the
kernels $H^{\mu\nu,\alpha\beta}(x,x')$ and
$D^{\mu\nu,\alpha\beta}(x,x')$ remain intact, i.e., they are
respectively even and odd in $x,x'$. The last term in the CTP
effective action gives the imaginary part of the effective action
and the kernel $N(x,x')$ is symmetric.

Continuing our general observations from this CTP effective
action, using the connection between this thermal CTP effective
action to the influence functional \cite{SuEtal88,CalHu94} via an
equation in the schematic form  (\ref{ctpif}). We see that the
nonlocal imaginary term containing the kernel
$N^{\mu\nu,\alpha\beta}(x,x')$ is responsible for the generation
of the stochastic noise term in the Einstein-Langevin equation and
the real non-local term containing kernel
$D^{\mu\nu,\alpha\beta}(x,x')$ is responsible for the non-local
dissipation term. To derive the Einstein-Langevin equation we
first construct the stochastic effective action (\ref{stochastic
eff action}). We then derive the equation of motion, as shown
earlier in (\ref{eq of motion}), by taking its functional
derivative with respect to $[h_{\mu\nu}]$ and equating it to zero.
With the identification of noise and dissipation kernels, one can
write down a linear, non-local relation of the form, \be N(t-t') =
~\int~d(s -s')K(t-t',s-s')\gamma(s -s') \label{FDR}, \te where
$D(t,t')=-\partial_{t'}\gamma (t,t')$.  This is the general
functional form of a fluctuation-dissipation relation and $K(t,s)$
is called the fluctuation-dissipation kernel
\cite{CalLeg83,GraSchIng88,HuPazZha92,HuPazZha93}. In the present
context this relation depicts the backreaction of thermal Hawking
radiance for a black hole in quasi-equilibrium.


\subsubsection{Einstein-Langevin equation}

In this section we show how  a semiclassical Einstein-Langevin
equation  can be derived from the previous thermal CTP effective
action. This equation depicts the stochastic evolution of the
perturbations of the black hole under the influence of the
fluctuations of the thermal scalar field.

The influence functional ${\cal F}_{\rm IF} \equiv \exp (iS_{\rm
IF})$ previously introduced in Eq. (\ref{influence functional})
can be written in terms of the the CTP effective action $S_{\rm
eff} ^\beta [h^\pm_{\mu\nu}]$ derived in equation (\ref{CTPbh})
using Eq.~(\ref{ctpif}).  The Einstein-Langevin equation follows
from taking the functional derivative of the stochastic effective
action (\ref{stochastic eff action}) with respect to
$[h_{\mu\nu}](x)$ and imposing $[h_{\mu\nu}](x) = 0$. This leads
to
\begin{eqnarray}
 \! \!\!\!\!&&\!\!\!\!\! {1\over 16\pi G_N}
   \int d^4x'\ \kl{L}{(o)}{x-x'} h_{\alpha\beta}(x')
  +{1\over2}\ T^{\mu\nu}_{(\beta)}
  +\int d^4x'\ \left( \kl{H}{}{x-x'}
   \right.
   \nonumber\\
   &&\ \ \ \ \ \ \ \ \ \ \ \ \ \ \ \
  \left.
                     -\kl{D}{}{x-x'}
               \right) h_{\alpha\beta}(x')
  +\xi^{\mu\nu}(x)
         = 0.
\end{eqnarray}
where
\begin{equation}
   \langle \xi^{\mu\nu}(x) \xi^{\alpha\beta}(x') \rangle_j
        \ = \ \kl{N}{}{x-x'},
   \label{eq:correlation}
\end{equation}
In the far field limit this equation should reduce to that
obtained by Campos and Hu \cite{CamHu98,CamHu99}: For gravitational
perturbations $h^{\mu\nu}$ defined in (\ref{eq:def bar h}) under
the harmonic gauge $\bar h^{\mu\nu}_{\,\,\,\,\, ,\nu} = 0$, their
Einstein-Langevin equation is given by
\begin{eqnarray}
  \!\!\!\!\!&&\!\!\!\!\! \Box\bar h^{\mu\nu}(x)
         + {1 \over 16 \pi G_N^2}
               \left\{ T^{\mu\nu}_{(\beta)}
                      +2P_{\rho\sigma,\alpha\beta}
                       \int d^4x'\ \left( \kl{H}{}{x-x'}
                       \right.
       \right.
   \nonumber\\
   &&\ \ \ \ \ \ \ \ \ \ \ \ \ \ \ \
  \left.
                       \left.
                   -\kl{D}{}{x-x'}
                                   \right)\bar h^{\rho\sigma}(x')
                      +2\xi^{\mu\nu}(x)
               \right\} = 0,
\end{eqnarray}
where the tensor $P_{\rho\sigma,\alpha\beta}$ is given by
\begin{equation}
   P_{\rho\sigma,\alpha\beta}
        \ = \ {1\over2}\left( \eta_{\rho\alpha}\eta_{\sigma\beta}
                             +\eta_{\rho\beta}\eta_{\sigma\alpha}
                             -\eta_{\rho\sigma}\eta_{\alpha\beta}
                       \right).
\end{equation}
The expression for  $P_{\rho\sigma,\alpha\beta}$ in the near
horizon limit of course cannot be expressed in such a simple form.
Note that this differential stochastic equation includes a
non-local term responsible for the dissipation of the
gravitational field and a noise source term which accounts for
the fluctuations of the quantum field . Note also that this
equation in combination with the correlation for the stochastic
variable (\ref{eq:correlation}) determine the two-point
correlation for the stochastic metric fluctuations $\langle \bar
h_{\mu\nu}(x) \bar h_{\alpha\beta}(x') \rangle_\xi$
self-consistently.

As we have seen before and here, the Einstein-Langevin equation is
a dynamical equation governing the dissipative evolution of the
gravitational field under the influence of the fluctuations of
the quantum field, which, in the case of black holes, takes the
form of thermal radiance. From its form we can see that even for
the quasi-static case under study the back reaction of Hawking
radiation on the black hole spacetime has an innate dynamical
nature.

For the far field case making use of the explicit forms available
for the noise and dissipation kernels Campos and Hu \cite{CamHu98,CamHu99}
formally proved the existence of a fluctuation-dissipation
relation at all temperatures between the quantum fluctuations of
the thermal radiance and the dissipation of the gravitational
field. They also showed the formal equivalence of this method with
linear response theory for lowest order perturbations of a
near-equilibrium system, and how the response functions such as
the contribution of the quantum scalar field to the thermal
graviton polarization tensor can be derived. An important quantity
not usually obtained in linear response theory, but of equal
importance, manifest in the CTP stochastic approach is the noise
term arising from the quantum and statistical fluctuations in the
thermal field. The example given in this section shows that the
back reaction is intrinsically a dynamic process described (at
this level of sophistication) by the Einstein-Langevin equation.
By comparison, traditional linear response theory calculations
cannot capture the dynamics as fully and thus cannot provide a
complete description of the backreaction problem.




\subsubsection{Comments}

As remarked earlier, except for the near-flat case, an analytic form
of the Green function is not available. Even the Page approximation
\cite{Pag82} which gives unexpectedly good results for the stress
energy tensor has been shown to fail in the fluctuations of the
energy density \cite{PhiHu03}. Thus using such an approximation for
the noise kernel will give unreliable results for the
Einstein-Langevin equation.  If we confine ourselves to Page's
approximation and derive the equation of motion without the
stochastic term, we expect to recover York's semiclassical Einstein's
equation if one retains only the zeroth order contribution, i.e, the
first two terms in the expression for the CTP effective action in Eq.
(\ref{CTPbh}). Thus, this offers a new route to arrive at York's
semiclassical Einstein's equations. (Not only is it a derivation of
York's result from a different point of view, but it also shows how
his result arises as an appropriate limit of a more complete
framework, i.e, it arises when one averages over the noise.) Another
point worth noting is that a non-local dissipation term arises from
the fourth term in Eq. (\ref{CTPbh}) in the CTP effective action
which is absent in York's treatment. This difference exists primarily
due to the difference in the way backreaction is treated,  at the
level of iterative approximations on the equation of motion as in
York, versus the treatment at the effective action level as in the
influence functional approach. In York's treatment, the Einstein
tensor is computed to first order in perturbation theory, while
$\langle T_{\mu\nu}\rangle$ on the right hand side of the
semiclassical Einstein equation is replaced by the zeroth order term.
In the influence functional treatment the full effective action is
computed to second order in perturbation, and hence includes the
higher order non-local terms.

The other important conceptual point that comes to light from this
new approach is that related to the Fluctuation-Dissipation Relation.
In the quantum Brownian motion analog (e.g.,
\cite{CalLeg83,GraSchIng88,HuPazZha92,HuPazZha93} and references
therein), the dissipation of the energy of the Brownian particle as
it approaches equilibrium and the fluctuations at equilibrium are
connected by the Fluctuation-Dissipation relation. Here the
backreaction of quantum fields on black holes also consists of two
forms -- dissipation and fluctuation or noise -- corresponding to the
real and imaginary parts of the influence functional as embodied in
the dissipation and noise kernels. A fluctuation-dissipation relation
has been shown to exist for the near flat case by Campos and Hu
\cite{CamHu98,CamHu99} and is expected to exist between the noise and
dissipation kernels for the general case, as it is a categorical
relation \cite{CalLeg83,GraSchIng88,HuPazZha92,HuPazZha93,Banff}.
Martin and Verdaguer have also proved the existence of a
fluctuation-dissipation relation when the semiclassical background is
a stationary spacetime and the quantum field is in thermal
equilibrium. Their result was then extended to a conformal field in a
conformally stationary background \cite{MarVer99a}. As discussed
earlier the existence of a fluctuation-dissipation relation for the
black hole case has been suggested by some authors previously
\cite{CanSci77,Sci79,SciCanDeu81,Mottola}. This relation and the
relevant physical quantities contained therein, such as the black
hole susceptibility function which characterizes the statistical
mechanical and dynamical responses of a black hole interacting with
its quantum field environment, will allow us to study the
\textit{nonequilibrium} thermodynamic properties of the black hole,
and through it perhaps the microscopic structure of spacetime.

There are limitations of a technical nature in the quasi-static case
studied, as mentioned above, i.e., there is no reliable approximation
to the Schwarzschild thermal Green's function to explicitly compute
the noise and dissipation kernels. Another technical limitation of
this example is the following. Although we have allowed for
backreaction effects to modify the initial state in the sense that
the temperature of the Hartle-Hawking state gets affected by the
backreaction,  our analysis is essentially confined to a
Hartle-Hawking thermal state of the field. It does not directly
extend to a more general class of states, for example to the case
where the initial state of the field is in the Unruh vacuum. To study
the dynamics of a radiating black hole under the influence of a
quantum field and its fluctuations a different model and approach are
needed which we now discuss.

\subsection{Metric Fluctuations of an Evaporating Black Hole}
\label{sec8.3}


At the semiclassical gravity level of description, black hole
evaporation results from the backreaction of particle production in
Hawking effect. This is believed to be valid at least before the
Planckian scale is reached \cite{Bar81,Mas95}. However, as is
explained above, semiclassical gravity
\cite{BirDav82,Wal94,FlaWal96} is a mean field description that
neglects the fluctuations of the spacetime geometry. A number of
studies have suggested the existence of large fluctuations near black
hole horizons \cite{Sor95,Sor96,CasEtal97,Mar05} (and even
instabilities \cite{MazMot04}) with characteristic time-scales much
shorter than the black hole evaporation time. For example, Casher et
al \cite{CasEtal97} and Sorkin \cite{Sor96,SorSud99} have
concentrated on the issue of fluctuations of the horizon induced by a
fluctuating metric.  Casher et al \cite{CasEtal97} considers the
fluctuations of the horizon induced by the ``atmosphere" of high
angular momentum particles near the horizon, while Sorkin
\cite{Sor96,SorSud99} calculates fluctuations of the shape of the
horizon induced by the quantum field fluctuations under a Newtonian
approximations. A relativistic generalization of this vein is given
by Marolf ~\cite{Mar05}.  Both groups of authors came to the
conclusion that horizon fluctuations become large at scales much
larger than the Planck scale (note Ford and Svaiter \cite{ForSva97}
later presented results contrary to this claim). Though these works
do deal with backreaction, the fluctuations considered do not arise
as an explicit stochastic noise term as in stochastic gravity.
Either states which are singular on the horizon (such as the Boulware
vacuum for Schwarzschild spacetime) were explicitly considered, or
fluctuations were computed with respect to those states and found to
be large near the horizon. Whether these huge fluctuations are of a
generic nature or an artifact from the consideration of states
singular on the horizon is an issue that deserves further
investigation.  By contrast, the fluctuations for states regular on
the horizon were estimated in Ref.~\cite{WuFor99} and found to be small
even when integrated over a time of the order of the evaporation
time. These apparently contradictory claims and the fact that most
claims on black hole horizon fluctuations were based on qualitative
arguments and/or semi-quantitative estimates indicate that a more
quantitative and self-consistent description is needed. This is what
stochastic gravity theory can provide.

Such a program of research has been pursued rigorously by Hu and
Roura (HR) \cite{HuRou06b,HuRou07} In contrast to the claims made
before, they find that even for states regular on the horizon the
accumulated fluctuations become significant by the time the black
hole mass has changed substantially, but well before reaching the
Planckian regime. This result is different from those obtained in
prior studies, but in agreement with earlier work by
Bekenstein~\cite{Bek84}. The apparent difference from the
conclusions drawn in the earlier work of Hu, Raval and Sinha
~\cite{Vishu}, which was also based on stochastic gravity, will be
explained later. We begin with the evolution of the mean geometry.

\subsubsection{Evolution of Mean Geometry of an Evaporating Black Hole}

Backreaction of the Hawking radiation emitted by the black hole on
the dynamics of spacetime geometry has been studied in some detail
for spherically symmetric black holes \cite{Bar81,Mas95}. For
a general spherically-symmetric metric there always exists a system
of coordinates in which it takes the form
\begin{equation}
ds^2 = - e^{2 \psi(v,r)} ( 1 - 2 m(v,r)/r ) dv^2 + 2 e^{\psi(v,r)} dv
dr + r^2 \left( d\theta^2 + \sin^2 \theta d\varphi^2 \right)
\label{metric1}.
\end{equation}
In general this metric exhibits an \emph{apparent horizon}, where the
expansion of the outgoing radial null geodesics vanishes and which
separates regions with positive and negative expansion for those
geodesics, at those radii that correspond to (odd degree) zeroes of
the $vv$ metric component. We mark the location of the apparent
horizon by $r_{AH}(v)=2M(v)$, where $M(v)$ satisfies the equation
$2m(2M(v),v)=2M(v)$.

The non-zero components of the semiclassical Einstein equation
associated with the metric in Eq.~(\ref{metric1}) become
\begin{eqnarray}
\frac{\partial m}{\partial v} &=& 4 \pi r^2 \langle T_v^r \rangle
\label{einstein2a},\\
\frac{\partial m}{\partial r} &=& - 4 \pi r^2 \langle T_v^v \rangle
\label{einstein2b},\\
\frac{\partial \psi}{\partial r} &=& 4 \pi r \langle T_{rr} \rangle
\label{einstein2c},
\end{eqnarray}
where in the above and henceforth we use $\langle T_{\mu \nu} \rangle
$ to denote the renormalized or regularized vacuum expectation value
of the stress energy tensor $\langle \hat{T}_{\mu \nu} [g] \rangle
_{ren}$ and employ Planckian units (with $m_{p}^2=1$).

Solving Eqs.~(\ref{einstein2a})-(\ref{einstein2c}) is not easy.
However, one can introduce a useful adiabatic approximation in the
regime where the mass of the black hole is much larger than the
Planck mass, which is in any case a necessary condition for the
semiclassical treatment to be valid. What this entails is that when
$M \gg 1$ (remember that we are using Planckian units) for each value
of $v$ one can simply substitute $\langle  T_{\mu \nu} \rangle$ by
its ``parametric value'' -- by this we mean the expectation value of
the stress energy tensor of the quantum field in a Schwarzschild
black hole with a mass corresponding to $M(v)$ evaluated at that
value of $v$. This is in contrast to its dynamical value, which
should be determined by solving self-consistently the semiclassical
Einstein equation for the spacetime metric and the equations of
motion for the quantum matter fields.
This kind of approximation introduces errors of higher order in
$L_{H} \equiv B/M^2$ ($B$ is a dimensionless parameter that depends
on the number of massless fields and their spins and accounts for
their corresponding grey-body factors; it has been estimated to be of
order $10^{-4}$ \cite{Pag76}), which are very small for black holes
well above Planckian scales. These errors are due to the fact that
$M(v)$ is not constant and that, even for a constant $M(v)$, the
resulting static geometry is not exactly Schwarzschild because the
vacuum polarization of the quantum fields gives rise to a
non-vanishing $\langle \hat{T}_{ab} [g] \rangle _{ren}$
\cite{Yor85}.

The expectation value of the stress tensor for Schwarzschild
spacetime has been found to correspond to a thermal flux of radiation
(with $\langle T_v^r \rangle = L_{H} / (4 \pi r^2)$) for large radii
and of order $L_{H}$ near the horizon
\cite{Can80,Pag82,HowCan84,How84,AndHisSam95}. This shows
the consistency of the adiabatic approximation for $L_{H} \ll 1$: the
right-hand side of Eqs.~(\ref{einstein2a})-(\ref{einstein2c})
contains terms of order $L_{H}$ and higher, so that the derivatives
of $m(v,r)$ and $\psi(v,r)$ are indeed small. We note that the
natural quantum state for a black hole formed by gravitational
collapse is the Unruh vacuum, which corresponds to the absence of
incoming radiation far from the horizon. The expectation value of the
stress tensor operator for that state is finite on the future horizon
of Schwarzschild, which is the relevant one when identifying a region
of the Schwarzschild geometry with the spacetime outside the
collapsing matter for a black hole formed by gravitational collapse.

One can use the $v$ component of the stress-energy conservation
equation 
\begin{equation}
\frac{\partial \left( r^2 \langle T_v^r \rangle \right)}{\partial r}
+ r^2 \frac{\partial \langle T_v^v \rangle}{\partial v} = 0
\label{conservation1},
\end{equation}
to relate the $\langle T_v^r \rangle$ components on the horizon and
far from it. Integrating Eq.~(\ref{conservation1}) radially, one gets
\begin{equation}
(r^2 \langle T_v^r \rangle) (r=2M(v),v) = (r^2 \langle T_v^r \rangle)
(r \approx 6M(v),v) + O(L_{{H}}^2), \label{conservation2}
\end{equation}
where we considered a radius sufficiently far from the horizon, but
not arbitrarily far (\emph{i.e.} $2M(v) \ll r \ll M(v)/L_{H}$).Hence,
Eq.~(\ref{conservation2}) relates the positive energy flux radiated
away far from the horizon and the negative energy flux crossing the
horizon. Taking into account this connection between energy fluxes
and evaluating Eq.~(\ref{einstein2a}) on the apparent horizon, we
finally get the equation governing the evolution of its size:
\begin{equation}
\frac{d M}{d v} = - \frac{B}{M^2} \label{einstein3}.
\end{equation}
Unless $M(v)$ is constant, the event horizon and the apparent horizon
do not coincide. However, in the adiabatic regime their radii are
related, differing by a quantity of higher order in $L_{H}$:
$r_{EH}(v) = r_{AH}(v) \, (1 + O(L_{H}))$.

\subsubsection{Spherically-symmetric induced fluctuations}

We now consider metric fluctuations around a background metric
$g_{ab}$ that corresponds to a given solution of semiclassical
gravity. Their dynamics is governed by the Einstein-Langevin equation
\cite{HuMat95,HuSin95,CamVer96,LomMaz97}
\begin{equation}
G_{ab}^{(1)}\left[ g+h\right] =\kappa \left\langle \hat{T}_{ab}^{(1)}
[g+h] \right\rangle _{ren} +\kappa \, \xi_{ab}\left[ g\right]
\label{einst-lang1},
\end{equation}
(The superindex $(1)$ indicates that only the terms linear in the
metric perturbations should be considered) Here $\xi_{ab}$ is a
Gaussian stochastic source with vanishing expectation value and
correlation function $\langle \xi_{ab} (x) \xi_{cd} (x') \rangle_\xi
= (1/2) \langle \{ \hat{t}_{ab} (x), \hat{t}_{cd} (x') \} \rangle$
(with $\hat{t}_{ab} \equiv \hat{T}_{ab} - \langle \hat{T}_{ab}
\rangle$), known as the noise kernel and denoted by $N_{abcd}(x,x')$.

As explained earlier the symmetrized two-point function consists of
two contributions: \emph{intrinsic} and \emph{induced} fluctuations.
The intrinsic fluctuations are a consequence of the quantum width of
the initial state of the metric perturbations, and they are obtained
in stochastic gravity by averaging over the initial conditions for
the solutions of the homogeneous part of Eq.~(\ref{einst-lang1})
distributed according to the reduced Wigner function associated with
the initial quantum state of the metric perturbations. On the other
hand, the induced fluctuations are due to the quantum fluctuations of
the matter fields interacting with the metric perturbations, and they
are obtained by solving the Einstein-Langevin equation using a
retarded propagator with vanishing initial conditions.

In this section we study the spherically-symmetric sector
[\emph{i.e.}, the monopole contribution, which corresponds to $l=0$,
in a multipole expansion in terms of spherical harmonics
$Y_{lm}(\theta,\phi)$] of metric fluctuations for an evaporating
black hole.  Restricting one's attention to the spherically-symmetric
sector of metric fluctuations necessarily implies a partial
description of the fluctuations because, contrary to the case for
semiclassical gravity solutions, even if one starts with
spherically-symmetric initial conditions, the stress tensor
fluctuations will induce fluctuations involving higher multipoles.
Thus, the multipole structure of the fluctuations is far richer than
that of spherically-symmetric semiclassical gravity solutions, but
this also means that obtaining a complete solution (including all
multipoles) for fluctuations rather than the mean value is much more
difficult. For spherically symmetric fluctuations only induced
fluctuations are possible. The fact that intrinsic fluctuations
cannot exist can be clearly seen if one neglects vacuum polarization
effects, since Birkhoff's theorem forbids the existence of
spherically-symmetric free metric perturbations in the exterior
vacuum region of a spherically-symmetric black hole that keep the ADM
mass constant. (This fact rings an alarm in the approach taken in
Ref.~\cite{YorBjo05} to the black hole fluctuation problem. The degrees
of freedom corresponding to spherically-symmetric perturbations are
constrained by the Hamiltonian and momentum constraints both at the
classical and quantum level. Therefore, they will not exhibit quantum
fluctuations unless they are coupled to a quantum matter field.) Even
when vacuum polarization effects are included, spherically-symmetric
perturbations, characterized by $m(v,r)$ and $\psi(v,r)$, are not
independent degrees of freedom. This follows from
Eqs.~(\ref{einstein2a})-(\ref{einstein2c}), which can be regarded as
constraint equations.

Considering only spherical symmetry fluctuations is a simplification
but it should be emphasized that it gives more accurate results than
two-dimensional dilaton-gravity models resulting from simple
dimensional reduction \cite{Tri93,StrTri93,LomMazRus99}. This
is because we project the solutions of the Einstein-Langevin equation
just at the end, rather than considering only the contribution of the
$s$-wave modes to the classical action for both the metric and the
matter fields from the very beginning. Hence, an infinite number of
modes for the matter fields with $l \neq 0$ contribute to the $l = 0$
projection of the noise kernel, whereas only the $s$-wave modes for
each matter field would contribute to the noise kernel if dimensional
reduction had been imposed right from the start, as done in
Refs.~\cite{Par01,Par01b,Par02} as well as in
studies of two-dimensional dilaton-gravity models.

The Einstein-Langevin equation for the spherically-symmetric sector
of metric perturbations can be obtained by considering linear
perturbations of $m(v,r)$ and $\psi(v,r)$, projecting the stochastic
source that accounts for the stress tensor fluctuations to the $l=0$
sector, and adding it to the right-hand side of
Eqs.~(\ref{einstein2a})-(\ref{einstein2c}). We will focus our
attention on the equation for the evolution of $\eta(v,r)$, the
perturbation of $m(v,r)$:
\begin{equation}
\frac{\partial (m + \eta)}{\partial v} = - \frac{B}{(m + \eta)^2} + 4
\pi r^2 \xi_v^r + O \left(L_{H}^2 \right) \label{einst-lang2},
\end{equation}
which reduces, after neglecting terms of order $L_{H}^2$ or higher,
to the following equation to linear order in $\eta$:
\begin{equation}
\frac{\partial \eta}{\partial v} = \frac{2 B}{m^3} \eta + 4 \pi r^2
\xi_v^r \label{einst-lang3}.
\end{equation}
It is important to emphasize that in Eq.~(\ref{einst-lang2}) we
assumed that the change in time of $\eta(v,r)$ is sufficiently slow
so that the adiabatic approximation employed in the previous section
to obtain the mean evolution of $m(v,r)$ can also be applied to the
perturbed quantity $m(v,r)+\eta(v,r)$. This is guaranteed as long as
the term corresponding to the stochastic source is of order $L_{H}$
or higher, a point which will be discussed below.


A more serious issue raised by HR is that in most previous
investigations \cite{Bek84,WuFor99} of the problem of metric
fluctuations driven by quantum matter field fluctuations of states
regular on the horizon (as far as the expectation value of the stress
tensor is concerned) most authors assumed the existence of
correlations between the outgoing energy flux far from the horizon
and a negative energy flux crossing the horizon. (See, however,
Refs.~\cite{Par01b,Par02}, where those correlators were
shown to vanish in an effectively two-dimensional model.)  In
semiclassical gravity, using energy conservation arguments, such
correlations have been confirmed for the expectation value of the
energy fluxes, provided that the mass of the black hole is much
larger than the Planck mass. However, a more careful analysis by HR
shows that no such simple connection exists for energy flux
fluctuations. It also reveals that the fluctuations on the horizon
are in fact divergent. This requires one modifies the classical
picture of the event horizon from a sharply defined three-dimensional
hypersurface to that possessing a finite  width, i.e., a fluctuating
geometry. One needs to find an appropriate way of probing the metric
fluctuations near the horizon and extracting physically meaningful
information. It also testifies to the necessity of a complete
reexamination of all cases afresh and that an evaluation of the noise
kernel near the horizon seems unavoidable for the consideration of
fluctuations and back-reaction issues.

Having registered this cautionary note, Hu and Roura \cite{HuRou07}
first make the assumption that a relation between the fluctuations of
the fluxes exists, so as to be able to compare with earlier work.
They then show that this relation does not hold and discuss the
essential elements required in understanding not only the
mathematical theory but also the operational meaning of metric
fluctuations.

\paragraph{Assuming that there is a relation between fluctuations}

Since the generation of Hawking radiation is especially sensitive to
what happens near the horizon, from now on we will concentrate on the
metric perturbations near the horizon and consider $\eta(v) =
\eta(v,2M(v))$. This means that possible effects on the Hawking
radiation due to the fluctuations of the potential barrier for the
radial mode functions will be missed by our analysis. Assuming that
the fluctuations of the energy flux crossing the horizon and those
far from it are exactly correlated, from Eq.~(\ref{einst-lang3}) we
have
\begin{equation}
\frac{d \eta(v)}{d v} = \frac{2 B}{M^3(v)} \eta(v) + \xi(v)
\label{einst-lang4},
\end{equation}
where $\xi(v) \equiv (4 \pi r^2\, \xi_v^r) (v,r \approx 6M(v))$. The
correlation function for the spherically-symmetric fluctuation
$\xi(v)$ is determined by the integral over the whole solid angle of
the $N^{r\;r}_{\;v\;v}$ component of the noise kernel, which is given
by $(1/2) \langle \{ \hat{t}_v^r (x), \hat{t}_v^r (x') \} \rangle$.
The $l=0$ fluctuations of the energy flux of Hawking radiation far
from a black hole formed by gravitational collapse, characterized
also by $(1/2) \langle \{ \hat{t}_v^r (x), \hat{t}_v^r (x') \}
\rangle$ averaged over the whole solid angle, have been studied in
Ref.~\cite{WuFor99}. Its main features are a correlation time of order
$M$ and a characteristic fluctuation amplitude of order $\epsilon_0 /
M^4$ (this is the result of smearing the stress tensor two-point
function, which diverges in the coincidence limit, over a period of
time of the order of the correlation time). The order of magnitude of
$\epsilon_0$ has been estimated to lie between $0.1 B$ and $B$
\cite{Bek84,WuFor99}.  For simplicity, we will consider
quantities smeared over a time of order $M$. We can then introduce
the Markovian approximation $(\epsilon_0 / M^3(v)) \delta(v-v')$,
which coarse-grains the information on features corresponding to
time-scales shorter than the correlation time $M$. Under those
conditions $r^2 \xi^r_v$ is of order $1/M^2$ and thus the adiabatic
approximation made when deriving Eq.~(\ref{einst-lang2}) is
justified.

The stochastic equation (\ref{einst-lang4}) for $\eta$ can be solved
in the usual way and the correlation function for $\eta(v)$ can then
be computed. Alternatively one can obtain an equation for $\langle
\eta^2 (v) \rangle_\xi$ by first multiplying Eq.~(\ref{einst-lang4})
by $\eta(v)$ and then taking the expectation value. This brings out a
term $\langle \eta (v) \xi (v) \rangle_\xi$ on the right hand side.
For delta-correlated noise (the Stratonovich prescription is the
appropriate one here), it is equal to one half the time-dependent
coefficient multiplying the delta function $\delta (v-v')$ in the
correlator $\langle \xi (v) \xi (v') \rangle_\xi$, which is given by
$\epsilon_0 / M^3(v)$ in our case. Finally, changing from the $v$
coordinate to the mass function $M(v)$ for the background solution,
we obtain
\begin{equation}
\frac{d}{dM} \langle \eta^2 (M) \rangle_\xi = - \frac{4}{M} \langle
\eta^2 (M) \rangle_\xi - \frac{(\epsilon_0 / B)}{M} \label{fluct3}.
\end{equation}
The solutions of this equation are given by
\begin{equation}
\langle \eta^2 (M) \rangle_\xi = \langle \eta^2 (M_0) \rangle_\xi
\left(\frac{M_0}{M}\right)^4 +\frac{\epsilon_0}{4 B}
\left[\left(\frac{M_0}{M}\right)^4 - 1\right] \label{fluct4}.
\end{equation}
Provided that the fluctuations at the initial time corresponding to
$M=M_0$ are negligible (much smaller than $\sqrt{\epsilon_0 / 4B}
\sim 1$), the fluctuations become comparable to the background
solution when $M \sim M_0^{2/3}$. Note that fluctuations of the
horizon radius of order one in Planckian units do not correspond to
Planck scale physics because near the horizon $\Delta R = r - 2M$
corresponds to a physical distance $L \sim \sqrt{M \, \Delta R}$, as
can be seen from the line element for Schwarzschild, $ds^2 = -
(1-2M/r) dt^2 + (1-2M/r)^{-1} dr^2 + r^2 (d\theta^2 + \sin^2 \theta
d\varphi^2)$, by considering pairs of points at constant $t$. So
$\Delta R \sim 1$ corresponds to $L \sim \sqrt{M}$, whereas a
physical distance of order one is associated with $\Delta R \sim
1/M$, which corresponds to an area change of order one for spheres
with those radii. One can, therefore, have initial fluctuations of
the horizon radius of order one for physical distances well above the
Planck length for a black hole with a mass much larger than the
Planck mass. One expects that the fluctuations for states that are
regular on the horizon correspond to physical distances not much
larger than the Planck length, so that the horizon radius
fluctuations would be much smaller than one for sufficiently large
black hole masses. Nevertheless, that may not be the case when
dealing with states which are singular on the horizon, with estimated
fluctuations of order $M^{1/3}$ or even $\sqrt{M}$
\cite{CasEtal97,Mar05,MazMot04}.

The result of HR for the growth of the fluctuations of the size of
the black hole horizon agrees with the result obtained by Bekenstein
in Ref.~\cite{Bek84} and implies that, for a sufficiently
massive black hole (with a few solar masses or a supermassive black
hole), the fluctuations become important before the Planckian regime
is reached.

This growth of the fluctuations which was found by Bekenstein and
confirmed here via the Einstein-Langevin equation seems to be in
conflict with the estimate given by Wu and Ford in Ref.~\cite{WuFor99}.
According to their estimate, the accumulated mass fluctuations over a
period of the order of the black hole evaporation time ($\Delta t
\sim M_0^3$) would be of the order of the Planck mass. The
discrepancy is due to the fact that the first term on the right-hand
side of Eq.~(\ref{einst-lang4}), which corresponds to the perturbed
expectation value $\langle \hat{T}_{ab}^{(1)} [g+h] \rangle _{ren}$
in Eq.~(\ref{einst-lang1}), was not taken into account in
Ref.~\cite{WuFor99}. The larger growth obtained here is a consequence of
the secular effect of that term, which builds up in time (slowly at
first, during most of the evaporation time, and becoming more
significant at late times when the mass has changed substantially)
and reflects the unstable nature of the background solution for an
evaporating black hole.

As for the relation between HR's results reported here and earlier
results of Hu, Raval and Sinha in Ref.~\cite{Vishu}, there should not
be any discrepancy since both adopted the stochastic gravity
framework and performed their analysis based on the Einstein-Langevin
equation. The claim in Ref.~\cite{Vishu} was based on a qualitative
argument that focused on the dynamics of the stochastic source alone.
If one adds in the consideration that the perturbations around the
mean are unstable for an evaporating black hole, their results agree.

All this can be qualitatively understood as 
follows. Consider an evaporating black hole with initial mass $M_0$
and suppose that the initial mass is perturbed by an amount $\delta
M_0 = 1$. The mean evolution for the perturbed black hole (without
taking into account any fluctuations) leads to a mass perturbation
that grows like $\delta M = (M_0/M)^2 \, \delta M_0 = (M_0/M)^2$, so
that it becomes comparable to the unperturbed mass $M$ when $M \sim
M_0^{2/3}$, which coincides with the result obtained above. Such a
coincidence has a simple explanation: the fluctuations of the Hawking
flux slowly accumulated during most of the evaporating time, which
are of the order of the Planck mass, as found by Wu and Ford, give a
dispersion of that order for the mass distribution at the time when
the instability of the small perturbations around the background
solution start to become significant.

\paragraph{When no such relation exists and consequences}

For conformal fields in two dimensional spacetimes, HR showed that
the correlations between the energy flux crossing the horizon and the
flux far from it vanish. The correlation function for the outgoing
and ingoing null energy fluxes in an effectively two-dimensional
model was explicitly computed in
Refs.~\cite{Par01b,Par02} and it was also found to
vanish. On the other hand, in four dimensions the correlation
function does not vanish in general and correlations between outgoing
and ingoing fluxes do exist near the horizon (at least partially).

For black hole masses much larger than the Planck mass, one can use
the adiabatic approximation for the background mean evolution.
Therefore, to lowest order in $L_{H}$ one can compute the
fluctuations of the stress tensor in Schwarzschild spacetime. In
Schwarzschild, the amplitude of the fluctuations of $r^2 \langle
T_v^r \rangle $ far from the horizon is of order $1/M^2$ ($= M^2 /
M^4$) when smearing over a correlation time of order $M$, which one
can estimate for a hot thermal plasma in flat space
\cite{CamHu98,CamHu99}
(see also Ref.~\cite{WuFor99} for a computation of the fluctuations of
$r^2 \langle T_v^r \rangle $ far from the horizon). The amplitude of
the fluctuations of $r^2 \langle T_v^r \rangle $ is thus of the same
order as its expectation value. However, their derivatives with
respect to $v$ are rather different: since the characteristic
variation times
for the expectation value and the fluctuations are $M^3$ and $M$
respectively, $\partial (r^2 \langle T_v^r \rangle) / \partial v$ is
of order $1/M^5$ whereas $\partial (r^2 \xi^r_v) / \partial v$ is of
order $1/M^3$. This implies an additional contribution of order
$L_{H}$ due to the second term in Eq.~(\ref{conservation1}) if one
radially integrates the same equation applied to stress tensor
fluctuations (the stochastic source in the Einstein-Langevin
equation). Hence, in contrast to the case of the mean value, the
contribution from the second term in Eq.~(\ref{conservation1}) cannot
be neglected when radially integrating since it is of the same order
as the contributions from the first term, and one can no longer
obtain a simple relation between the outgoing energy flux far from
the horizon and the energy flux crossing the horizon.

What then?  Without this convenience (which almost all earlier
researchers had taken for granted) to get a more precise depiction we
need to compute the noise kernel near the horizon. However, as shown
by Hu and Phillips earlier \cite{PhiHu03} when they examine the
coincidence limit of the noise kernel, and confirmed by the careful
analysis of Hu and Roura using  smearing functions \cite{HuRou07},
the noise kernel smeared over the horizon is divergent, and so are
the induced metric fluctuations. Hence, one cannot study the
fluctuations of the horizon as a three-dimensional hypersurface for
each realization of the stochastic source because the amplitude of
the fluctuations is infinite, even when restricting one's attention
to the $l=0$ sector. Instead one should regard the horizon as
possessing a finite effective width due to quantum fluctuations. In
order to characterize its width one must find a sensible way of
probing the metric fluctuations near the horizon and extracting
physically meaningful information, such as their effect on the
Hawking radiation emitted by the black hole. How to probe metric
fluctuations is an issue at the root base which needs be dealt with
in all discussions of metric fluctuations.

\subsubsection{Summary and prospects}

The work of Hu and Roura \cite{HuRou07} based on the stochastic
gravity program found that the spherically-symmetric fluctuations of
the horizon size of an evaporating black hole become important at
late times, and even comparable to its mean value when $M \sim
M_0^{2/3}$, where $M_0$ is the mass of the black hole at some initial
time when the fluctuations of the horizon radius are much smaller
than the Planck length (remember that for large black hole
masses this can still correspond to physical distances much larger
than the Planck length, as explained in Sec.~\ref{sec3}). This is
consistent with the result previously obtained by Bekenstein in
Ref.~\cite{Bek84}.

It is important to realize that for a sufficiently massive black
hole, the fluctuations become significant well before the Planckian
regime is reached. More specifically, for a solar mass black hole
they become comparable to the mean value when the black hole radius
is of the order of $10 {nm}$, whereas for a supermassive black hole
with $M \sim 10^7 M_\odot$, that happens when the radius reaches a
size of the order of $1 {mm}$. One expects that in those
circumstances the low-energy effective field theory approach of
stochastic gravity should provide a reliable description.

Due to the nonlinear nature of the back-reaction equations, such as
Eq.~(\ref{einst-lang2}), the fact that the fluctuations of the
horizon size can grow and become comparable to the mean value implies
non-negligible corrections to the dynamics of the mean value itself.
This can be seen by expanding Eq.~(\ref{einst-lang2}) (evaluated on
the horizon) in powers of $\eta$ and taking the expectation value.
Through order $\eta^2$ we get
\begin{eqnarray}
\frac{d (M(v) + \langle \eta(v) \rangle_\xi)}{d v} &=& - \left
\langle \frac{B}{(M(v) + \eta(v))^2} \right \rangle_\xi
\nonumber \\
&=& - \frac{B}{M^2(v)} \left[ 1 - \frac{2}{M(v)} \langle \eta (v)
\rangle_\xi + \frac{3}{M^2(v)} \langle \eta^2 (v) \rangle_\xi + O
\left( \frac{\eta^3}{M^3} \right) \right]  \label{rad_correction}.
\end{eqnarray}
When the fluctuations become comparable to the mass itself, the third
term (and higher order terms) on the right-hand side is no longer
negligible and we get non-trivial corrections to
Eq.~(\ref{einstein3}) for the dynamics of the mean value. These
corrections can be interpreted as higher order radiative corrections
to semiclassical gravity that include the effects of metric
fluctuations on the evolution of the mean value.

Finally we remark on the relation of this finding with earlier
well-known results. Does the existence of the significant deviations
for the mean evolution mentioned above invalidate the earlier results
by Bardeen and Massar based on semiclassical gravity in
Refs.~\cite{Bar81,Mas95}? First, those deviations start to
become significant only after a period of the order of the
evaporation time when the mass of the black hole has decreased
substantially. Second, since fluctuations were not considered in
those references, a direct comparison cannot be established.
Nevertheless,we can compare the average of the fluctuating ensemble.
Doing so exhibits an evolution that deviates significantly when the
fluctuations become important. However, if one considers a single
member of the ensemble at that time, its evolution will be accurately
described by the corresponding semiclassical gravity solution until
the fluctuations around that particular solution become important
again, after a period of the order of the evaporation time associated
with the new initial value of the mass at that time.


\subsection{Other work on metric fluctuations but without
backreaction}

In closing we mention some work on metric fluctuations where no
backreaction is considered. Barrabes et al
\cite{BarFroPar99,BarFroPar00} have considered the propagation of
null rays and massless fields in a black hole fluctuating geometry
and have shown that the stochastic nature of the metric leads to a
modified dispersion relation and helps to confront the
trans-Planckian frequency problem. However, in this case the
stochastic noise is put in by hand and does not naturally arise from
coarse graining as in a quantum open systems approach, in terms of
which stochastic gravity can be interpreted. It also does not take
backreaction into account. It will be interesting to explore how a
stochastic black hole metric, arising as a solution to the
Einstein-Langevin equation, hence fully incorporating backreaction,
would affect the trans-Planckian problem.

As mentioned earlier, Ford and his collaborators
\cite{ForSva97,ForWu,WuFor99} have also explored the issue of metric
fluctuations in detail and in particular have studied the
fluctuations of the black hole horizon induced by metric
fluctuations. However, the fluctuations they considered are in the
context of a fixed background and do not relate to the backreaction.


Another work on metric fluctuations with no backreaction is that of
Hu and Shiokawa \cite{HuShi98}, who study effects associated with
electromagnetic wave propagation in a Robertson-Walker universe and
the Schwarzschild spacetime with a small amount of given metric
stochasticity. They find that time-independent randomness can
decrease the total luminosity of Hawking radiation due to multiple
scattering of waves outside the black hole and gives rise to event
horizon fluctuations and fluctuations in the Hawking temperature. The
stochasticity in a background metric in their work is assumed rather
than derived (as induced by quantum field fluctuations). But it is
interesting to compare their results with that obtained in stochastic
gravity with backreaction, so one can begin to get a sense of the
different sources of stochasticity and their weights (see, e.g.,
\cite{stogra} for a list of possible sources of stochasticity.)

In a subsequent paper Shiokawa \cite{Shio} showed that the scalar
and spinor waves in a stochastic spacetime behave similarly to
the electrons in a disordered system. Viewing this as a quantum
transport problem, he expressed the conductance and its
fluctuations in terms of a nonlinear sigma model in the closed
time path formalism and showed that the conductance fluctuations
are universal, independent of the volume of the stochastic region
and the amount of stochasticity. This result can have significant
importance in characterizing the mesoscopic behavior of
spacetimes resting between the semiclassical and the quantum
regimes.

\section{Concluding Remarks}

In the first part of this review on the fundamentals of the theory we
have given two routes to the establishment of stochastic gravity and
derived a general (finite) expression for the noise kernel. In the
second part we gave three applications, the correlation functions of
gravitons in a perturbed Minkowski metric, structure formation in
stochastic gravity theory and the outline of a program for the study
of black hole fluctuations and backreaction.
We have also discussed the problem of the validity of semiclassical gravity,
a central issue which stochastic gravity is in a unique position to address.

We have also pointed out a number of ongoing research related to the topics
discussed in this review.
Such as the equivalence of the correlation functions for the
metric perturbations obtained using the Einstein-Langevin equations
and the quantum correlation functions that follow from a purely quantum
field theory calculation up to leading order in the large $N$ limit.
Or the calculation of the spectrum of the metric fluctuations
in inflationary models driven by the trace anomaly due to conformally coupled fields.
The related problem of the runaway solutions in the backreaction equations. The issue of the
coincidence limit in the noise kernel for the black hole fluctuations.

Theoretically, stochastic gravity is at the frontline of the
`bottom-up' approach to quantum gravity
\cite{grhydro,stogra,kinQG,Hu07}. pathway or angle starting from the
well-defined and well-understood theory of semiclassical gravity.
Structurally, as can be seen from the issues discussed and the
applications given, stochastic gravity has a very rich constituency
because it is based on quantum field theory and nonequilibrium
statistical mechanics in a curved spacetime context. The open systems
concepts and the closed-time-path / influence functional methods
constitute an extended framework suitable for treating the
backreaction and fluctuations problems of dynamical spacetimes
interacting with quantum fields. We have seen applications to
cosmological structure formation and black hole backreaction from
particle creation. A more complete understanding of the backreaction
of Hawking radiation in a fully dynamical black hole situation will
enable one to address fundamental issues such as the black hole end
state and information loss puzzles. The main reason why this program has
not progressed as swiftly as desired is due more to technical rather
than programatic difficulties (such as finding reasonable analytic
approximations for the Green function or numerical evaluation of
mode-sums near the black hole horizon). Finally, the multiplex
structure of this theory could be used to explore new lines of
inquiry and launch new programs of research, such as {\it
nonequilibrium} black hole thermodynamics and the microscopic
structures of spacetime.

\section{Acknowledgements}

The materials presented here originated from research work of BLH
with Antonio Campos, Nicholas Phillips, Alpan Raval, Albert Roura and
Sukanya Sinha, and of EV with Rosario Martin and Albert Roura. We
thank them as well as Daniel Arteaga, Andrew Matacz,  Tom Shiokawa,
and Yuhong Zhang for fruitful collaboration and their cordial
friendship since their Ph. D. days. We enjoy lively discussions with
our friends and colleagues Esteban Calzetta, Diego Mazzitelli and
Juan Pablo Paz whose work in the early years contributed toward the
establishment of this field. We acknowledge useful discussions with
Paul Anderson, Larry Ford, Ted Jacobson, Emil Mottola, Renaud
Parentani, Raphael Sorkin and Richard Woodard. This work is supported
in part by NSF grant PHY06-01550, the MEC Research projects
FPA-2004-04582C02 and FPA-2007-66665C02 and by DURSI 2005SGR00082.


\end{document}